\documentclass[a4paper,12pt,titlepage]{report}
\usepackage[utf8]{inputenc}
\usepackage{amsmath}
\usepackage{amsmath,empheq}
\usepackage{amsfonts}
\usepackage{amssymb}

\usepackage{mathrsfs}
\usepackage{latexsym}
\usepackage{footnote}
\usepackage{setspace}
\usepackage{mathtools}
\usepackage{multirow}
\usepackage{longtable}
\usepackage{bibentry}

\usepackage{cmap}
\usepackage[T1]{fontenc}
\usepackage{textcomp}
\usepackage[margin=2.5cm]{geometry}
\usepackage{graphicx}
\usepackage{scalerel}
\usepackage{color}
\usepackage{caption}
\usepackage{subcaption}
\usepackage{enumitem}
\usepackage{stackengine}
\usepackage{wallpaper}

\usepackage[tight]{units}
\usepackage{fancyhdr}
\usepackage{bbold}
\usepackage{booktabs}
\usepackage{siunitx}
\usepackage{cite}
\usepackage{subcaption}
\usepackage[labelformat=parens,labelsep=quad,skip=3pt]{caption}
\usepackage[toc,page]{appendix}

\usepackage[printonlyused]{acronym}

\usepackage{cancel} 
\usepackage{hyperref}
\hypersetup{colorlinks=true, linkcolor=blue, citecolor=blue, filecolor=blue, pagecolor=blue, urlcolor=red,
            pdfauthor={Nome do Autor},
            pdftitle={Título do Projeto},
            pdfsubject={Assunto do Projeto},
            pdfkeywords={palavra-chave, palavra-chave, palavra-chave},
            pdfproducer={Latex},
            pdfcreator={pdflatex}}

\usepackage{pdfpages} 

\begin{document}

\pagenumbering{roman}\setcounter{page}{2}\pagestyle{plain}

\begin{titlepage}
\begin{center}

\includegraphics[scale=0.5]{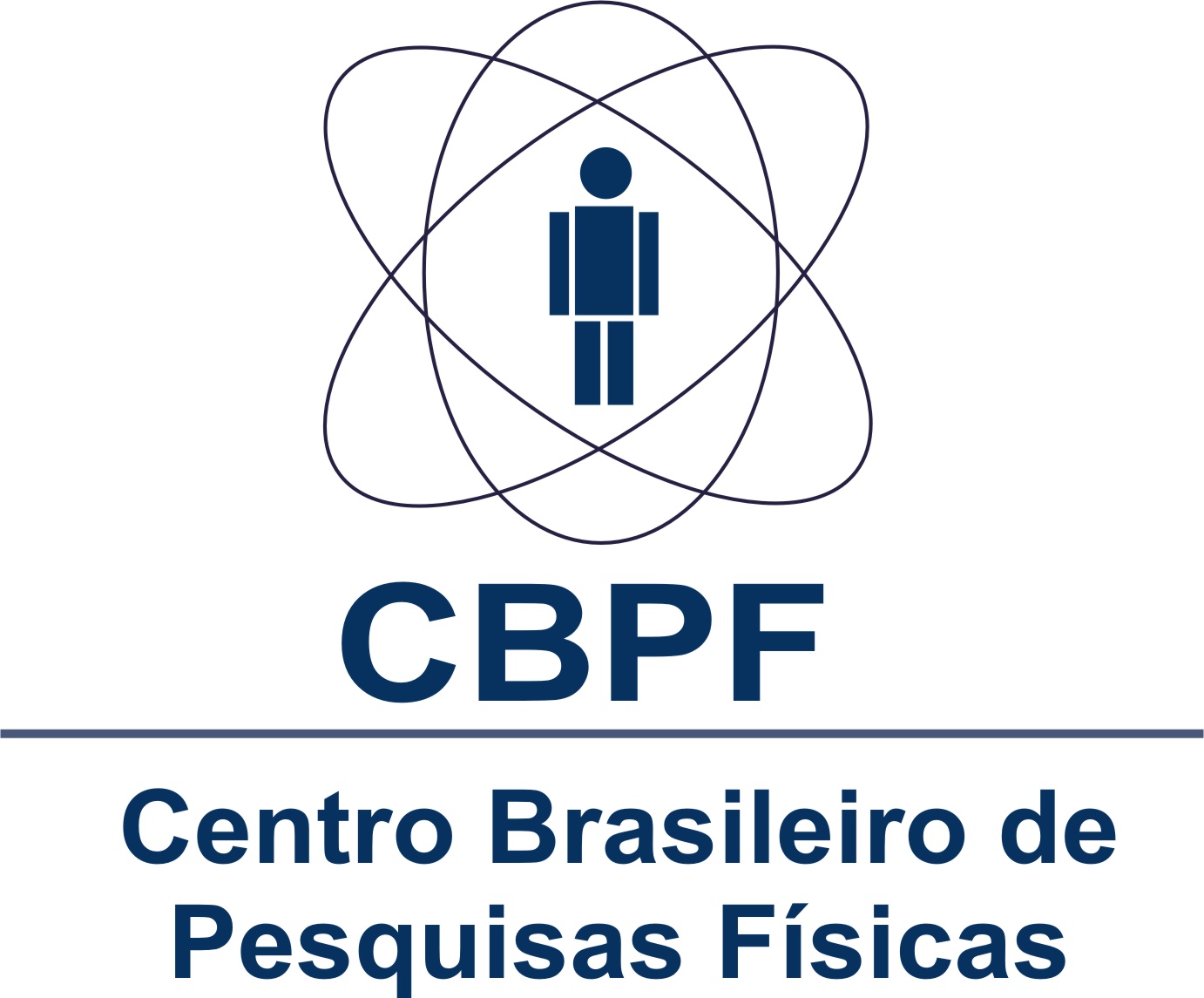}\\[2.0cm]

\Huge{\bfseries{Search for new physics in light of interparticle potentials and a very heavy dark matter candidate}}\\[5.0cm]

\large{\textbf{Felipe Almeida Gomes Ferreira}}\\[0.7cm]
\large{\textbf{Supervisor: Prof. Carsten Hensel}}\\[2.0cm]

\vfill

\large{\bf{Rio de Janeiro, July 2019}}

\end{center}
\end{titlepage}

\newpage




\includepdf[page=-]{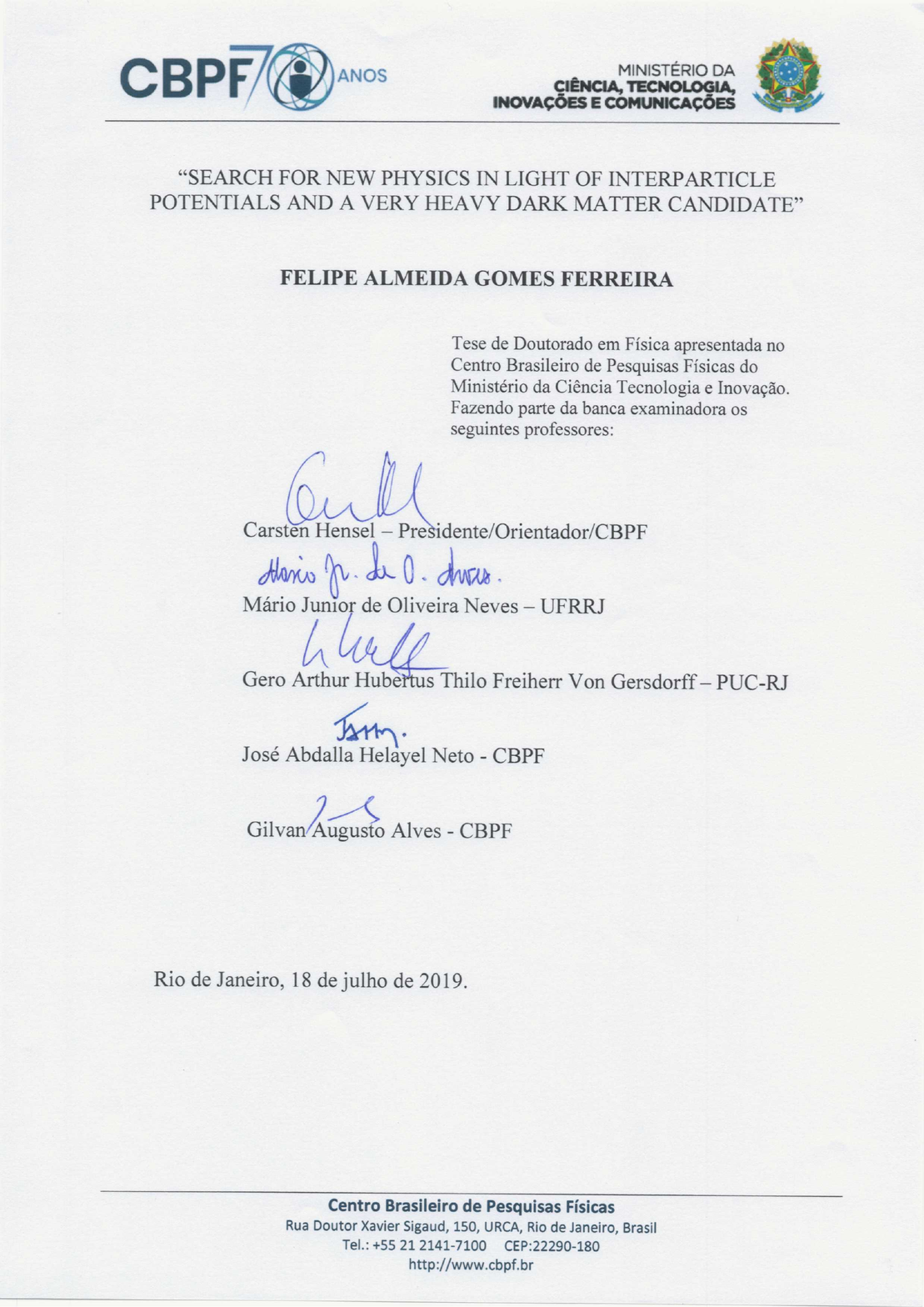}

\begin{center}
\noindent\rule{16cm}{2.0pt}
\end{center}

\begin{center}
\Large \textbf{\Huge List of Publications}
\end{center}

\begin{center}
\noindent\rule{16cm}{2.0pt}
\end{center}

\bigskip
\bigskip
\bigskip
\bigskip

This thesis is based on the following scientific articles:

\begin{itemize}
\item Manuel~Drees and Felipe~A.~Gomes Ferreira,
\textbf{A very heavy sneutrino as viable thermal dark matter candidate in $U(1)'$ extensions of the MSSM},
\href{https://doi.org/10.1007/JHEP04(2019)167}{JHEP04(2019)167}
\href{http://arxiv.org/abs/arXiv:1711.00038}{[arXiv:1711.00038]}.

\item F.~A.~Gomes~Ferreira, P.~C.~Malta, L.~P.~R.~Ospedal and J.~A.~Helayël-Neto, \\
\textbf{Topologically Massive Spin-1 Particles and Spin-Dependent Potentials},
\href{https://doi.org/10.1140/epjc/s10052-015-3470-1}{Eur.Phys.J. C75 (2015) no.5, 238}
\href{http://arxiv.org/abs/arXiv:1411.3991}{[arXiv:1411.3991]}.
\end{itemize}

\newpage

\begin{center}
\section*{\Huge \bf Abstract}
\end{center}
\addcontentsline{toc}{section}{Abstract} 
\indent
\bigskip
\bigskip
\bigskip
\bigskip

It is generally well known that the Standard Model of particle physics is not the ultimate theory of fundamental interactions as it has inumerous unsolved problems, so it must be extended. Deciphering the nature of dark matter remains one of the great challenges of contemporary physics. Supersymmetry is probably the most attractive extension of the SM as it can simultaneously provide a natural solution to the hierarchy problem and unify the gauge couplings at the GUT scale in such a way that doesn't affect its low-energy phenomenology. Furthermore, the lightest supersymmetric particle is one of the most popular candidates for the dark matter particle.
 
In the first part of this thesis we study the interparticle potentials generated by the interactions between spin-1/2 sources that are mediated by spin-1 particles in the limit of low momentum transfer. We investigate different representations of spin-1 particle to see how it modifies the profiles of the interparticle potentials and we also include in our analysis all types of couplings between fermionic currents and the mediator boson. The comparison between the well-known case of the Proca field and that of an exchanged spin-1 boson (with gauge-invariant mass) described by a 2-form potential mixed with a 4-vector gauge field is established in order to pursue an analysis of spin- as well as velocity-dependent profiles of the interparticle potentials. We discuss possible applications and derive an upper bound on the product of vector and pseudo-tensor coupling constants. The spin- and velocity-dependent interparticle potentials that we obtain can be used to explain effects possibly associated to new macroscopic forces such as modifications of the inverse-square law and possible spin-gravity coupling effects.

The second part of this thesis is based on the dark matter phenomenology of well-motivated $U(1)'$ extensions of the Minimal Supersymmetric Standard Model. In these models the right-handed sneutrino is a good DM candidate whose dark matter properties are in agreement with the present relic density and current experimental limits on the DM-nucleon scattering cross section. The RH sneutrino can annihilate into lighter particles via the exchange of massive gauge and Higgs bosons through s-channel processes. In order to see how heavy can the RH sneutrino be as a viable thermal dark matter candidate we explore its DM properties in the parameter region that minimize its relic density via resonance effects and thus allows it to be a heavier DM particle. We found that the RH sneutrino can behave as a good DM particle within minimal cosmology even with masses of the order of tens of TeV, which is much above the masses that viable thermal DM candidates usually have in most of dark matter particle models.

\bigskip
\bigskip
\bigskip
\bigskip
\bigskip
\bigskip
\bigskip
\bigskip
\bigskip
\bigskip
\bigskip
\bigskip
\bigskip
\bigskip
\bigskip
\bigskip
\bigskip
\bigskip
\bigskip
\bigskip
\bigskip
\bigskip
\bigskip
\bigskip
\bigskip
\bigskip
\bigskip
\bigskip
\bigskip
\bigskip
\bigskip
\bigskip
\bigskip
\bigskip
\bigskip
\bigskip
\bigskip
\bigskip
\bigskip
\bigskip
\bigskip
\bigskip

\textbf{Keywords:} Spin-dependent potentials; cosmology of theories beyond the Standard Model, Supersymmetric Standard Model, supersymmetry phenomenology.

\newpage

\begin{center}
\section*{\Huge \bf Resumo}
\end{center}
\addcontentsline{toc}{section}{Resumo} 
\indent
\bigskip
\bigskip
\bigskip
\bigskip

É geralmente bem conhecido que o Modelo Padrão da física de partículas não é a teoria final das interações fundamentais, pois tem inúmeros problemas não resolvidos, portanto, deve ser estendido. Decifrar a natureza da matéria escura continua sendo um dos grandes desafios da física contemporânea. A Supersimetria é provavelmente a mais atraente extensão do Modelo Padrão pois pode simultaneamente fornecer uma solução natural para o problema da hierarquia e unificar os acoplamentos de calibre na escala de Grande Unificação de uma tal maneira que não afeta a sua fenomenologia de baixas energias. Além disso, a partícula supersimétrica mais leve é um dos candidatos mais populares para a partícula de matéria escura.

Na primeira parte desta tese estudamos os potenciais interpartículas gerados pelas interações entre correntes de spin 1/2 que são mediadas por partículas de spin 1 no limite de baixa transferência de momento. Investigamos diferentes representações do mediador de spin 1 para ver como ele modifica os perfis dos potenciais interpartículas e também incluímos em nossa análise todos os tipos de acoplamentos entre correntes fermiônicas e o bóson mediador. A comparação entre o caso bem conhecido do campo de Proca e o de um bóson de spin-1 (com massa invariante de calibre) descrito pela mistura de uma 2-forma com um campo quadrivetorial de calibre é estabelecida a fim de buscar uma análise dos perfis dependentes de spin e de velocidade dos potenciais interpartículas. Discutimos possíveis aplicações e derivamos um limite superior no produto das constantes de acoplamento vetoriais e pseudo-tensoriais. Os potenciais interpartículas dependentes de spin e de velocidade que obtivemos podem ser usados para explicar efeitos possivelmente associados a novas forças macroscópicas, tais como modificações na lei do inverso do quadrado e possíveis efeitos de acoplamento do tipo spin-gravidade.

A segunda parte desta tese é baseada na fenomenologia da matéria escura de extensões $U(1)'$ do Modelo Padrão Supersimétrico Mínimo bem motivadas. Nestes modelos, o sneutrino de mão direita é um bom candidato para a matéria escura cujas propriedades da matéria escura estão de acordo com a densidade relíquia atual e os atuais limites experimentais obtidos para a seção de choque do espalhamento entre nucleons e matéria escura. O sneutrino pode se aniquilar em partículas mais leves através da troca de bosons de calibre massivos e bosons de Higgs através de processos de canal s. Para ver o quão pesado o sneutrino de mão direita pode ser como candidato viável de matéria escura, nós exploramos suas propriedades de matéria escura na região do espaço de parâmetros que minimiza sua densidade relíquia via efeitos de ressonância e assim permite que ele seja uma partícula de matéria escura mais pesada. Descobrimos que, no contexto do modelo padrão de cosmologia, o sneutrino de mão direita pode se comportar como uma boa partícula de matéria escura mesmo com massas da ordem de dezenas de TeV, o que está bem acima das massas que os candidatos de matéria escura do tipo WIMPs geralmente têm na maioria dos modelos de partículas de matéria escura.

\bigskip
\bigskip
\bigskip
\bigskip
\bigskip
\bigskip
\bigskip
\bigskip
\bigskip
\bigskip
\bigskip
\bigskip
\bigskip
\bigskip
\bigskip
\bigskip
\bigskip
\bigskip
\bigskip
\bigskip
\bigskip
\bigskip
\bigskip
\bigskip
\bigskip
\bigskip
\bigskip
\bigskip
\bigskip
\bigskip
\bigskip
\bigskip
\bigskip
\bigskip
\bigskip
\bigskip
\bigskip
\bigskip
\bigskip
\bigskip
\bigskip
\bigskip

\textbf{Palavras-chave:} Potenciais dependentes de spin; cosmologia de teorias além do Modelo Padrão, Modelo Padrão Supersimétrico, fenomenologia de supersimetria.

\begin{center}
\chapter*{Acknowledgements}
\end{center}
\addcontentsline{toc}{section}{Acknowledgements}
\indent

I am immensely grateful to Carsten Hensel for having me under his supervision and for giving me the most possible freedom concerning my research activities. I would like to thank my brazilian advisor José Abdalla Helayël-Neto for his supervision and for organizing many interesting courses that helped me a lot throughout my master and PhD studies. I thank my partners Pedro Malta and Leonardo Ospedal for their collaborations in our research activities in CBPF. Thanks to all my other colleagues from CBPF: Celio Marques, Pedro Costa, Yuri Müller, Lais Lavra, Erich Cavalcanti, Ivana Cavalcanti, Gabriela Cerqueira, Max Jáuregui and Mylena Pinto Nascimento for sharing happy moments during this long battle.  

I thank Manuel Drees for accepting me as a long-term visiting PhD student in his group to work on very interesting research projects. I thank him specially for his supervision and for the productive scientific discussions we had. I also would like to thank the Bethe Center for Theoretical Physics (University of Bonn) for the hospitality and for the organization of all the very interesting conferences and workshops of which I attended. Many thanks goes to my BCTP colleagues Annika Reinert, Andreas Trautner, Manuel Krauss, Victor Martin Lozano and Fabian Fischbach for the pleasure of their friendly conversations and for helping me to solve the problems I had during my stay in Bonn city.

This work has been funded mostly by the Brazilian Coordination for the Improvement of Higher Education Personnel (CAPES) whom I'd like to thank a lot for all the financial support. And finally, of course, I wish to thank my parents for all the support they provided to me and for the inspiration advices that enabled me to obtain my education.

\newpage

\begin{center}
\section*{\Huge \bf List of Abbreviations}
\end{center}
\bigskip
\bigskip
\bigskip

\begin{tabular}{p{4cm}p{12cm}}

\textbf{AMSB} & \textbf{A}nomaly-\textbf{M}ediated \textbf{S}upersymmetry \textbf{B}reaking \\ \\ 

\textbf{BSM} & \textbf{B}eyond the \textbf{S}tandard \textbf{M}odel \\ \\

\textbf{CDM} & \textbf{C}old \textbf{D}ark \textbf{M}atter \\ \\

\textbf{CMB} & \textbf{C}osmic-\textbf{M}icrowave \textbf{B}ackground \\ \\

\textbf{DM} & \textbf{D}ark \textbf{M}atter \\ \\

\textbf{EWSB} & \textbf{E}lectro\textbf{W}eak \textbf{S}ymmetry \textbf{B}reaking \\ \\

\textbf{FRW} & \textbf{F}riedmann-\textbf{R}obertson-\textbf{W}alker \\ \\

\textbf{GMSB} & \textbf{G}auge-\textbf{M}ediated \textbf{S}upersymmetry \textbf{B}reaking \\ \\

\textbf{GUT} & \textbf{G}rand \textbf{U}nified \textbf{T}heories  \\ \\

\textbf{HDM} & \textbf{H}ot \textbf{D}ark \textbf{M}atter \\ \\

\textbf{$\boldsymbol{\Lambda}$CDM} & \textbf{L}ambda \textbf{C}old \textbf{D}ark \textbf{M}atter \\ \\

\textbf{LH} & \textbf{L}eft-\textbf{H}anded \\ \\

\textbf{LHC}  & \textbf{L}arge \textbf{H}adron \textbf{C}ollider \\ \\

\textbf{LSP}  & \textbf{L}ightest \textbf{S}upersymmtric \textbf{P}article \\ \\

\textbf{MACHO} & \textbf{MA}ssive \textbf{C}ompact \textbf{H}alo \textbf{O}bject \\ \\

\textbf{MC} & \textbf{M}onte-\textbf{C}arlo \\ \\

\end{tabular}

\newpage

\begin{tabular}{p{4cm}p{12cm}}

\textbf{MSSM} & \textbf{M}inimal \textbf{S}upersymmetric \textbf{S}tandard \textbf{M}odel \\ \\

\textbf{NMSSM} & \textbf{N}ext-to-\textbf{MSSM} \\ \\

\textbf{NLSP}  &  \textbf{N}ext-to-\textbf{LSP} \\ \\

\textbf{QCD} & \textbf{Q}uantum \textbf{C}hromo\textbf{D}ynamics \\ \\

\textbf{QED} & \textbf{Q}uantum \textbf{E}lectro\textbf{D}ynamics \\ \\

\textbf{RG} & \textbf{R}enormalization \textbf{G}roup \\ \\

\textbf{RGE} & \textbf{R}enormalization \textbf{G}roup \textbf{E}quation \\ \\

\textbf{RH} & \textbf{R}ight-\textbf{H}anded \\ \\

\textbf{RHSN} & \textbf{R}ight-\textbf{H}anded \textbf{SN}eutrino \\ \\

\textbf{RW} & \textbf{R}obertson-\textbf{W}alker \\ \\

\textbf{SM} & \textbf{S}tandard \textbf{M}odel \\ \\

\textbf{SUGRA} & \textbf{SU}per\textbf{GR}avity \\ \\

\textbf{SUSY} & \textbf{SU}per\textbf{SY}mmetry \\ \\

\textbf{SYM} & \textbf{S}upersymmetric \textbf{Y}ang-\textbf{M}ills \\ \\

\textbf{VEV} & \textbf{V}acuum \textbf{E}xpectation \textbf{V}alue   \\ \\

\textbf{WIMP}  & \textbf{W}eakly \textbf{I}nteracting \textbf{M}assive \textbf{P}article \\ \\

\end{tabular}

\tableofcontents

\newpage\pagenumbering{arabic}\setcounter{page}{1}\pagestyle{plain}

\begingroup       
\setstretch{1}

\chapter{Introduction}
\indent 

The Standard Model (SM) of elementary particle physics provides a very remarkably and successful description of presently known phenomena of high energy physics. In the last decades the SM has been extensively and successfully tested in many different experiments associated to great particle colliders such as the Large Hadron Collider (LHC), Tevatron and the Large Electron-Positron Collider (LEP). Complementary to the high energy phenomena probed at the LHC, many other low-energy experiments report excellent agreement with all the predictions of the SM. 


The SM is a renormalizable quantum field theory defined in a four-dimensional spacetime that respects Poincaré invariance whose gauge interactions are based on the following gauge group
\begin{equation} 
G_{\text{SM}} = SU(3)_{C}\times SU(2)_{L}\times U(1)_{Y},
\end{equation}
where $C$ stands for color, $L$ for left in the sense of chirality and $Y$ for hypercharge. The presence of the $SU(2)_{L}$ gauge group indicates that the SM is a chiral gauge theory, which means that a left-handed field transforms in a different way compared to its right-handed counterpart. As a result, the left-handed charged leptons $e_{L i}$ and their corresponding neutrinos $\nu_{L i}$ form a doublet representation of the $SU(2)_{L}$, while the right-handed charged leptons $e_{R i}$ are singlets. The quark fields are also separated in left-handed and right-handed chiral components, whereas right-handed neutrino fields $\nu_{R i}$ are not included in the particle content of the SM. The quarks and leptons are divided in three families with $i = (1, 2, 3)$. The gauge bosons are the force carriers of the three fundamental interactions that are described by the SM, which are the electromagnetic, the weak and the strong forces. The particle content of the SM is shown in Figure \ref{SMParticlecontent} including the well-known spin, masses and electric charges of each particle.

\begin{figure}[h!]
\centering
\includegraphics[width=0.60\textwidth]{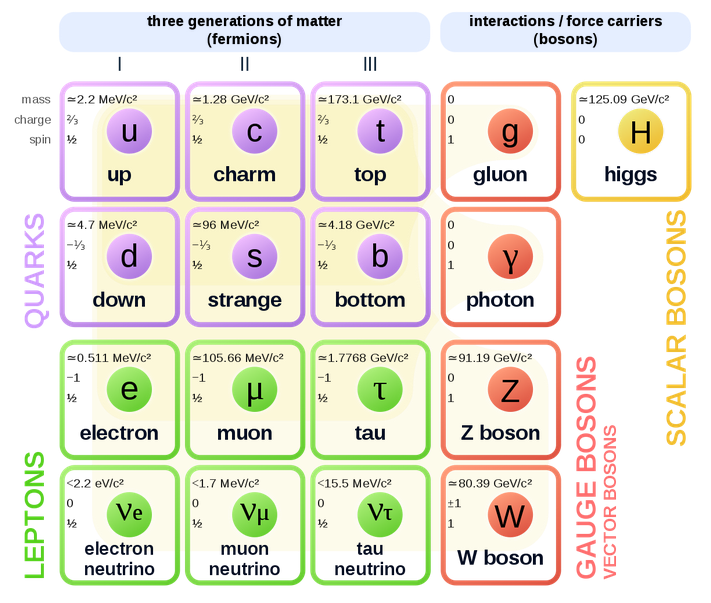}
\caption{\label{fig:overleaf} Content of particles of the SM with their corresponding masses, spin and electric charges.} \label{SMParticlecontent}
\end{figure}

If the gauge symmetry were exact, not only the gauge bosons but also the matter fermions would remain massless, as including mass terms for the vector bosons and Dirac mass terms for the fermions would explicitly break the gauge symmetry. This problem is solved by the method of spontaneous symmetry breaking proposed in the Brout-Englert-Higgs mechanism \cite{Englert:1964et, Higgs:1964pj}, in which a neutral component of the Higgs doublet acquires a non-zero vacuum expectation value. Because of this the electroweak gauge group $SU(2)_{L}\times U(1)_{Y}$ breaks to the Abelian electromagnetic gauge symmetry $U(1)_{\text{em}}$ that describes all the familiar electromagnetic interactions. As a consequence, the $W^{\pm}$ and $Z_{0}$ bosons acquire masses proportional to the VEV of the Higgs field $v$, while the physical quarks and charged leptons of the SM acquire masses proportional to the product of $v$ and their corresponding Yukawa coupling. The new massive scalar particle discovered by the ATLAS and CMS Collaborations of the LHC in 2012 is compatible with the Higgs boson predicted by the SM \cite{Aad:2012tfa,Chatrchyan:2012xdj}, which was the only missing piece in the construction of this model. Therefore, now the SM is complete and all its parameters are numerically determined.

Although the SM is in excellent agreement with almost all available experimental results, it is {\em not} the ultimate particle physics theory. There are several problems that cannot be solved by the SM, such as:

\begin{itemize}
   
\item {\bf Neutrinos Masses}: currently there are many experimental evidences showing that neutrinos oscillate from one flavor to another and, as a consequence, should have masses \cite{Fukuda:1998mi, Gonzalez-Garcia:2014bfa}. The generation of neutrino masses in the theory cannot be done without extending the Standard Model. One way to generate mass terms for the neutrinos is by adding right-handed neutrinos to the particle content and produce Dirac mass terms via the Yukawa interaction with Higgs field. Another one is via the see-saw mechanism \cite{Minkowski:1977sc,Mohapatra:1979ia,Schechter:1980gr} where Majorana masses for the neutrinos are generated through the Weinberg operator \cite{Weinberg:1979sa};

\item {\bf Dark Matter Candidate}: cosmological and astrophysical observations provide strong evidences that there is a new form of non-baryonic matter in the Universe which composes approximately 80\% of the its matter content. So far this mysterious form of matter has only been observed indirectly via  gravitational effects on the galaxy rotation curves or via gravitational lensing effects because it does not interact with light, therefore it is known as {\em dark matter}. The SM does not provide a viable dark matter candidate consistent with its known properties; 

\item {\bf Hierarchy Problem}: the Higgs mass must receive quantum corrections to get close to 125 GeV. To include quantum corrections on the Higgs mass one must introduce a cut-off scale $\Lambda$ in such a way that the quantum corrections to $m^{2}_{h^{0}}$ are proportional to $\Lambda^{2}$. Above the electroweak symmetry breaking scale the next natural scale of new physics that we know is the Planck scale $M_{P} \simeq 10^{19}$ GeV, which implies on extremely large corrections to the SM Higgs mass. This problem could be avoided if a new Physics scale appears around the TeV region.

\item {\bf Gravity}: The SM is not able to include a quantum description of gravity, which is a phenomenon that appears to be relevant only at energy scales close to the Planck scale. This points towards the direction that the SM is not the ultimate theory of Nature but rather an effective description of an ultraviolet one. In this case string theory appears to be the best candidate to describe quantum gravity in connection to the SM.

\end{itemize}

Most macroscopic phenomena that we know originate either from gravitational or electromagnetic interactions. There has been some experimental effort over the past decades towards the improvement of low-energy measurements of the inverse-square law, with fairly good agreement between theory and experiment \cite{Adelberger:1992ph,Adelberger:2003zx}. The equivalence principle has also been recently tested to search for a possible spin-gravity coupling \cite{Tarallo:2014oaa}. On the other hand, a number of scenarios beyond the Standard Model (BSM) motivated by high-energy phenomena predict very light, weakly interacting sub-eV particles (WISPs) that could generate new long-range forces, such as axions \cite{Moody:1984ba}, SUSY-motivated particles \cite{Fayet:1986vz} or paraphotons \cite{Okun:1982xi,Holdom:1985ag,Dobrescu:2006au,Accioly:2014cea}. 

The discovery of a new, though feeble, fundamental force would represent a remarkable advance. Besides the Coulomb-like ``monopole-monopole" force, it is also possible that spin- and velocity-dependent forces arise from monopole-dipole and dipole-dipole (spin-spin) interactions. Those types of behavior are closely related to two important aspects of any interacting field theory: matter-mediator interaction vertices and the propagator of intermediate particles. Part I of this thesis is  mainly concerned with this issue and its consequences on the shape of the potential between two fermionic sources. This discussion is also of relevance in connection with the study, for example, of the quarkonium spectrum, for which spin-dependent terms in the interaction potential may contribute considerable corrections \cite{Badalian:2008sv}. Other sources (systems) involving neutral and charged particles, with or without spin, have been considered by Holstein \cite{Holstein:2008fs}.

Supersymmetry (SUSY) is one of the best motivated theories to describe new physics beyond the Standard Model (SM) at TeV scale. It introduces a useful space-time symmetry that relates bosons and fermions which can be used to cancel the quadratic divergencies that appears in the radiative corrections of the masses of scalar bosons, providing thus a natural solution to the hierarchy problem of the SM. It allows the gauge couplings to unify at a certain grand unified scale in the vicinity of the Planck scale \cite{Dimopoulos:1981yj,Ibanez:1981yh,Marciano:1981un,Einhorn:1981sx,Amaldi:1991cn,Langacker:1991an,Ellis:1990wk} and this can be seen as a clear hint that SUSY is the next step towards a grand unified theory (GUT) \cite{Mohapatra:1999vv}. One of the most interesting features of low energy supersymmetric models is that, when conserving a discrete symmetry that appears in these models called R-parity, the lightest supersymmetric particle (LSP) is absolutely stable and behaves as a realistic weakly interacting massive particle (WIMP) dark matter candidate \cite{Ellis:1983ew,Jungman:1995df} which is able to account for the observed cold dark matter relic density recently measured and precisely analyzed by ${\it Planck}$ Collaboration $\Omega_{\text{DM}} h^{2}=0.1188\pm 0.0010$ \cite{Ade:2015xua}.

The Minimal Supersymmetric Standard Model (MSSM) is the simplest supersymmetric extension of the SM which contains the smallest number of new particles in a consistent way with supersymmetry and the particle content of the SM. It is constructed as a renormalizable $\mathcal{N}=1$ super Yang–Mills theory also based on the gauge group $SU(3)_{C}\times SU(2)_{L}\times U(1)_{Y}$. Although the MSSM has been the mostly studied supersymmetric model it  does not solve all the open problems of the SM. The tree-level mass for the SM-like Higgs boson in the MSSM is limited to the $Z^{0}$ boson mass via the relation $m_{h^{0}}|_{\text{tree}} < m_{Z^{0}} |\cos{2 \beta}|$ \cite{Inoue:1982ej}. Thus, large quantum corrections are necessary to take into account in this model to obtain the actual value for the Higgs mass $m_{h^{0}}= 125$ GeV, which requires a heavy third generation of squarks and large stop mixing \cite{Arbey:2011ab,Carena:2011aa,Fan:2014txa}. It turns out that this implies a hierarchy in the mass distribution of the squarks and thus it doesn't seem to be a natural prediction of the SM Higgs mass.

Therefore, in order to solve the problems of the MSSM, it is useful to consider extensions of it. For concreteness we work with $U(1)'$ extensions of the MSSM which originate from the breaking of the $E_{6}$ GUT gauge group. These models, which are referred to as UMSSM \cite{Langacker:1998tc,King:2005jy,Belanger:2011rs,Belanger:2015cra,Belanger:2017vpq}, contain three right--handed neutrino superfields plus an extra gauge boson $Z'$ and an additional SM singlet Higgs with mass $\simeq M_{Z'}$, together with their superpartners. Models of this kind were first studied more than 30 years ago in the
wake of the first ``superstring revolution'' \cite{London:1986dk,
  Hewett:1988xc}. This framework allows to study a wide range of $U(1)'$ groups, since $E_6$ contains {\em two} $U(1)$ factors beyond
the SM gauge group. In comparison to the MSSM, the prediction of the mass of the lightest CP-even Higgs boson at tree-level also receives contributions from an F- and a D-term in such a way that it is not necessary to have large loop corrections to explain the correct SM Higgs mass. Moreover, the scalar members
$\tilde \nu_{R,i}$ of the right--handed neutrino superfields make good WIMP candidates
\cite{Lee:2007mt, Cerdeno:2008ep, Allahverdi:2009ae, Khalil:2011tb,
  Belanger:2011rs, Belanger:2015cra, Belanger:2017vpq}. 

In the UMSSM the right sneutrino is charged under the extra $U(1)'$ gauge symmetry; it can therefore annihilate into lighter particles via gauge interactions. In particular, for $M_{\tilde \nu_R} \simeq M_{Z'}/2$ the sneutrinos can annihilate by the exchange of (nearly) on--shell gauge or Higgs bosons. We focus on this region of parameter space. For some charge assignment we find viable thermal $\tilde \nu_R$ dark matter for mass up to $\sim 43$ TeV. This is very heavy compared to most of the masses seen in the literature of WIMP-type dark matter candidates. Our result can also be applied to other models of spin$-0$ dark matter candidates annihilating through the resonant exchange of a scalar particle. These models cannot be tested at the LHC, nor in present or near--future direct detection experiments, but could lead to visible indirect detection signals in future Cherenkov telescopes.

In chapter \ref{chapterDARKMATTER} we review the dark matter problem in the context of cosmology and astrophysics. First we discuss some of the main observational evidences for the existence of DM in the Universe and explain the necessary properties that a particle must have to become a viable dark matter candidate. We focus in the category of Weakly Interacting Massive
Particles (WIMPs), which has been the most studied particle dark matter candidate in the literature. Then we introduce basic concepts of standard cosmology which are essential to further describe in the next section the thermal production of DM in the Early Universe and how the dark matter decoupled from the original thermal plasma to give the relic density measured nowadays. This thesis is divided in two parts which are organized as follows:

\begin{itemize}

\item {\bf Part I}: in chapter \ref{chapterPAPER1} we investigate the role played by particular field representations of an intermediate massive spin-1 boson in the context of interparticle potentials between fermionic sources. We show that changing the representation of the spin-1 mediator one obtains different profiles of velocity- and spin-dependent interparticle potentials in the limit of low momentum transfer \cite{Ferreira:2014era}; 

\item {\bf Part II}: in chapter \ref{chapterSUPERSYMMETRY} we start by motivating supersymmetry and then we review its theoretical framework. After a brief summary of the MSSM we introduce the phenomenological tools mostly used to explore other models of particle physics beyond the SM in great detail. In chapter \ref{chapterUMSSM} we describe the theoretical framework of the well-motivated $U(1)'$ extensions of the MSSM and discuss its particle content, with a particular emphasis on the gauge, Higgs, sneutrino and neutralino sectors. In chapter \ref{ChapterUPPERLIMIT} we firstly describe the calculation of the relic density of the right-handed sneutrino and explain our procedure to minimize it. Secondly, we present the results of our numerical analysis for the dark matter phenomenology of the right-handed sneutrino in the general UMSSM and also discuss prospects of probing such scenarios experimentally. The last two chapters are based on the article \cite{Ferreira:2017osm}.

\end{itemize}

Finally in chapter \ref{FINALCONCLUSIONS} general conclusions about the most important results of this thesis are given. Two Appendices follow: in the Appendix \ref{AppendixA}, the list of all relevant vertices in the low-energy limit used in chapter \ref{chapterPAPER1} are presented; next, in the Appendix \ref{AppendixB}, we present the multiplicative algebra of a set of relevant spin operators that appear in the attainment of a set of propagators that we computed in Section \ref{CHAP3section3.4}.



\chapter{Dark Matter}
\label{chapterDARKMATTER}
\indent 

Currently, only 5\% of the energy content of the Universe is made of ordinary matter such as atoms which make stars, planets and us. All the rest is dark and unknown composed of dark matter and dark energy where invisible dark matter makes up 27\% of the matter content of the Universe. Dark matter is a hypothetical form of matter which has been postulated to explain certain phenomena which cannot be explained by ordinary matter. So far the existence of dark matter is mostly inferred from observation of gravitational effects on visible matter and background radiation and not through its direct nor indirect detection. For a deeper review of the subject see refs. \cite{Bergstrom:2000pn,Bertone:2004pz,Bertone:2010zza}. 

In this chapter we introduce the basic concepts of dark matter phenomenology. In particular, we will first discuss some of the main observational evidences of the existence of DM in our Universe. Then we describe with more details the aspects of the most attractive category of dark matter candidate known as WIMPs (Weakly Interacting Massive Particles) and how they were thermally produced in the era of the early Universe.

\section{Dark matter evidence}
\indent

There are several observational evidences from astrophysics and cosmology that imply the existence of a non-luminous form of matter in the Universe. The first one came from Fritz Zwicky's work \cite{Zwicky:1933gu} in 1933 where, applying the virial theorem to estimate the mass of clusters of galaxies, he found that the magnitude of mass in the Coma Cluster was about two orders bigger than the visually observable mass. The next one came with the investigation of rotation curves of some spiral galaxies by Rubin and her collaborators \cite{Rubin:1970zza, Rubin:1980zd, Rubin:1985ze}. They worked with a new sensitive
spectrograph that was able to obtain the velocity curve of certain spiral galaxies with a higher level of accuracy. As a result, they also concluded that most of the mass of these galaxies is also not in luminous stars. Here I give a brief resume of their analysis.

In a spiral galaxy the mass distribution of the luminous matter is modelled by a disk and a bulge. If we assume Newton's laws of gravity, the circular velocity of a star of the galaxy located at a distance $r$ from the galaxy'center is given by

\begin{equation} 
v_{\text{circ}}(r) = \sqrt{\frac{G M(r)}{r}},
\end{equation}
where $M(r)$ is the mass enclosed inside a radius $r$. Assuming all the mass is concentrated in the galactic disc $R_{disc}$, for distances larger than the disc ($r > R_{disc}$) $M(r)$ should remain constant, which leads to a circular velocity $v_{circ}(r) \propto 1/\sqrt{r}$. However, according to cosmological observations, the velocity distribution is approximately flat far away from the center of the galaxy, as shown in Figure \ref{FigVELOCITYDISTRIBUTION}. The behaviour of the observed high velocities cannot be explained by taking into account only the visible mass which is proportional to light emitted by stars. This can be well explained if there is a spherical halo around the galaxy with $M(r) \propto r$ in which case most of the mass of the galaxy would be concentrated in the dark region of this halo. This shows that there must exist some non-visible form of matter in the Universe known as \textit{dark matter}. Similar results were also obtained in the cases of other galaxies with different mass distributions.

\begin{figure}[h!]
\centering
\includegraphics[width=0.60\textwidth]{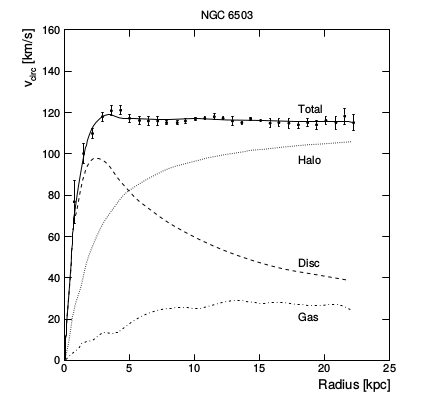}
\caption{\label{fig:overleaf} Measurements of the velocity rotation curves of the galaxy NGC 6503 \cite{Begeman:1991iy} with different contributions coming from the galactic disc, dark matter halo and gas. The data points that fit the total curve take into account all type of contributions.} \label{FigVELOCITYDISTRIBUTION}
\end{figure}

In general, depending on which scale we are looking at, different methods of noting and measuring directly or indirectly the presence of non-visible matter can be employed. The following observations strengthen the fact that there is a significant amount of dark matter in the Universe:

\begin{itemize}

\item {\bf Gravitational Lensing}: according to Einstein's gravity theory of general relativity, the curvature of space-time caused by matter gives rise to a deflection of light rays. Since the deflection angle is proportional to the mass of the object that causes the deflection and behaves as a lens, this is a good tool for estimating directly the mass of large astrophysical objects, from planets and upwards to galaxy clusters. The analysis of the gravitational lensing data \cite{Wittman:2000tc} indicates that there is a lot of dark matter;
  
\item {\bf Cosmic Microwave Background (CMB)}: at some moment in the epoch of the Early Universe, matter and radiation formed a hot plasma in thermal equilibrium and, as the Universe expanded and cooled, the photons eventually began to propagate freely. Today these photons are interpreted as CMB, which is characterized in a good approximation by a thermal black body spectrum with a mean temperature of $T= (2.7255 \pm 0.0006)$ K \cite{Patrignani:2016xqp}. However, there have been observed some anisotropies in the temperature distribution of the CMB. The COBE, WMAP and Planck satellites measured the angular power spectrum of the thermal anisotropies in increasing order of precision and the final results obtained by Planck Collaboration \cite{Aghanim:2018eyx} are shown in Figure \ref{FigPLANCK2018SPECTRUM}. The positions and relative magnitudes of the peaks are used to fit the parameters of specific models of cosmology. As a result, the best fit favors the existence of cold dark matter in the Universe;

\item {\bf Large Scale Structure Formation}: since gravitational interaction is attractive,  large structures such as stars, galaxies and clusters of galaxies have been formed because of huge gravitational collapses acting in opposition to the expansion of the Universe. The amount of DM in the matter power spectrum has to be sufficiently large to create enough gravitational fields that can overcome the electromagnetic pressure of the baryonic matter, and thus eventually allow the formation of large scale structures on the Universe.

\begin{figure}[h!]
\centering
\includegraphics[width=0.75\textwidth]{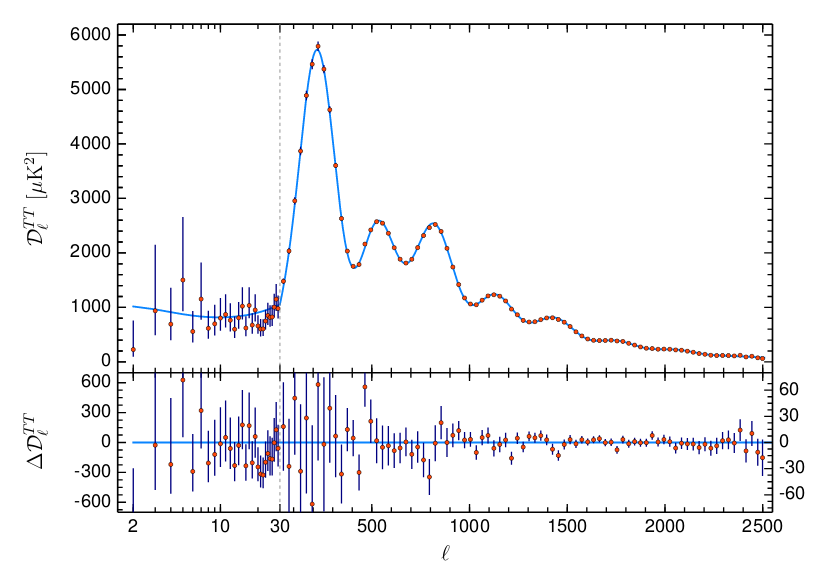}
\caption{\label{fig:overleaf} Temperature power spectrum of the CMB anisotropies. The red points represents the experimental data obtained by the Planck satellite \cite{Aghanim:2018eyx} including error bars, while the blue line gives the best fit of the standard model of cosmology, which is the $\Lambda$CDM model, to the Planck data.} \label{FigPLANCK2018SPECTRUM}
\end{figure}

\end{itemize}

The observations cited above show that there is an important amount of evidence that points DM as the dominant component of the matter spectrum of the Universe. DM communicates with its environment essentially via gravitational interaction. However, we still do not know the nature of dark matter. One of the first proposals to explain this problem was considering the hypothesis that DM could be baryonic matter in the form of MAssive Compact Halo Objects (MACHOs) such as black holes or brown dwarfs. However, the study of baryonic production in the Big Bang Nucleosynthesis contradicts MACHOs as DM \cite{Dar:1995gh}. Currently, the observations from astrophysics and cosmology give higher support for non-baryonic particle-like DM candidate instead of a sizeable object. Hence, it is believed that dark matter is composed of a new type of elementary particle that appears in models of new physics beyond the SM \cite{Bertone:2004pz,Jungman:1995df}. Several different DM candidates have been proposed in the literature, for reviews see Ref.\cite{Bergstrom:2000pn}.

\section{Particle dark matter: WIMPs}
\indent

Relativistic particles are candidates to Hot Dark Matter (HDM), whereas non-relativistic particles are candidates to Cold Dark Matter (CDM). If we compare the relativistic to the non-relativistic nature of the DM, it turns out that HDM is incompatible with data from Large Scale Structure observations \cite{Abazajian:2005xn,dePutter:2012sh,Lukash:2012tq}, which constrain the allowed average velocity of the DM particles from above. For this reason, relativistic HDM particles cannot dominate the constitution of DM. CDM particles is the option that mostly fulfills DM constraints and, as a consequence, this scenario is the one that is considered in the standard model of cosmology. If there is a particle that plays the role for the DM candidate it should respect the following requirements:

\begin{itemize} [label=$\star$]

\item It must be electrically neutral and must interact very weakly with photons. Otherwise it could have emitted light and been seen in astrophysical observations;

\item It must have no colour charge. Otherwise it would hadronise and behave as a baryonic form of matter; 

\item It must be stable or long-lived with a lifetime that exceeds the age of the Universe. Otherwise, it would not have the relic density that we observe nowadays;

\item It must be in agreement with current experimental constraints and observations.

\end{itemize}

One of the main candidates for a non-baryonic cold dark matter particle that agrees with all these requirements is the Weakly Interacting Massive Particle (WIMP), a class of neutral stable particles that interacts with SM particles only via weak interactions. There are two main motivations that made WIMPs become the favoured and most popular category of DM candidates. The first one, as we will discuss in Sec.~\ref{DMThermalProduction}, is based on the fact that the simplest production mechanism for massive dark matter relics from the early Universe in standard cosmology automatically supports the weak scale. To obtain the correct relic abundance the DM particle should have a self-annihilation thermally averaged cross section of the order of $10^{-26} {\rm \text{ cm}}^3/{\rm \text{s}}$ \cite{Steigman:2012nb, Drees:2015exa}. The second one is that a roughly weak-scale annihilation cross section indicates that both the annihilation of WIMPs at the current temperature of the Universe and the elastic scattering of WIMPs on heavy target nuclei might be detectable; the former goes under indirect detection experiments, while the latter constitutes techniques of direct detection.

\section{Elements of standard cosmology}
\indent

Considering the fact that on sufficiently large scales the properties of the Universe are the same for all observers, one arrives at the {\it cosmological principle} of standard cosmology: the Universe is homogeneous and isotropic. A solution of the Einstein's General Relativity (GR) Equations in agreement with this principle yields a spacetime metric in the so-called Friedmann-Robertson-Walker (FRW) form. In spherical coordinates this metric is
\begin{equation}
ds^{2}=g_{\mu\nu} dx^{\mu}dx^{\nu}=dt^{2}-a(t)^{2}\left[\frac{dr^{2}}{1-Kr^{2}}+r^{2}
\left(d\theta^{2}+\sin^{2}\theta d\phi^{2}\right)\right],  \label{FRW metric}
\end{equation}
where $a(t)$ is the scale factor of the Universe and $K$ is the curvature parameter which is defined for three different cases: 0 for a flat Universe, +1 for a closed Universe and -1 for an open Universe. The scale factor is a function that depends only on the time coordinate $t$ and it is used to determine the behaviour of the evolution of the Universe. 

The behaviour of the dynamics of the scale factor depends on the energy and matter content of the Universe, which is represented by the energy-momentum tensor $T_{\mu\nu}$ that appears on the right-hand side of Einstein's equations of GR
\begin{equation}
R_{\mu\nu}-\frac{1}{2} R g_{\mu\nu}= 8\pi G \left(T_{\mu\nu} + \Lambda g_{\mu\nu}\right).    \label{Einstein eqs}
\end{equation}
Here $R_{\mu\nu}$ and $R$ are the Ricci tensor and the Ricci scalar respectively, whereas $\Lambda$ is the cosmological constant. With the assumption that the matter-energy content of the Universe is described by a perfect homogenous and isotropic fluid with a time-dependent total energy density $\rho(t)$ and a time-dependent pressure $p(t)$, we obtain the following energy-momentum tensor 
\begin{equation}
T^{\mu}_{\nu}=\text{diag}(\rho,-p,-p,-p).  \label{energy-momentum tensor}
\end{equation}
Using eqs.(\ref{FRW metric}), (\ref{Einstein eqs}) and (\ref{energy-momentum tensor}) one can obtain the so-called Friedmann-Lemaître equations:
\begin{eqnarray}
H^{2} & = & \left(\frac{\dot{a}}{a}\right)^{2}=\frac{8\pi G}{3}\rho+ \frac{\Lambda}{3}-\frac{K}{a^{2}}   \label{FriedmannEq1} \\
\frac{\ddot{a}}{a} & = & -\frac{4\pi G}{3}\left(\rho+3 p\right) + \frac{\Lambda}{3}   \label{FriedmannEq2},
\end{eqnarray}
where $H$ is the Hubble parameter and the dots represent derivatives with respect to time. The first equation describes the evolution of the expansion of the Universe, while the second one determines if the expansion is accelerated or decelerated at a certain time $t$. 

Combining eq.(\ref{FriedmannEq1}) with eq.(\ref{FriedmannEq2}) one arrives at the adiabatic equation:
\begin{equation}
\frac{d}{dt}(\rho a^{3})+p\frac{d}{dt}(a^{3})=0  \quad\quad\quad \Rightarrow \quad\quad\quad \dot{\rho}=-3H(\rho+p),    \label{FriedmannTermodynamics}
\end{equation}
which is the relativistic version of the first law of thermodynamics $T dS=dE+ pdV$ in a Universe with constant entropy. The solution of eq.(\ref{FriedmannTermodynamics}) gives the equations of state of the fluids that compose the Universe, where the equations of state for radiation, matter and vacuum energy (cosmological constant) are given below

\begin{equation}
\rho \propto a^{-3(1+\omega)} \quad \Rightarrow \quad
   \left\{
  \begin{array}{l l}
    p=\frac{1}{3} \rho \quad & \rightarrow \quad \rho \propto a^{-4} \quad\quad\quad\quad \text{ for radiation}\\
    p=0 \quad & \rightarrow \quad \rho \propto a^{-3} \quad\quad\quad\quad \text{ for matter}\\
    p=-\rho \quad & \rightarrow \quad \rho \propto \text{constant} \quad\quad \text{for vacuum energy}. \\
  \end{array} \right.
\end{equation}

Eq.(\ref{FriedmannEq1}) can be used to determine the critical energy density $\rho_{c}$
\begin{equation}
\rho_{c}=\frac{3 H^{2}}{8\pi G},    \label{criticaldensity}
\end{equation}
which is obtained by supposing that the Universe is flat and neglecting the cosmological constant ($\Lambda =0$). The cosmological data obtained in these last decades tell us that our Universe is (practically) flat \cite{Spergel:2006hy}. Usually it is more convenient to express the energy densities of each components of the Universe as being its density divided by the critical density
\begin{equation}
\Omega_{i} \equiv \frac{\rho_{i}}{\rho_{c}},    
\end{equation}
where $i$ stands for matter ($m$) or radiation ($r$). Noting that the specific cases of the curvature and the cosmological constant are 
\begin{equation}
\Omega_{K}= -\frac{K}{a^{2}H^{2}},  \quad\quad \Omega_{\Lambda}= \frac{\Lambda}{3H^{2}},  
\end{equation}
we can write eq.(\ref{FriedmannEq1}) as
\begin{equation}
\Omega_{m} + \Omega_{r} + \Omega_{\Lambda}=\Omega_{total}=1 -\Omega_{K},   \label{sum densities}
\end{equation}
where we used $\rho= \rho_{m}+\rho_{r}$. According to the recent data obtained by the Planck satellite \cite{Aghanim:2018eyx}, we have
\begin{equation}
\Omega_{m}= 0.315 \pm 0.007, \quad\quad \Omega_{\Lambda}=0.6847 \pm 0.0073, \quad\quad \Omega_{K}=0.001 \pm 0.002. 
\end{equation}
The curvature density is negligible and, using eq.(\ref{sum densities}) we can see that the radiation density of our Universe today is also very small. This means that the dominant components of the Universe are the dark energy and the matter content, and thus the radiation does not play a significant role in the present configuration of our Universe.


\section{Dark matter thermal production}
\label{DMThermalProduction}
\indent

From a thermal point of view, a dark matter particle can be classified in two types: relativistic or {\it hot} and non-relativistic or {\it cold}. A particle $X$ of mass $m_{X}$ at temperature $T$ is hot if $T \gg m_{X}$ and cold if $T \ll m_{X}$. At temperatures much higher than the WIMP mass, the colliding particle-antiparticle pairs of the plasma had sufficient energy to create WIMP pairs in an efficient way. And also, the inverse reactions of WIMPs annihilating into pairs of SM particles were as efficient as the WIMP-producing processes in such a way that the WIMPs were in thermal equilibrium. 

The Early Universe can be seen as a hot plasma of particles interacting with each other in thermal equilibrium where WIMPs were annihilating and being produced in collisions between these particles of the thermal plasma during the radiation-dominated era.  However, as the Universe expanded and cooled, two things happened: 1) the SM particles produced by the annihilation of WIMPs no longer had sufficient kinetic energy (thermal energy) to reproduce WIMPs through interactions and 2) the expansion of the Universe diluted the number of all particles in such a way that the interactions were no longer occuring as in the epoch of the Early Universe. In other words, at some point the density of some massive particles became too low to support the frequent interactions of the plasma and, as a consequence, the conditions for thermal equilibrium were violated. At this point, particles are said to ``freeze-out'' and their number density remains constant. Freeze-out occurs when the expansion rate of the Universe overtakes the annihilation rate.

In order to describe the physical processes that occurred in the hot Universe, we must determine the thermal distribution of each particle in the thermal plasma. For a particle $X$ in thermal equilibrium with temperature $T_{X}$ with an energy $E_{X}$ and a chemical potential $\mu_{X}$, the thermal distribution fuction is   

\begin{equation} 
f_{X}(\vec{p},\vec{x},t)=\frac{1}{e^{\frac{(E_{X}-\mu_{X})}{T_{X}}} \pm 1}, 
\end{equation}
with $E_{X}^{2} = |\vec{p}|^{2}+m_{X}^{2} \ $. The sign is positive for a Fermi-Dirac particle (fermion) and negative for a Bose-Einstein particle (boson). Using the cosmological principle and the Robertson-Walker (RW) metric, the distribution function is simplified to a function that depends only on the absolute value of the momentum of the particle and on time, hence $f(\vec{p},\vec{x},t)=f(|\vec{p}|,t)$. Noting that each particle $X$ has $g_{X}$ internal degrees of freedom, the number density $n_{\scaleto{X}{4pt}}$, the energy density $\rho_{\scaleto{X}{4pt}}$ and the pressure $p_{\scaleto{X}{4pt}}$ of particles $X$ can be defined as

\begin{eqnarray}
n_{\scaleto{X}{4pt}}(t) & = & \frac{g_{X}}{(2 \pi)^{3}} \int f_{X}(|\vec{p}|,t) d^{3} p \label{numberdens} \\
\rho_{\scaleto{X}{4pt}}(t) & = & \frac{g_{X}}{(2 \pi)^{3}} \int E_{X}(|\vec{p}|)f_{X}(|\vec{p}|,t) d^{3} p \\
p_{\scaleto{X}{4pt}}(t) & = & \frac{g_{X}}{(2 \pi)^{3}} \int \frac{|\vec{p}|^{2}}{3 E_{X}(|\vec{p}|)} f_{X}(|\vec{p}|,t) d^{3} p .
\end{eqnarray}


The annihilation rate of the DM particles can be defined as 
\begin{equation} 
\Gamma_{ann, \scaleto{X}{4pt}}= \langle \sigma_{ann} v \rangle n_{eq, \scaleto{X}{4pt}}, 
\end{equation}
where $\sigma_{ann}$ is the DM annihilation cross-section, $v$ is the relative velocity of the DM particles, the angle brackets denote an average over the DM thermal distribution and $n_{eq, \scaleto{X}{4pt}}$ is the number density of the dark matter particle in chemical equilibrium. For non-relativistic particles, using the Maxwell-Boltzmann approximation on eq.(\ref{numberdens}) we obtain 

\begin{equation}
n_{eq, \scaleto{X}{4pt}}= \frac{g_{\scaleto{X}{4pt}}}{(2 \pi)^{3}} \int e^{-\frac{E_{\scaleto{X}{4pt}}}{T}} d^{3} p = g_{\scaleto{X}{4pt}}\left(\frac{m_{\scaleto{X}{3.5pt}} T}{2 \pi}\right)^{3/2} e^{-m_{\scaleto{X}{3pt}}/T} , \text{   for  } T\ll m_{\scaleto{X}{4pt}}.
\end{equation}

The evolution of the number density $n_{\scaleto{X}{4pt}}$ of dark matter $X$ is governed by the Boltzmann equation \cite{kt,Dodelson:2003ft}
\begin{equation} 
\frac{d n_{\scaleto{X}{4pt}}}{dt}= -3 H n_{\scaleto{X}{4pt}}- \langle \sigma_{ann} v \rangle (n_{\scaleto{X}{4pt}}^{2}-n_{eq, \scaleto{X}{4pt}}^{2}),  \label{BoltzmannEq} 
\end{equation}
where $H$ is the Hubble parameter and $\langle \sigma_{ann} v \rangle$ is the {\it thermally averaged} annihilation cross-section times the relative velocity of the dark matter $X$. The first term on the right-hand side of eq.(\ref{BoltzmannEq}) shows the effect of the expansion of the Universe, while the second term represents the change in the number density of the dark matter due to annihilations and creations. The WIMP number density $n_{\scaleto{X}{4pt}}$ that appears in eq.(\ref{BoltzmannEq}) is the sum of the number densities of each species $i$ that eventually annihilates into the DM particle,
\begin{equation} 
 n_{\scaleto{X}{4pt}} = \sum_{\substack{l=1}}^{N} n_{\scaleto{l}{5pt}},
\end{equation}
where $N$ is the total number of such species. The thermally averaged annihilation cross-section contains the essencial pieces of particle physics that are necessary to calculate the WIMP number density, and it must be computed in the context of a specific BSM model.

It is customary to solve eq.(\ref{BoltzmannEq}) by introducing the yield $Y \equiv n_{\scaleto{X}{4pt}}/s$ and a new dimensionless variable $x= m_{\scaleto{X}{5pt}}/T$, where $s$ is the entropy density and $T$ is the photon temperature. In terms of these new variables, eq.(\ref{BoltzmannEq}) simplifies to \cite{kt,Bertone:2004pz}
\begin{equation} 
\frac{d Y}{dx} = -\frac{\langle \sigma_{ann} v \rangle s}{Hx}\left(Y^{2}-Y^{2}_{eq}\right).   \label{BoltzmannEqY}
\end{equation} 
The WIMP dark matter relic density is defined as the ratio of the WIMP mass density and to the critical density
\begin{equation} 
\Omega_{\scaleto{X}{4pt}} h^{2} =\frac{\rho_{\scaleto{X}{4pt}}}{\rho_{c}} h^{2}=\frac{n_{\scaleto{X}{4pt}} m_{\scaleto{X}{4.5pt}}}{\rho_{c}} h^{2}.
\end{equation} 
The present day relic density is obtained by taking the value of the yield at $T \rightarrow 0$
\begin{equation} 
\Omega_{\scaleto{X}{4pt}} h^{2} =\frac{Y(x) s(x) m_{\scaleto{X}{4.5pt}}}{\rho_{c}(x)} h^{2} \bigg|_{x \rightarrow \infty}= \frac{Y_{0} s_{0} m_{\scaleto{X}{4.5pt}}}{\rho_{c, 0}} h^{2} 
\end{equation}
where $h=H_{0}/(100 \text{km} \cdot \text{s}^{-1} \cdot \text{Mpc}^{-1}) \approx 0.678$ is the dimensionless Hubble parameter \cite{Ade:2015xua}, $s_{0} \approx 2.9 \times 10^{-3}$ is the present day entropy density \cite{kt} and $\rho_{c, 0} \approx 8 \times 10^{-47} \ h^{2} \ \text{GeV}^{4}$ \cite{Patrignani:2016xqp} is the present critical density. The numerical solution of eq.(\ref{BoltzmannEqY}) shows that at high temperatures $Y$ is close to its equilibrium value $Y_{eq.}$ and, as the temperature decreases, $Y_{eq.}$ becomes exponentially suppressed in such a way that $Y$ can no longer track its equilibrium value and then levels off to a frozen-out constant value. In the scenario of standard cosmology, the WIMP freeze-out temperature is obtained for $10 \lesssim x\lesssim 30$. The evolution of the WIMP abundance per comoving volume is shown in Figure~\ref{FigRELICDENSITY}, which also shows that for higher annihilation cross section the freeze-out time is later.  

\begin{figure}[h!]
\centering
\includegraphics[width=0.60\textwidth]{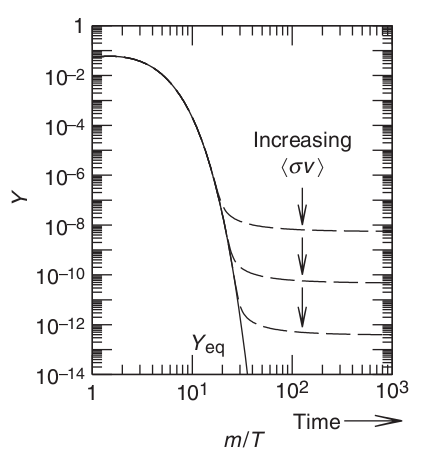}
\caption{\label{fig:overleaf} The dynamics of the WIMP comoving number density during the freeze-out epoch. The dashed curves show the current abundance, while the solid curve is the equilibrium abundance.} \label{FigRELICDENSITY}
\end{figure}

\subsection{Precise calculation}
\indent

The standard way to calculate the present day abundance of dark matter is by solving the Boltzmann equation, expand $\sigma_{ann} v$ in powers of $v^{2}$ 
\begin{equation} 
\sigma_{ann} v = a + bv^{2} + ...,
\end{equation}
and then take the thermal average of this expansion using the Maxwell-Boltzman velocity distribution. The first term of this expansion comes from an s-wave annihilation (L=0) while the second one comes from s- and p-wave annihilation (L=1). There are cases in which $a$ is dominant where $\sigma v$ would be energy independent, but there are also cases, for example for a Majorana particle, where the s-wave annihilation is helicity suppressed and thus the p-wave term must be taken into account. 

This expansion can be done when $\langle \sigma_{ann} v \rangle$ varies slowly with the energy. However, there are some special cases in which this does not work \cite{Gondolo:1990dk,Griest:1990kh}, and thus needs deeper attention: 

\begin{itemize}

\item {\bf s-channel resonance}: when the DM annihilates into other particles via an s-channel process in which the mediator mass is approximately $M_{S} \approx 2m_{X}$. Since the annihilation cross section is not a smooth function of the Mandelstam variable s in the vicinity of the pole of an s-channel diagram, the velocity expansion of  $\langle \sigma_{ann} v \rangle$ fails;
  
\item {\bf Annihilation thresholds}: when the mass of the DM particle is close to the threshold of the annihilation channel $X + \overline{X} \rightarrow A+B$ with $2 m_{\scaleto{X}{5pt}} \approx m_{A} + m_{B}$. In this case, the velocity expansion of $\langle \sigma_{ann} v \rangle$ diverges at the threshold energy;

\item {\bf Coannihilations}: when there are other exotic particles $X_{i}'$ in the model that can contribute to the calculation of the annihilation cross sections through processes like
\[
\\ X + X_{i}' \rightarrow SM+SM \quad\quad\quad\quad \text{or} \quad\quad\quad\quad X_{i}'+ \overline{X}_{i}' \rightarrow SM+SM.
\]
Coannihilation effects are significantly relevant for the relic density calculation when the exotic particle and the DM are almost degenerate in mass, with a mass difference of about $m_{\scaleto{X_{i}'}{7pt}}-m_{\scaleto{X}{5pt}} \lesssim 0.1 m_{\scaleto{X}{5pt}}$.

\end{itemize}
All these special cases are now taken into account in the numerical tool MicrOMEGAs \cite{Belanger:2001fz,Belanger:2014hqa,Belanger:2014vza}, which was used in Part II of this thesis to study the DM properties of the right-handed sneutrino. The obtained value for the DM relic density from MicrOMEGAs can then be compared to the current observed value \cite{Ade:2015xua}
\begin{equation} 
\Omega_{\rm DM} h^2 = 0.1188\pm 0.0010\,,
\end{equation}
to see if a point of the parameter space of a model is in agreement or not with current cosmological observations. MicrOMEGAs code also calculates with good precision the thermally averaged annihilation cross section at different temperatures and the rates for direct and indirect detection of dark matter of a specific model.





\part{Interparticle Potentials}

\chapter{Topologically Massive Spin-1 Particles and Spin-Dependent Potentials}
\label{chapterPAPER1}
\indent 
 
\section{Introduction}
\indent

Propagators are read off from the quadratic part of a given Lagrangean density and depend on intrinsic attributes of the fields, such as their spin. Most of the literature is concerned with spin-1 bosons in the $\left(\frac{1}{2}, \frac{1}{2}\right)$-representation of the Lorentz group (e.g., photon). Here, we would like to address the following questions: for two different fields representing the same sort of (on-shell) spin-1 particle, which role does a particular representation play in the final form of the interaction? Is the form of the mass term (corresponding to some specific mass-generation mechanism) determinant for the macroscopic characterization of the interparticle potential?

The amplitude for the elastic scattering of two fermions is sensitive to the fundamental, microscopic, properties of the intermediate boson. Our work  sets out to study the potential generated by the exchange of two different classes of neutral particles: a Proca (vector) boson and a rank-2 anti-symmetric tensor, the Cremer-Scherk-Kalb-Ramond (CSKR) field \cite{Kalb:1974yc,Cremmer:1973mg}, mixed to another vector boson, i.e., the $\left\{ A_{\mu}, B_{\nu \kappa}\right\}$-system with a topological mixing term. Two-form gauge fields are typical of off-shell SUGRA multiplets in four and higher dimensions \cite{Teitelboim:1985ya,Teitelboim:1985yc,Gates:1980ay,Binetruy:2000zx,deWit:2002vz} and the motivation to take them into consideration is two-fold:

i)	They may be the messenger, or the remnant, of some Physics beyond the Standard Model. This is why we are interested in understanding whether we may find out the track of a 2-form gauge sector in the profile of spin-dependent potentials.

ii)	In four space-time dimensions, a pure on-shell rank-2 gauge potential actually describes
 a scalar particle. However, off-shell it is not so. This means that the quantum 
fluctuations of a rank-2 gauge field may induce a new pattern of spin-dependence. Moreover, its mixing 
with an Abelian gauge potential sets up a different scenario to analyse potentials 
induced by massive vector particles.

Our object of interest is a neutral massive spin-1 mediating particle, which we might identify as a sort of massive photon. Such a particle is extensively discussed in the literature, dubbed as $Z^{0'}$-particle. In the review articles 
of Ref. \cite{Hewett:1988xc,Leike:1998wr,Langacker:2008yv}, the authors present an exhaustive list of different $Z^{0'}$-particles and phenomenological constraints on their masses and couplings. In this chapter, we shall be studying interaction potentials between fermionic currents as induced by $Z^{0'}$ virtual particles; their effects are then included in the interparticle potentials we are going to work out. Therefore, the velocity- and spin-dependence of our potentials appear as an effect of the interchange of a virtual $Z^{0'}$-particle.

We exploit a variety of couplings to ordinary matter in order to extract possible experimental signatures 
that allow to distinguish between the two types of mediation in the regime of low-energy 
interactions. Just as in the usual electromagnetic case, where the 4-potential is subject to gauge-fixing 
conditions to reduce the number of degrees of freedom (d.o.f.), we shall also impose 
gauge-fixing conditions to the $\left\{A_{\mu}, B_{\nu \kappa} \right\}$-system in order to ensure that only the spin-1 d.o.f. survives. From the physical side, we expect those potentials to exhibit a polynomial correction (in powers 
of $1/r$) to the well-known $e^{-m_{0}r}/r$ Yukawa potential. This implies that a laboratory aparatus with 
typical dimensions of $\sim mm$ could be used to examine the interaction mediated by 
massive bosons with $m_{0} \sim 10^{-3} eV$.

Developments in the measurement of macroscopic interactions between unpolarized and 
polarized objects \cite{Adelberger:1992ph,Adelberger:2003zx}\cite{Hunter:2013hza,Ledbetter:2012xd,Heckel:2013ina,Kotler:2014qxa} are able to 
constrain many of the couplings between electrons and nucleons (protons and neutrons), so 
that we can concentrate on more fundamental questions, such as the impact of the particular field 
representation of the intermediate boson in the fermionic interparticle potential. To 
this end, we discuss the case of monopole-dipole interactions in order to directly compare the Proca and $\left\{A_{\mu}, B_{\nu \kappa} \right\}$-mechanisms. We shall also present bounds on the vector/pseudo-tensor couplings that 
arise  from a possible application to the study of the hydrogen atom.
 
We would like to point out that our main contribution here is actually to associate
different field representations (which differ from each other by their respective 
off-shell d.o.f.) to the explicit spin-dependence in the particle potentials we derive. 
Rather than focusing on the constraints on the parameters, we aim at an understanding of 
the interplay between different field representations for a given spin and spin-spin 
dependence of the potentials that appear from the associated field-theoretic models. This shall be explicitly highlighted in the end of Section \ref{CHAP3subsection3.4.2}. We anticipate here however that
four particular types of spin- and velocity-dependences show up only in the topologically massive case we
discuss here. The Proca-type massive exchange do exclude these four terms, as it shall
become clear in Section \ref{CHAP3subsection3.4.2}.

\section{The Cremmer-Scherk-Kalb-Ramond field}  \label{CSKR}
\indent

In 1974, Cremmer and Scherk proposed a new mass generation mechanism \cite{Cremmer:1973mg} different than the Higgs mechanism which was used in the context of dual
models applied to strong interactions. This mechanism of mass generation is based on a pair of fields, namely, a 4-vector field and a 2-rank anti-symmetric tensor field which are connected via a topological mixing term. Few days later, Kalb and Ramond introduced a similar system of fields to study the equations of motion of the classical interaction between strings \cite{Kalb:1974yc}. In this section we introduce the $\left\{A_{\mu}, B_{\nu \kappa} \right\}$-system of the topologically massive fields and explain its properties.

The Proca vector field transforms under the $\left(\frac{1}{2},\frac{1}{2}\right)$-representation of the Lorentz group and its Lagrangean is the simplest extension leading to a massive intermediate vector boson, but it is not the only one. A massive spin-1 particle can also be described through a gauge-invariant formulation: a vector and a tensor fields connected by a mixing topological mass term \cite{Cocuroci:2013bga}. Both the vector $A_{\mu}$ and the tensor $B_{\mu\nu}$ are gauge fields described by the following Lagrangean:
\begin{equation}
\mathcal{L}_0 = - \frac{1}{4} \, F_{\mu \nu}^2 + \frac{1}{6} \, G_{\mu \nu \kappa 
}^2 + \frac{m_0}{\sqrt{2}} \,  \epsilon^{\mu \nu \alpha \beta} \, A_\mu \partial_\nu B_{\alpha \beta}, 
\label{L_B}
\end{equation}
where the field-strength tensor for the vector field is given by $F_{\mu \nu} = \partial_\mu A_\nu - \partial_\nu A_\mu $ and the field-strength for the anti-symmetric tensor is $G_{\mu \nu \kappa } = \partial_\mu B_{\nu \kappa } + \partial_\nu 
B_{ \kappa \mu} + \partial_\kappa B_{\mu \nu }$. The origin of the term ``topological'' lies in the fact that this term does not depend on the metric of the space-time, which implies that it does not contribute to obtain the energy-momentum tensor.

The action is invariant under the independent local Abelian gauge transformations given by
\begin{eqnarray}
A'_\mu & = & A_\mu + \partial_\mu \alpha  \label{Agauge} \\
B'_{\mu \nu} & = & B_{\mu \nu} + \partial_\mu \beta_\nu - \partial_\nu \beta_\mu 
\label{Bgauge}
\end{eqnarray}
and, because of the anti-symmetry of $B_{\mu\nu}$, the vector function $\beta_{\mu} (x)$ can also suffer the following gauge transformation that does not affect the primary gauge transformations:
\begin{equation} 
\beta_{\mu}(x)' = \beta_{\mu}(x) + \partial_{\mu} f(x). \label{betatransformation}
\end{equation}
As one can note, from the original four degrees of freedom of the vector field $A_{\mu}$ one is eliminated by choosing a gauge in eq.\ \eqref{Agauge} and, from the original six degrees of freedom of the 2-rank anti-symmetric tensor field $B_{\mu\nu}$, three are eliminated by the combination of eqs.\ \eqref{Bgauge} and \eqref{betatransformation}. 

It can be shown that together with the equations of motion, the pair $\{A_{\mu}, B_{\nu \kappa}\}$ carries three (on-shell) degrees of freedom, being, therefore, equivalent to a massive vector field. It is interesting to note that, contrary to the typical Proca case, the topological mass term does not break gauge invariance, so that no spontaneous symmetry breakdown is invoked. 

Before moving to the analysis of the field equations, let's briefly resume the behavior of the model in the massless limit with $m_{0}=0$. In this limit, the fields $A^{\mu}$ and $B^{\mu\nu}$ are decoupled and behave as free fields that obey the equations $\partial F^{\mu\nu}=0$ and $\partial_{\mu}G^{\mu\nu\kappa}=0$. These equations of motion together with the gauge transformations and the Bianchi identities shows us that $A^{\mu}$ is equivalent to the electromagnetic massless photon with two degrees of freedom. And if we solve the equations for $G$ using $G^{\mu\nu\kappa}= \epsilon_{\mu\nu\kappa\lambda} \partial^{\lambda}\varphi$, where $\varphi$ is a scalar function, we see that the free $B^{\mu\nu}$ describes effectively only one massless scalar degree of freedom carried by $\varphi$.

Now let's return to the topologically massive model described by the Lagrangian \eqref{L_B}. After applying the variational principle to this Lagrangian, we obtain the following set of field equations

\begin{eqnarray}
\partial_{\mu} F^{\mu\nu} + \sqrt{2}m_{0}\tilde{G}^{\nu} & = & 0    \label{FirstCSKRequation}
\\  
\partial_{\mu}G^{\mu\nu\kappa} - \frac{m_{0}}{\sqrt{2}}\tilde{F}^{\nu\kappa} & = & 0,  \label{SecondCSKRequation} 
\end{eqnarray}
where $\widetilde{G}^\nu = \frac{1}{6}\epsilon^{\nu\alpha\beta\delta}G_{\alpha\beta\delta}$ is the dual field-strength of $B^{\mu\nu}$. By operating on these equations with $\epsilon \partial$ and using the Bianchi identities, one can extract two wave equations for the dual field-strengths, which are
 
\begin{eqnarray}
\left(\Box + m_{0}^{2} \right)\tilde{G}^{\mu} & = & 0   
\\  
\left( \Box + m_{0}^{2} \right)\tilde{F}^{\mu\nu} & = & 0.   
\end{eqnarray}
Note that, as anticipated, the excitations described by the fields are massive excitations with mass $m_{0}$. A good way to see how the degrees of freedom are shared by the $A^{\mu}$ and $B^{\mu\nu}$ fields is by writing $\tilde{G}^{\nu} = \partial^{\nu}\xi + \frac{m_{0}}{\sqrt{2}} A^{\nu}$ to solve eq.\ \eqref{SecondCSKRequation} and rewrite eq.\ \eqref{FirstCSKRequation} as 

\begin{equation}
\left(\Box + m_{0}^{2} \right) A^{\mu} - \partial^{\mu} \left(\partial_{\nu}A^{\nu} - \sqrt{2} m_{0}\xi \right).
\label{NewFirstCSKRequation}
\end{equation}

If we choose $\partial_{\nu}A^{\nu} = \sqrt{2} m_{0}\xi$ as our gauge-fixing condition, we see that the equation above describes a free massive vector boson with mass $m_{0}$. Additionally, this gauge choice shows that the longitudinal component of $A^{\mu}$ is described by a scalar degree of freedom that originally belongs to $B^{\mu\nu}$. This dynamical transfer of degrees of freedom between the gauge fields $A^{\mu}$ and $B^{\mu\nu}$ happened thanks to the topological mixing term. Note that here the vector field acquired a mass without breaking the gauge symmetry. 

\section{Methodology}    \label{CHAP3section3.2}
\indent

Let us first establish the kinematics of our problem. We are dealing with two fermions, 1 and 2, which scatter elastically. If we work in the center of mass frame (CM), we can assign them momenta as indicated in Fig.\ \eqref{Fig3.1} below, where $\vec{q}$ is the momentum transfer and $\vec{p}$ is the average momentum of fermion 1 before and after the scattering. 
\begin{figure}[h!]
\centering
\includegraphics[width=0.5\textwidth]{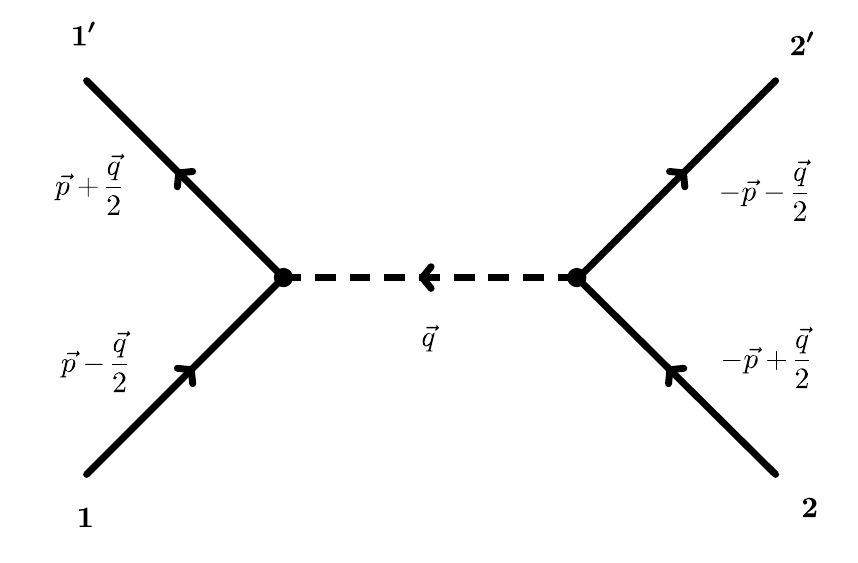}
\caption{\label{fig:overleaf}Basic vertex structure and momentum assignments.} \label{Fig3.1}
\end{figure}

Given energy conservation and our choice of reference frame, one can readily show that $\vec{p} \cdot \vec{q} = 0$ and that $q^{\mu}$ is space-like: $q^{2} = -\vec{q}^{\,2}$. The amplitude will be expressed in terms of $\vec{q}$ and $\vec{p}$ and we shall keep only terms linear in $|\vec{p}|/m_{1,2}$. It will also include the spin of the particles involved.

According to the first Born approximation, the two-fermion potential can be obtained from the Fourier transform of the tree-level momentum-space amplitude with respect to the momentum transfer $\vec{q}$
\begin{equation} 
V(r ,v) = - \int \frac{d^3\vec{q}}{(2\pi)^3} \, e^{i \vec{q} \cdot \,\vec{r}} \, \mathcal{A} ( \vec{q} , m \vec{v} ), 
\end{equation}
where $\vec{r}$, $r$ and $v = |\vec{p}|/m_{1,2}$ are the relative position vector, its modulus and average velocity of the fermions, respectively. The long-range behaviour is related to the non-analytical pieces of the amplitude in the non-relativistic limit \cite{Feinberg:1989ps}. We evaluate the fermionic currents up to first order in $|\vec{p}|/m_{1,2}$ and $|\vec{q}|/m_{1,2}$, as indicated in the Appendix \ref{AppendixA} (an important exception is discussed in Section \ref{CHAP3subsection3.4.2} in connection with the mixed propagator $\langle A_{\mu} B_{\nu\kappa} \rangle$ since, in that case, contact terms arise).

We restrict ourselves to tree-level amplitudes since we are considering weakly interacting particles, thus carrying tiny coupling constants that suppress higher-order diagrams. The typical outcome are Yukawa-like potentials with extra $1/r$ contributions which also depend on the spin of the sources, as well as on their velocity. Contrary to the usual Coulomb case, spin- and velocity-dependent terms are the rule, not exception.

\section{The pure spin-1 case: the Proca field}     \label{CHAP3section3.3}
\indent

In order to establish the comparison between the two situations that involve a massive spin-1 particle, we start off by quickly reviewing the simplest realization of a neutral massive vector particle, the Proca field $A_{\mu}(x)$, described by the Lagrangean
\begin{equation} 
\mathcal{L}_{Proca} = - \frac{1}{4} \, F_{\mu \nu}^2 + \frac{1}{2} \, m_0^2 \, 
A_\mu^2 . \label{L_proca} 
\end{equation} 

Since we are concerned with the interaction mediated by such a field, it is necessary to calculate its propagator, $\langle A_\mu A_\nu\rangle$. The Lagrangean above can be suitably rewritten as $\frac{1}{2} A^{\mu} \mathcal{O}_{\mu \nu} A^{\nu}$, in which the operator $\mathcal{O}_{\mu \nu}$, essentially the inverse of the propagator, is $\mathcal{O}_{\mu \nu} = \left( \Box + m_0^2 \right) \, 
\theta_{\mu \nu} + m_0^2 \omega_{\mu \nu}$, where we introduced the transverse and longitudinal projection operators defined as
\begin{eqnarray}
\theta_{\mu \nu} & \equiv & \eta_{\mu \nu} - \frac{\partial_\mu \partial_\nu}{\Box}, \\
\omega_{\mu \nu} & \equiv & \frac{\partial_\mu \partial_\nu}{\Box}, \label{projproc} 
\end{eqnarray}
which satisfy $\theta^{2} = \theta$, $\omega^{2} = \omega$, $\theta\omega = 0$ and $\theta + \omega = 1$. Due to these simple algebraic properties it is easy to invert $\mathcal{O}_{\mu \nu}$ and, transforming to momentum space, we finally have
\begin{equation} 
\langle A_\mu A_\nu\rangle = - \frac{i}{k^2 - m_0^2} \left( \eta_{\mu 
\nu} - \frac{ k_\mu k_\nu }{m_0^2} \right), \label{propproc} 
\end{equation}
from which we proceed to the calculation of the potentials.

Let us solve in more detail the case of two fermionic vector currents interacting via the Proca field. Using the parametrization of Fig.\eqref{Fig3.1} and applying the Feynman rules, we get 
\begin{eqnarray*} 
i \mathcal{A}_{V-V}^{Proca} & = & \bar{u}(p + q/2) \left\{ i g^V_1 
 \gamma^\mu  \right\} u(p - q/2) \langle A_\mu A_\nu\rangle \times \\ \nonumber
& \times & \bar{u}(- p - q/2) \left\{ i g^V_2 \gamma^\nu  \right\} 
u(- p + q/2)
\end{eqnarray*}
with $g^{V}_1$ and $g^{V}_2$ refering to the coupling constants. The equation above can be put in a simpler form as below
\begin{equation} 
\mathcal{A}_{V-V}^{Proca} =  i \, J_{1}^\mu \, \langle  A_\mu A_\nu \rangle \, 
J_{2}^\nu. 
\end{equation} 

If we use that $q^{0}=0$ and current conservation, we find that the amplitude is $\mathcal{A}_{V-V}^{Proca} =  -\frac{1}{\vec{q}^{\, 2} + m_0^2} J_1^\mu \, J_{2 \mu}$ and, according to eq.\ \eqref{vector_c_2}, we have $J_1^i \, J_{2i} \sim \mathcal{O}( v^2/c^2)$. Therefore, only the term $J^0_1 \, J_{20} \approx g^V_1 g^V_2 \delta_1 \delta_2$ contributes to the scattering amplitude, thus giving
\begin{equation} 
\mathcal{A}_{V-V}^{Proca} =  -g^{V}_1  g^{V}_2 \, \frac{\delta_1 
\delta_2}{\vec{q}^{\,2} + m_0^2}, \label{eq7}
\end{equation}
where $\delta_i$ ($i=1,2$ labels the particles) is such that $\delta_i=+1$ if the $i$-th particle experiences no spin flip in the interaction, and $\delta_i=0$ otherwise. In the eq.\ \eqref{eq7} above, the global term $\delta_1 \delta_2$ is present to indicate that the amplitude is non-trivial only if both particles do not flip their respective spins. If one of them changes its spin the potential vanishes. This means that this interaction only occurs with no spin flip. In what follows, we shall come across situations where only a single $\delta_i$ appears, thus justifying the effort to keep the $\delta_i$ explicit. 

Finally, we take the Fourier transform in order to obtain the potential between two static (vector) currents, 
\begin{equation} 
V^{Proca}_{V-V} =  \frac{g^{V}_1  g^{V}_2 \delta_1 \delta_2}{4 \pi}  \, \frac{e^{- m_0 r}}{r}, \label{eq8} 
\end{equation}
which displays the well-known exponentially suppressed repulsive Yukawa behaviour typical of a massive $s=1$ boson exchange. In our notation, the potential is indicated as $V_{v_{1} - v_{2}}$, where $v_{1,2}$ refer to the vertices related to the particles 1 and 2. In the case above, the subscripts $V$ stand for vector currents. As already announced, the typical decay length is $1/m_0$ and we expect that very light bosons will be measurable for (laboratory) macroscopic distances, e.g. for masses of $\sim 10^{-3} eV$, we have ranges of $d \sim 1 mm$.

Following the same procedure, we can exploit other situations, namely: vector with pseudo-vector currents and two pseudo-vector currents. The results are cast in what follows:
\begin{eqnarray}
V^{Proca}_{V-PV} & = &  -\frac{g_1^V g_2^{PV}}{4 \pi } \Big\{  \vec{p} \,
\cdot \langle \vec{\sigma}\rangle_2  \, \frac{\delta_1}{r} \left[ \frac{1}{m_1} + 
\frac{1}{m_2} \right] + \nonumber \\
& + & \frac{\left( 1 + m_0 r \right)}{2 m_1 r^2} \, \left[\langle \vec{\sigma}\rangle_1 \times \langle \vec{\sigma}\rangle_2 \right] \cdot \hat{r} \,  \Big\} \, e^{-m_0 r} 
\end{eqnarray}
\begin{equation}
V^{Proca}_{PV-PV} =  -\frac{g_1^{PV} g_2^{PV}}{4 \pi } \, 
\langle \vec{\sigma}\rangle_1 \cdot \langle \vec{\sigma}\rangle_2  \, \frac{e^{-  m_0 
r}}{r},\end{equation}
and we notice that all kinds of spin-dependent interactions appear while the $r$ factors are limited to $r^{-2}$. It is also easy to see that $V^{Proca}_{PV-PV}$ and $V^{Proca}_{V-PV}$ are even and odd against a parity transformation, respectively. In the next section, we shall conclude that a richer class of potentials is generated if the massive spin-1 Abelian boson exhibits a gauge-invariant mass that comes from the mixing between a one- and a two-form potentials.

\section{The topologically massive spin-1 case}    \label{CHAP3section3.4}
\indent

Even though the Proca field and the mixed $\{A_{\mu}, B_{\nu \kappa}\}$-system describe both an on-shell spin-1 massive particle, these two cases are significantly different when considered off-shell. Our topologically massive spin-1 system displays 6 d.o.f. when considered off-shell (since gauge symmetry allows us to eliminate 4 compensating modes), whereas the Proca field carries 4 off-shell d.o.f. (the subsidiary condition, which is an on-shell statement, eliminates one d.o.f.). It is the on-shell spin-1 massive boson corresponding to the mixed $\{A_{\mu}, B_{\nu \kappa}\}$-system that we refer to as our $Z^{0'}$-type particle. Its exchange between external fermionic currents gives rise to the classes of interparticle potentials we wish to calculate and discuss in this project.

On the other hand, since the potential evaluation is an off-shell procedure, we consider relevant to compare both situations bearing in mind that the potential profiles may indicate - if we are able to set up an experiment - whether a particular mechanism is preferable in the case of a specific physical system. Characteristic aspects of the potentials in these two situations might select one or other mechanism in some possible physical scenario, therefore being able to distinguish between different BSM models.

Our goal is to investigate the potentials between fermions induced by the exchange of the mixed vector and tensor fields and compare the spin-, velocity- and distance-dependence against the Proca case. To do that, we need, first of all, to derive the whole set of propagators.


\subsection{The propagators}   \label{CHAP3subsection3.4.1}
\indent

As in Section \ref{CHAP3section3.3}, it is important to obtain suitable spin operators in order to 
obtain the propagators of the model. The spin operators that act on an 
anti-symmetric 2-form are
\begin{equation} 
\left( P^1_b \right)_{\mu \nu  , \, \rho \sigma } \equiv \frac{1}{2} 
\left( 
\theta_{\mu \rho} \, \theta_{\nu \sigma} - \theta_{\mu \sigma} \, \theta_{\nu \rho}
\right) \label{proj_B_1} 
\end{equation}
\begin{equation} 
\left( P^1_e \right)_{\mu \nu  , \, \rho \sigma } \equiv \frac{1}{2} 
\left( \theta_{\mu \rho} \, \omega_{\nu \sigma} + \theta_{\nu \sigma} \, \omega_{\mu \rho} -
\theta_{\mu \sigma} \, \omega_{\nu \rho} - \theta_{\nu \rho} \, \omega_{\mu \sigma}
\right) \label{proj_B_2} 
\end{equation}
which are anti-symmetric generalizations of 
the projectors $\theta_{\mu\nu}$ and $\omega_{\mu\nu}$ 
\cite{Ferreira:2000pi,proj_proca_1,Kuhfuss:1986rb}. The comma indicates that we have anti-symmetry 
in changes $\mu \leftrightarrow \nu$ or $ \rho \leftrightarrow \sigma$. The algebra 
fulfilled by these operators is collected in the Appendix \ref{AppendixB}. We quote them since they are very useful in the extraction of the propagators from Lagrangean \eqref{L_B}.

Adding up the gauge-fixing terms to the Lagrangean \eqref{L_B},
\begin{equation} 
\mathcal{L}_{g.f.} = \frac{1}{2 \alpha} \, \left( \partial_\mu A^\mu \right)^2 + 
\frac{1}{2 \beta} \, \left( \partial_\mu B^{\mu \nu} \right)^2 \, ,\end{equation}
yields the full Lagrangean $\mathcal{L} = \mathcal{L}_0 + \mathcal{L}_{g.f.}$. In terms 
of the spin
operators, $\mathcal{L}$ can be cast in a more compact form as:

\begin{equation} 
\mathcal{L} = \frac{1}{2} \, \left( \begin{array}{cc} A^\mu & B^{\kappa \lambda }  
 \end{array}\right) \left( \begin{array}{cc} P_{\mu \nu} & Q_{\mu \rho \sigma }  \\
R_{\kappa \lambda \nu } & \mathbf{S}_{\kappa \lambda , \, \rho \sigma } 
\end{array}\right)
\left( \begin{array}{c} A^\nu \\ B^{\rho \sigma}  \end{array} \right),
\label{matrix_op} 
\end{equation}
where we identify
\begin{eqnarray} 
P_{\mu \nu} & \equiv & \Box \theta_{\mu \nu} - \frac{\Box}{\alpha} 
\omega_{\mu \nu} \\
Q_{\mu \rho \sigma } & \equiv & m_0 \, S_{\mu \rho \sigma}/\sqrt{2} \\
R_{\kappa \lambda \nu } & \equiv & - m_0 \, S_{\kappa \lambda \nu }/\sqrt{2} \\
\mathbf{S}_{\kappa \lambda , \, \rho \sigma } & \equiv & - \Box \left( 
P_b^1 \right)_{\kappa 
\lambda , \, \rho \sigma } - \frac{\Box}{2 \beta} \left( P_e^1 \right)_{\kappa \lambda 
, \, \rho \sigma }. 
\end{eqnarray}

With the help of Appendix \ref{AppendixB}, we invert the matrix operator in $(\ref{matrix_op})$ 
and read off the $ \langle A_\mu A_\nu \rangle$, $\langle A_{\mu} B_{\kappa \lambda 
}\rangle $ and $\langle B_{\mu \nu} B_{\kappa \lambda }\rangle $ momentum-space 
propagators, which turn out to be given as below: 

\begin{equation} 
\langle A_\mu A_\nu\rangle = - \frac{i}{k^2 - m_0^2} \, \eta_{\mu \nu} + i\left( 
\frac{1}{k^2 - m_0^2}  +  \frac{\alpha}{k^2} \right) \, \frac{k_\mu 
k_\nu}{k^2} \label{prop_AA} 
\end{equation}
\begin{equation} 
\langle B_{\mu \nu} B_{\kappa \lambda }\rangle =  \frac{i}{k^2 -  m_0^2} \, \left( 
P_b^1 \right)_{\mu \nu , \, \kappa \lambda } + \frac{ 2i \beta}{k^2} \, \left( P_e^1 
\right)_{\mu \nu , \, \kappa \lambda } \label{prop_BB} 
\end{equation}
\begin{equation} 
\langle A_\mu B_{\nu \kappa }\rangle =  \frac{m_{0}/\sqrt{2}}{k^2 \left( k^2 -  m_{0}^{2} 
\right)} \, \epsilon_{\mu \nu \kappa \lambda } \, k^\lambda \label{prop_AB}. 
\end{equation}

From the propagators above, we clearly understand that the massive pole $k^2= m_0^2$, 
present in $(\ref{prop_AA})$-$(\ref{prop_AB})$, actually describes the spin-1 massive
excitation carried by the set $\{A_{\mu}, B_{\nu \kappa}\}$.

In contrast to the off-shell regime of the so-called BF-model \cite{Birmingham:1991ty}, our 
non-diagonal $\langle A_\mu B_{\nu \kappa }\rangle$-propagator exhibits a massive pole 
and it cannot be considered separately from the $\langle A_\mu A_{\nu }\rangle$- 
and $\langle A_{\mu} B_{ \kappa \lambda}\rangle$-propagators: only the full set of 
fields together correspond to the 3 d.o.f. of the on-shell massive spin-1 boson we 
consider in our study.

Different from the point of view adopted in Ref. \cite{Allen:1990gb}, where the authors treat
the topological mass term as a vertex insertion (they keep the $ \langle A_\mu B_{\nu\rho} \rangle$- and
$ \langle B_{\mu \nu} B_{\kappa \lambda} \rangle$-propagators separately and with a trivial pole $k^2 =
0$), we consider it as a genuine bilinear term and include it in the
sector of 2-point functions. For that, we introduce the mixed spin operator $ S_{\mu \nu \kappa} $
in the algebra of operators and its final effect is to yield the mixed $ \langle A_\mu B_{\nu \kappa} \rangle$-propagator. The commom pole at $k^2 = m_{0}^2$ does not describe different particles, but a single
massive spin-1 excitation described by the combined $ \left\{ A_\mu , \, B_{\nu \kappa} \right\}$-fields, as already stated in the previous paragraph. Ref. \cite{Allen:1990gb} sums up the (massive) vertex insertions into the $ \langle A_\mu A_\nu \rangle$-propagator
which develops a pole at $k^2 = m^2$. They leave the $ \langle B_{\mu \nu} B_{\kappa \lambda} \rangle$-propagator aside because the $B_{\mu \nu}$-field does not interact with the fermions; the latter
are minimally coupled only to $A_\mu$.

On the other hand, in Ref. \cite{Leblanc:1993gx}, the topological mass term that mixes
$A_\mu$ and $B_{\nu \kappa}$ is generated by radiative corrections induced by the 4-fermion
interactions. So, for the sake of their calculations, the authors work with a massless vector
propagator whose mass is dynamically generated. This is not what we do here. In a more recent paper \cite{Diamantini:2013yka}, again an induced topological mass term mixes $A_\mu$
and $ B_{\nu \kappa} $ but, in this case, it is a topological current that radiatively generates the
mass. 

We point out the seminal paper by Cremmer and Scherk \cite{Cremmer:1973mg}, where they show that, for the spectrum analysis, it is possible to take the field-strength $G_{\mu \nu \kappa}$ and its dual $\widetilde{G}_\mu$, as fundamental fields, thus enabling them to go into a new field basis where a Proca-like field emerges upon a field redefinition. We cannot follow this road here, for our $B_{\mu \nu}$  is coupled to a tensor and to a pseudo-tensor currents in the process of evaluating some of our potentials. This prevents us from adopting $\widetilde{G}_\mu$ as a fundamental field, as it is done in \cite{Cremmer:1973mg}; this would be conflicting with the locality of the action. But, for the sake of analysing the spectrum, Cremmer and Scherk's procedure works perfectly well.

Finally, we also point out the paper by Kamefuchi, O' Raifeartaigh and Salam \cite{Kamefuchi:1961sb} that discusses the conditions on field reshufflings which do not change the physical results, namely, the $S$-matrix elements. A crucial point is that the change of basis in field space does not yield non-local interactions. So, once again, we stress that, once both the $A_\mu$- and the $B_{\mu \nu}$-fields interact with external currents, a diagonalization in the (free) kinetic Lagrangean leads to non-local terms and we, therefore, would not be able to control the equivalence between the results worked out with the two choices of field bases.


\subsection{The potentials}   \label{CHAP3subsection3.4.2}
\indent

We have already discussed the procedure to obtain the spin- and velocity-dependent potentials in previous 
sections. Thus, we shall focus on the particular case in which we have the 
propagator $ \langle B_{\mu \nu} B_{\kappa \lambda }\rangle $ and two 
tensor currents. In the following, we adopt the same parametrization of Fig.\ \eqref{Fig3.1}. After applying the 
Feynman rules, we can rewrite the scattering amplitude for this process as
\begin{equation} 
\mathcal{A}^{\langle BB\rangle}_{T-T} = i J^{\mu \nu}_1 \langle B_{\mu \nu} B_{\kappa 
\lambda }\rangle J_2^{\kappa \lambda } \label{amp_BB_TT} 
\end{equation}
with the tensor currents given by eq.\ \eqref{tensor_c_0}. Substituting the propagator $(\ref{prop_BB})$ in eq.\ $(\ref{amp_BB_TT})$ and eliminating its longitudinal sector (due to current conservation), we have
\begin{equation} 
\mathcal{A}^{\langle BB\rangle}_{T-T} =  -\frac{1}{q^2 - m_0^2} J_1^{\mu \rho } J_{2 
\mu \rho }. \label{eq38}
\end{equation}

The product of currents leads to $J_1^{\mu \rho } J_{2 \mu \rho } = 2 J_1^{0i } J_{2 \,0i } + 
J_1^{ij} J_{2 \, ij }$. However, according to eq.\ $(\ref{tensor_c_1})$, we conclude 
that $J_1^{0i } J_{2 \,0i } \sim \mathcal{O}(v^2/c^2)$ does not contribute to the 
non-relativistic amplitude. The term $J_1^{ij} J_{2 \, ij }$ can be simplified by using eq.$\ (\ref{tensor_c_2})$ (with the appropriate changes to the second current), so that we get 
\begin{equation}
\mathcal{A}^{\langle BB\rangle}_{T-T} = \frac{1}{2} \, \frac{g^T_1 
g^T_2}{\vec{q}^{\, 2} + m_0^2} \, \langle \vec{\sigma}\rangle_1 \cdot \langle \vec{\sigma}\rangle_2.
\end{equation} 
Performing the well-known Fourier integral, we obtain the non-relativistic spin-spin potential, namely
\begin{equation} 
V^{\langle BB\rangle}_{T-T}=  - \frac{g_1^T g_2^T}{8 \pi} \, \langle \vec{\sigma}\rangle_1 \cdot 
\langle \vec{\sigma}\rangle_2 \frac{e^{- m_0 r}}{r},\label{Pot_T_T} 
\end{equation}
and, similarly, we find the interaction potentials between tensor and pseudo-tensor currents as well as two pseudo-tensors currents to be
\begin{eqnarray}
V^{\langle BB\rangle}_{T-PT} & = &  \frac{g_1^T g_2^{PT}}{8 \pi r} \, \Big\{ \left(\frac{1}{m_{1}} + \frac{1}{m_{2}}\right) \, \vec{p} \cdot \left( \langle \vec{\sigma}\rangle_1 
\times \langle \vec{\sigma}\rangle_2 \right) + \nonumber \\
& + & \frac{\left( 1 + m_0 r \right)}{2r} \, \left( \frac{\delta_2}{m_2}  \langle \vec{\sigma}\rangle_1 - 
\frac{\delta_1}{m_1} \langle \vec{\sigma}\rangle_2  \right) 
\cdot \hat{r} \,  \Big\} \, e^{- 
m_0 r} \\ 
V^{\langle BB\rangle}_{PT-PT} & = & \frac{g_1^{PT} g_2^{PT}}{8 \pi} \, 
\langle \vec{\sigma}\rangle_1 
\cdot \langle \vec{\sigma}\rangle_2 \frac{e^{-m_0 r}}{r}.\label{Pot_PT_PT}
\end{eqnarray}

It is worthy comparing the potentials $(\ref{Pot_T_T})$ and $(\ref{Pot_PT_PT})$. We observe that they differ by a relative minus sign. This means that they exhibit opposite behaviors for a given spin configuration: one is attractive and the other repulsive. The physical reason is that the $PT-PT$ and $T-T$ potentials stem from different sectors of the currents: the $PT-PT$ amplitude is composed by the $(0i)-(0j)$ terms of the currents; the $T-T$ amplitude, on the other hand, arises from the $(ij)-(kl)$ components, as it can be seen from eq.\ \eqref{amp_BB_TT}. 

In the light of that, we check the structure of the $\langle B_{\mu \nu} B_{\kappa \lambda }\rangle$-propagator and it becomes clear that, in the case of the $\langle B_{0i} B_{0j}\rangle$-mediator, an off-shell scalar mode is exchanged. In contrast, in the $\langle B_{ij} B_{kl}\rangle$-sector the only exchange is of a pure $s=1$ (off-shell) quantum. It is well-known, however, that the exchange of a scalar and a $s=1$ boson between sources of equal charges yields attractive and repulsive interactions, respectively, therefore justifying the aforementioned sign difference between Eqs.$\ (\ref{Pot_T_T})$ and $(\ref{Pot_PT_PT})$.

For the mixed propagator $\ \langle A_\mu B_{\kappa \lambda }\rangle$, eq.\ \eqref{prop_AB}, we have 
four possibilities envolving the following currents: 
vector with tensor, vector with pseudo-tensor, pseudo-vector with 
tensor and pseudo-vector with pseudo-tensor. The results are given below:
\begin{eqnarray}
V^{\langle AB\rangle}_{V-T} & = & \frac{ g_1^{V} g_2^{T} \delta_1}{4\pi\sqrt{2}  m_0 r^2} \, 
\left[ 1 - \left( 1 +  m_0 r \right)  \, 
e^{-m_0 r} \right]\langle \vec{\sigma}\rangle_2 \cdot \hat{r} \\
& & \nonumber \\
V^{\langle AB\rangle}_{PV-T} & =  & \frac{g_1^{PV}g_2^{T}}{4\pi\sqrt{2}m_0 \mu r^2} \big[1 - \left( 1  
 + m_0 r \right)e^{-m_0 r}\big]\left( \langle \vec{\sigma}\rangle_1 
\cdot \vec{p} \right)\left( \langle \vec{\sigma}\rangle_2 \cdot \hat{r} \right) \\
& & \nonumber \\
V^{\langle AB\rangle}_{PV-PT} & = & \frac{ g_1^{PV} g_2^{PT}}{\sqrt{2} m_0} \biggl\{  \frac{\delta_{2}}{2m_{1}m_{2}}\left[\delta^3(\vec{r}) + \frac{m_0^2}{4 \pi r}e^{-m_0 r}\right] \langle \vec{\sigma} \rangle_{1}\cdot \vec{p}  + \nonumber \\  
& + &  \frac{1}{4\pi r^{2}} \left[ 1 - \left( 1 + m_0 r \right)e^{-m_0 r} \right] \left(\langle \vec{\sigma}\rangle_2 \times \langle \vec{\sigma}\rangle_1 \right) \cdot \hat{r}   \biggr\}.
\end{eqnarray}

The richest potential is the one between vector and pseudo-tensor sources, given by
\begin{eqnarray}
V_{V-PT}^{^{\langle AB\rangle}} & = & \frac{g^{V}_{1}g^{PT}_{2}}{\sqrt{2}m_{0}} \biggl\{ \frac{\delta_{1} 
\delta_{2}}{2m_{2}} \left[\delta^3(\vec{r}) + \frac{m_{0}^{2}}{4\pi r} e^{-m_{0}r} \right] + \nonumber \\ 
& + & \frac{\delta_{1}}{4 \pi \mu r^{3}}  \left[1 - \left(1 + m_{0}r\right)e^{-m_0 r}\right]\vec{L} \cdot \langle \vec{\sigma}\rangle_2 + \nonumber \\
& + & \frac{1}{2m_{1}}\left[\delta^3(\vec{r}) + \frac{m_{0}^{2}}{4\pi r}e^{-m_{0} r} - \frac{1}{4\pi r^{3}}\left[ 1 + \left( 1 + m_{0}r \right)e^{-m_{0}r} \right] \right]\langle \vec{\sigma}\rangle_1 \cdot \langle \vec{\sigma}\rangle_2 + \nonumber \\
& + & \frac{1}{8\pi m_{1} r^{3}}\left[ 3 + \left( 3 + 3m_{0}r + m_{0}^{2}r^{2} \right)e^{-m_{0}r} \right] \left( \langle \vec{\sigma}\rangle_1 \cdot \hat{r} \right) \left( \langle \vec{\sigma}\rangle_2 \cdot 
\hat{r} \right) \biggr\} \label{LS}
\end{eqnarray}
where we have introduced the reduced mass of the fermion system $\mu^{-1} = m_{1}^{-1} + m_{2}^{-1}$ and $\vec{L} = \vec{r} \times \vec{p}$ stands for the orbital angular momentum. 

Naturally, the contact terms do not contribute to a macroscopic interaction. Nevertheless, they are significant in quantum-mechanical applications in the case of {\it s-waves} which can overlap the origin. This is a peculiarity of $\langle A_\mu B_{\kappa \lambda }\rangle$-sector due to the extra $q^{2}$-factor in the denominator, which forces us to keep terms of order $|\vec{q}|^{2}$ in the current products. 

For the propagator $\langle A_\mu A_\nu\rangle$, eq.\ $(\ref{prop_AA})$, we find the 
same results as the ones in the Proca situation, due to current conservation. This means that, even though the vector field appears now mixed with the $B_{\mu \nu}$-field with a gauge-preserving mass term, for the sake of the interaction potentials, the results are the same as in the Proca case as far as the $A_\mu$-field exchange is concerned. We mention in passing that the $V^{\langle BB\rangle}_{T-T}$, $V^{\langle BB\rangle}_{PT-PT}$, $V^{\langle AB\rangle}_{PV-T}$ and $V^{\langle AB\rangle}_{V-PT}$ potentials are even under parity, while $V^{\langle BB\rangle}_{T-PT}$, $V^{\langle AB\rangle}_{V-T}$ and $V^{\langle AB\rangle}_{PV-PT}$ are odd. This difference is due to the presence of a single factor of the momentum transfer in the mixed propagator, eq.\ \eqref{prop_AB}.

We point out that experiments with ferrimagnetic rare earth iron garnet test masses \cite{Leslie:2014mua} could be a possible scenario to distinguish  the two different mass mechanisms. In the Proca mechanism, we obtained the following spin- and velocity-dependence: 
$ \vec{p} \cdot \vec{\sigma} \, $, $ \,( \vec{\sigma}_1  \times \vec{\sigma}_2   ) \cdot \hat{r} $
and $\vec{\sigma}_1 \cdot \vec{\sigma}_2 $. These also appear in the gauge-preserving mass mechanism, but there we have additional profiles, given by $( \vec{\sigma}_1  \times \vec{\sigma}_2   ) \cdot \vec{p} $,
$\, \vec{\sigma} \cdot \hat{r} $, $ \, ( \vec{\sigma}_1  \cdot \vec{p}   )  ( \vec{\sigma}_2  \cdot \hat{r}   ) $ and
$\,  ( \hat{r}  \times \vec{p}   ) \cdot \vec{\sigma} $. The experiment provides six configurations $(C1,...,C6)$ by changing the relative polarization of the detector and the test mass (with respective spin polarizations and relative velocities). One of these configurations is interesting to our work, namely, the $C5$ is sensitive only to $\,  ( \hat{r}  \times \vec{p}   ) \cdot \vec{\sigma} $ dependence, which is only present in the gauge-preserving mass mechanism. For the other profiles we cannot distinguish the contributions of different mechanisms in this experiment. For example, the $C2$ configuration is sensitive to both $ \, ( \vec{\sigma}_1  \cdot \vec{p} )  ( \vec{\sigma}_2  \cdot \hat{r}) $ and $ \vec{\sigma}_1  \cdot \vec{\sigma}_2 $ dependences.

\section{Conclusions}
\indent

The model we investigated describes an extra Abelian gauge boson, a sort of $Z^{0'}$, which appears as a neutral massive excitation of a mixed $\{A_\mu, B_{\nu \kappa}\}$-system of fields. It may be originated from some sector of BSM physics, where the coupling between an Abelian field and the 2-form gauge potential in the SUGRA multiplet may yield the topologically massive spin-1 particle we are considering. To have detectable macroscopic effect, this intermediate particle should have a very small mass, of the order of meV. This would be possible in the class of phenomenological models with the so-called large extra dimensions.

It is clear that the considerable number of off-shell degrees of freedom of the $\{A_\mu, B_{\nu \kappa}\}$-model accounts for the variety of potentials presented above. In order to distinguish between the two models, a possible experimental set-up could consist of a neutral and a polarized source (1 and 2, respectively). Suppose the sources display all kinds of interactions (V, PV, T, etc). In this case, we must collect the terms proportional to $\langle \vec{\sigma}\rangle_2 \equiv \langle \vec{\sigma}\rangle$, namely  
\begin{eqnarray}
V^{Proca}_{mon-dip} & = & -\frac{g^{2}}{\mu} \frac{e^{-m_{0}r}}{r} \vec{p}\cdot \langle \vec{\sigma}\rangle \\
V^{\left\{A,B\right\}}_{mon-dip} & = & -\frac{g^{2} }{\mu} \frac{e^{-m_{0}r}}{r} \vec{p}\cdot \langle \vec{\sigma}\rangle + \nonumber \\
& - & \frac{g^{2}}{m_{1}} \frac{(1 + m_{0}r)e^{-m_{0}r}}{r^{2}} \hat{r} \cdot \langle \vec{\sigma}\rangle + \nonumber \\
& + & \frac{g^{2}}{m_{0}} \frac{\left[ 1 - \left(1 +  m_{0}r \right)e^{-m_{0}r} \right]}{r^{2}}
 \hat{r} \cdot \langle \vec{\sigma}\rangle + \nonumber \\
& - & \frac{g^{2}m_{0}}{m_{1}m_{2}} \frac{e^{-m_{0}r}}{r} \vec{p} \cdot \langle \vec{\sigma}\rangle \nonumber \\
& + & \frac{g^{2} }{\mu m_{0}} \frac{\left[1 - \left(1 + m_{0}r\right)e^{-m_0 r}\right]}{r^{3}} \left(\vec{r} \times \vec{p}\right) \cdot \langle \vec{\sigma}\rangle, \label{md}
\end{eqnarray}
where, for simplicity, we have omitted the labels in the coupling constants. In the macroscopic limit these would be effectively substituted by $g \rightarrow gN_{i}$, being $N_{i}$ the number of interacting particles of type $i$ in each source. If we consider the case in which the source 1 carries momentum so that $\vec{p} \,  // \, \langle \vec{\sigma} \rangle$, the last term above vanishes. Similarly, it is easy to see that the third term is essencially constant, while the fourth one is negligeable, since $m_{0}|\vec{p}|/m_{1}m_{2} \ll 1$ by definition. In Fig.\eqref{Fig3.2}, we plot the two resulting potentials.
\begin{figure}[h!]
\centering
\includegraphics[width=0.7\textwidth]{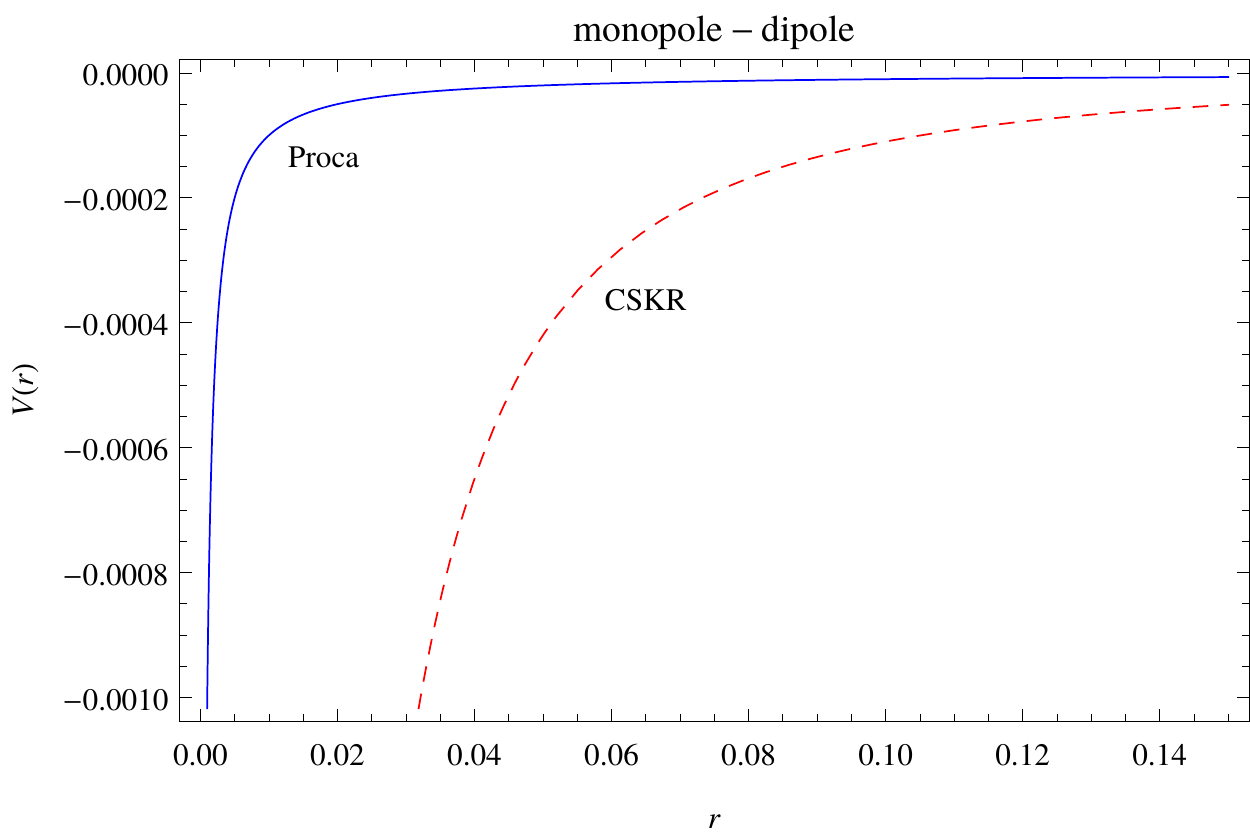}
\caption{\label{fig:overleaf}Monopole-dipole potentials with $m_{1}=m_e = 10^{5}$ eV, $m_{0}=10^{-3}$ eV and source 1 with velocity of order $v \simeq 10^{-6}$. The scale is irrelevant and coupling constants were not included for simplicity.} \label{Fig3.2}
\end{figure}

It would then be possible, in principle, to determine which field representation, Proca or $\{A_{\mu}, B_{\nu\kappa}\}$, better describes the interaction at hand. It is worth mentioning that this difference is regulated by the $1/m_1$ factor in the second term of eq.\ \eqref{md}, so that only the lightest fermions (i.e., electrons and not the protons or neutrons, provided that, in a macroscopic source, we can safely neglect the internal structure of the nucleons) would be able to contribute significantly. 

If we take the second line of eq.\ \eqref{LS}, for example, we notice a coupling of the angular momentum of the first fermion with the spin of the second. Such a spin-orbit coupling is also found in the hydrogen atom, contributing to its fine structure (with typical order of magnitude of $10^{-6¨} eV$). Supposing that the proton and electron are charged under the gauge symmetries leading to the $\left\{A_\mu B_{\nu \kappa}\right\}$-fields, we can calculate a correction to the energy levels of their bound state due to $\langle A_\mu B_{\kappa \lambda }\rangle$ exchange as a means of estimation for the $V-PT$ coupling constants as a function of $m_{0}$.
Expanding the exponential in $1-(1+m_{0}r)e^{-m_{0}r}$ and keeping only the leading term, the spin-orbit term simplifies to
\begin{eqnarray}
V_{V-PT}^{LS} & = & \frac{\sqrt{2}g^{V}_{1}g^{PT}_{2} m_{0}}{8\pi\mu}\frac{1}{r} {\bf L}\cdot{\bf S}
\end{eqnarray}
with ${\bf S}=\langle \vec{\sigma}\rangle_2 /2$. Applying first-order perturbation theory to this potential gives a correction to the energy of 
\begin{eqnarray}
\Delta E^{LS} = \frac{g^{V}_{1}g^{PT}_{2} m_{0}}{8\pi\sqrt{2}\mu (n^{2}a_{0})} X_{l},
\end{eqnarray} 
where $X_{l} = l$ for $j=l+1/2$ and $X_{l} = -(l+1)$ for $j=l-1/2$. As we are interested in an estimate, we suppose $|X_{l}|/n^{2} \sim 1$. Given that the reduced mass and the Bohr radius are $\mu \simeq m_{e} = 5.11 \times 10^{5}$ eV and $a_{0} = 2.69 \times 10^{-4}$ $\text{eV}^{-1}$, respectively, we can constrain $\Delta E^{LS}$ to be smaller than the current spectroscopic uncertainties of one part in $10^{14}$ \cite{Parthey:2011lfa}. We then obtain $|g^{V}g^{PT}| < 10^{-8}$ for a mass of order $m_{0} \sim 10^{-2}$ eV, which poses a less stringent, but consistent (in regard to the orders of magnitude of other couplings \cite{Dobrescu:2006au}), upper bound on the couplings. We see that this correction is much smaller than the typical spin-orbit contribution. This problem can be analyzed with more details in a more comprehensive study that applies atomic spectroscopy of both electronic and muonic hydrogen atoms.


\part{Supersymmetric Dark Matter}


\chapter{Supersymmetry}
\label{chapterSUPERSYMMETRY}
\indent

\section{Motivations}
\indent

Supersymmetry is probably the most fascinating theory to describe the new physics that may appear above the electroweak scale. Although the experimentalists have not yet detected signals of supersymmetric particles, there are strong reasons to believe that low energy supersymmetry is the next outcome of experimental and theoretical progress. The main reason why low energy supersymmetry has long been considered the best-motivated possibility for new physics at the TeV scale is that it can simultaneously solve numerous fundamental open questions of the SM: 

\begin{enumerate}

\item {\bf Hierarchy problem}: Supersymmetry provides a solution to the hierarchy problem \cite{Gildener:1976ai,Nilles:1982ik,Witten:1981nf,Kaul:1981hi} and explains how hierarchies arise. Adding a superpartner to each SM particle, SUSY provides new loop corrections to the SM Higgs boson that can cancel the quadratic divergent terms.

\item {\bf Electroweak symmetry breaking}: In the SM the spontaneous breaking of the electroweak symmetry is parameterized by the Higgs boson $h^{0}$ and its scalar potential $V(h^{0})$. However there are no symmetry principles to constrain the Higgs sector, and thus the Higgs field is put into the theory by hand. EWSB can occur in a natural way via a certain radiative mechanism in the context of SUSY theories \cite{Ibanez:1982fr,Ellis:1982wr,AlvarezGaume:1983gj}. 

\item {\bf Gauge coupling unification}: Although the SM unifies the electromagnetic and weak interactions at 246 GeV into the single electroweak interaction, it cannot unify this force with the strong QCD interaction. Because of its extended particle content, the MSSM can unify the gauge couplings of the SM at the GUT scale, which is $\mathcal{O}(10^{16}$ GeV) \cite{Amaldi:1991cn,Langacker:1991an,Ellis:1990wk}. The evolution of the coupling constants with the energy scale $Q$ is governed by their Renormalization Group Equations (RGEs). The one-loop RGEs for the SM gauge couplings $(g_{1}, g_{2}, g_{3})$ are

\begin{subequations}
\begin{align}
\begin{split}
\beta_{i} &= Q\frac{d }{d Q} g_{i}(Q)=\frac{1}{16\pi^{2}} b_{i}g_{i}^{3},    \quad\quad\quad\quad\quad  (i = 1,2,3)
\end{split}\\
\begin{split}  
 Q\frac{d}{d Q}\left(\frac{g_{i}^{2}}{4\pi}\right)^{-1} &= Q\frac{d}{d Q}\alpha_{i}^{-1} = -\frac{1}{2\pi} b_{i},     
\end{split}
\end{align}
\end{subequations}
where, according to GUT normalization, $g_{1}=\sqrt{5/3} g_{Y}$. These equations show that $\alpha_{i}^{-1}$ evolves linearly with log $(Q)$. The coefficients $b_{i}$ for the SM and MSSM are as follows \cite{Martin:1997ns} 

\begin{subequations}
\begin{align}
\begin{split}
\left(b_{1}, b_{2}, b_{3}\right)_{\text{SM}} &= \left(\frac{41}{10}, -\frac{19}{6}, -7\right),    
\end{split}\\
\begin{split}  
\left(b_{1}, b_{2}, b_{3}\right)_{\text{MSSM}} &= \left(\frac{33}{5}, 1, -3\right).     
\end{split}
\end{align}
\end{subequations}

Figure~\ref{GaugecouplingEvolution} compares the two-loop RG evolution of the gauge couplings in the SM and the MSSSM and shows the possibility of unifying all the three fundamental forces at a scale $Q_{\text{GUT}} \sim 10^{16}$ GeV in the context of SUSY. Embedding the MSSM gauge group in a larger gauge group of grand unification such as SU(5) \cite{Georgi:1974sy,Dimopoulos:1981zb,Dimopoulos:1981yj}, SO(10) \cite{Fritzsch:1974nn,Fukuyama:2012rw} or $E_{6}$ \cite{Hewett:1988xc} one can also unify all the matter content in restricted set of representations of the gauge group. 

\begin{figure}[h!]
\centering
\includegraphics[width=0.5\textwidth]{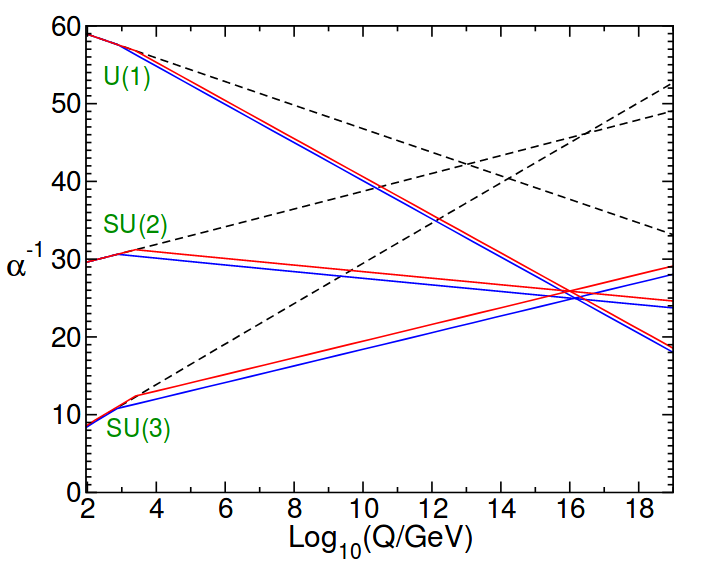}
\caption{\label{fig:overleaf}Two-loop renormalization group evolution of the inverse gauge couplings $\alpha_{i}^{-1}$ in the SM (dashed) and the MSSM (solid lines). Figure taken from Ref.\cite{Martin:1997ns}.} \label{GaugecouplingEvolution}
\end{figure}

\item {\bf Cosmological challenges}: Several problems appear when one tries to build cosmological models based solely on the SM particle content, such as the absence of CDM particle and a small baryon asymmetry generated at the electroweak phase transition. The Lightest Supersymmetric Particle (LSP) of the supersymmetric models can behave as a stable dark matter particle with the correct relic abundance if R-parity is preserved in the model. SUSY is able to link particle physics and cosmology not only through its possibility to have viable dark matter candidates but also by providing viable inflaton candidates \cite{Allahverdi:2006iq,Allahverdi:2006we,Allahverdi:2006cx}. In order to reproduce the observed baryon-antibaryon asymmetry, supersymmetric extensions of the SM can clearly satisfy the necessary Sakharov conditions \cite{Sakharov:1967dj} for successful baryogenesis. 

\item {\bf Gravitation}: Although gravitation is a fundamental force, it is not included in the SM. If SUSY is implemented as a local symmetry it automatically introduces a spin-2 particle known as {\it graviton} that mediates the gravitational interaction. The resulting gauge theory of local SUSY is called {\it supergravity} \cite{VanNieuwenhuizen:1981ae,Nilles:1983ge,Wess:1992cp,West:1990tg,Bailin:1994qt,Sohnius:1985qm,Cerdeno:1998hs}. Supergravity is also seen as the low-energy limit of superstring theories, which shows that it plays essential roles in the context of string theories.

\end{enumerate}

\section{Supersymmetry algebra and superfields}
\indent

Relativistic quantum field theories are invariant under the Poincaré group which contains all the symmetries of special relativity where space-time rotations are encoded in the generator $M_{\mu\nu}$ while space-time translations are realized by the generator $P_{\mu}$. The Lie algebra of the Poincaré group is defined by the following commutation relations

\begin{eqnarray}
\left[P_{\mu}, P_{\nu}\right] & = & 0, \label{eq 4.1}  
\\  
\left[M_{\mu\nu}, P_{\rho}\right] & = & i \left(P_{\mu}\eta_{\nu\rho}-P_{\nu}\eta_{\mu\rho} \right),  \label{eq 4.2} 
\\
\left[M_{\mu\nu}, M_{\rho\sigma}\right] & = & i \left(\eta_{\mu\sigma}M_{\nu\rho}-\eta_{\mu\rho}M_{\nu\sigma}+\eta_{\nu\rho}M_{\mu\sigma}-\eta_{\nu\sigma}M_{\mu\rho} \right).    \label{eq 4.3}
\end{eqnarray}
Generally, the Poincaré algebra cannot mix non-trivially with other usual Lie algebras defined only by commutation relations. In 1967 Coleman and Mandula \cite{Coleman:1967ad} proved a \emph{no-go theorem} which says that in a consistent interacting quantum field theory the most general bosonic symmetry that the S-matrix can have is a direct product of the Poincaré and internal symmetries. Introducing fermionic generators which satisfy anti-commutation relations indeed allows us to extend the symmetries of the quantum field theory. In 1975 Haag, Lopuszanski and Sohnius showed that a graded Lie algebra called super-Poincaré algebra is the most general extension of the Poincaré algebra \cite{Haag:1974qh}.

Supersymmetry is a spacetime symmetry that maps particles and fields of half-integer spin (fermion) into particles and fields of integer spin (boson). The generators $Q$ and $\overline{Q}$ of this symmetry are spinor operators that acts schematically as 

\begin{subequations}\label{grp}
\begin{align}
Q|\text{fermion} \rangle=|\text{boson} \rangle 
\quad\quad\quad\quad\quad\quad\quad\quad\quad
Q|\text{boson} \rangle=|\text{fermion} \rangle  
\label{eq 4.4a}
\\
 \overline{Q}|\text{fermion} \rangle= |\text{boson} \rangle 
\quad\quad\quad\quad\quad\quad\quad\quad\quad
\overline{Q}|\text{boson} \rangle=|\text{fermion} \rangle.  
\label{eq 4.4b}
\end{align}
\end{subequations}
In its most simple version where there is only $\mathcal{N}=1$ pair of supersymmetric generators $\left(Q_{\alpha},\overline{Q}_{\dot \alpha}\right)$ with $\left\{\alpha, \dot \alpha\right\}=\left\{1, 2\right\}$, the super-Poincaré algebra is defined by

\begin{eqnarray}
\left\{Q_{\alpha}, \overline{Q}_{\dot \beta}\right\} & = & 2(\sigma^{\mu})_{\alpha \dot \beta}P_{\mu} \label{eq 4.5}  \\  
\left\{Q_{\alpha}, Q_{\beta}\right\} & = & \left\{\overline{Q}_{\dot \alpha}, \overline{Q}_{\dot \beta}\right\} = 0  \label{eq 4.6}
\\     
\left[P_{\mu}, Q_{\alpha}\right]  & = &  \left[P_{\mu},  \overline{Q}_{\dot \alpha}\right] =0  \label{eq 4.7}
\\
\left[M_{\mu\nu}, Q_{\alpha}\right] & = & i(\sigma_{\mu\nu})_{\alpha}^{\ \beta}
Q_{\beta}  \label{eq 4.8} 
\\
\left[M_{\mu\nu}, \overline{Q}^{\dot \alpha}\right] & = & i(\overline{\sigma}_{\mu\nu})^{\dot \alpha}_{\ \dot \beta}\overline{Q}^{\dot \beta},    \label{eq 4.9}
\end{eqnarray}
where the Pauli matrices $\sigma^{i}$ are obtained from the following relations
\begin{subequations}
\begin{align}
\left(\sigma^{\mu\nu}\right)_{\alpha}^{\ \beta} & =  \frac{1}{4}\Big(\sigma_{\alpha \dot \alpha}^{\mu} \overline{\sigma}^{\nu \ \dot \alpha \beta}-\sigma_{\alpha \dot \alpha}^{\nu} \overline{\sigma}^{\mu\ \dot \alpha \beta}\Big)       
\\
\left(\overline{\sigma}^{\mu\nu}\right)^{\dot \alpha}_{\ \dot \beta} & =  \frac{1}{4}\Big(\overline{\sigma}^{\mu\ \dot \alpha \alpha} \sigma^{\nu}_{\alpha \dot \beta}-\overline{\sigma}^{\nu\ \dot \alpha \alpha} \sigma^{\mu}_{\alpha \dot \beta}\Big) 
\\
\sigma^{\mu} & = (\bf 1, \sigma^{i}) 
\\
\overline{\sigma}^{\mu} & = (\bf 1, -\sigma^{i}).
\end{align}
\end{subequations}
In extended versions of the SUSY algebra with $\mathcal{N}>1$ the supersymmetric field theories must also contain particles of higher spin. We consider in this thesis only unextended $\mathcal{N}=1$ supersymmetry because this is the only version that allows to define properly chiral fermions \cite{Nilles:1983ge,Wess:1992cp,West:1990tg,Bailin:1994qt,Sohnius:1985qm} and thus is the most phenomenologically interesting one.

To construct supersymmetric field theories consistent with the super-Poincaré algebra it is necessary to have fermionic coordinates $\theta^{\alpha}$ and $\overline{\theta}_{\dot \alpha}$ associated to the SUSY generators $Q_{\alpha}$ and $\overline{Q}_{\dot \alpha}$ together with the spacetime coordinates $x^{\mu}$ associated to the bosonic generator $P_{\mu}$. These fermionic coordinates are Grassmann variables that obey the relations
\begin{equation}
\left\{\theta^{\alpha}, \theta^{\beta}\right\}=\left\{\theta^{\alpha}, \overline{\theta}_{\dot \beta}\right\}=\left\{\overline{\theta}_{\dot \alpha}, \overline{\theta}_{\dot \beta}\right\}=0.    
\end{equation}  
The resulting space with 4 bosonic and 4 fermionic coordinates ${\bf X}=\left(x^{\mu},\theta^{\alpha},\overline{\theta}_{\dot \alpha}\right)$ is called superspace and all the fields ${\bf \Omega}(x^{\mu},\theta^{\alpha},\overline{\theta}_{\dot \alpha})$ defined in this extended space are called superfields. In the superspace a generic element $G(x,\omega,\theta,\overline{\theta})$ of the super-Poincaré group that realizes a finite SUSY transformation on the superfields can be written as 
\begin{equation}
G(x,\omega,\theta,\overline{\theta})=e^{i\left(\theta Q+\overline{\theta}\overline{Q}-x^{\mu}P_{\mu}-\frac{1}{2}\omega_{\mu\nu}M^{\mu\nu}\right)}.    
\end{equation}
More details about the differentiation and integration properties of the Grassmann variables of the superspace can be seen in \cite{Lykken:1996xt,Bilal:2001nv,Martin:1997ns}.

Infinitesimal SUSY transformations applied on superfields can be written as 
\begin{eqnarray}
\delta_{\epsilon,\overline{\epsilon}}\Omega(x,\theta,\overline{\theta}) & = & \left[\epsilon^{\alpha}\frac{\partial}{\partial \theta^{\alpha}}+\overline{\epsilon}^{\dot \alpha}\frac{\partial}{\partial \overline{\theta}^{\dot \alpha}}-i\left(\epsilon^{\alpha}\sigma_{\alpha \dot \beta}^{\mu}\overline{\theta}^{\dot \beta}-\theta^{\beta}\sigma_{\beta \dot \alpha}^{\mu}\overline{\epsilon}^{\dot \alpha}\right)\frac{\partial}{\partial x^{\mu}}\right]\Omega(x,\theta,\overline{\theta})
\nonumber \\
& = & i\left(\epsilon \mathcal{Q}+\overline{\epsilon}\overline{\mathcal{Q}}\right)\Omega(x,\theta,\overline{\theta}),   \label{eq 4.13}
\end{eqnarray}
where $\epsilon$ and $\overline{\epsilon}$ are Grassmann parameters. This shows that the SUSY generators have the following representation in the differential form
\begin{equation}
\mathcal{Q}_{\alpha}=-i \partial_{\alpha}-\sigma_{\alpha \dot \beta}^{\mu}\overline{\theta}^{\dot \beta}\partial_{\mu} \quad\quad\quad\quad\quad \overline{\mathcal{Q}}_{\dot \alpha}=i \partial_{\dot \alpha}+\theta^{\beta}\sigma_{\beta \dot \alpha}^{\mu}\partial_{\mu}.     \label{eq 4.14}
\end{equation}
It is useful to define SUSY-covariant derivatives that anticommute with the SUSY generators and thus are useful for writing down SUSY invariant expressions. They are given by
\begin{equation}
D_{\alpha}= \partial_{\alpha}+i\sigma_{\alpha \dot \beta}^{\mu}\overline{\theta}^{\dot \beta}\partial_{\mu} \quad\quad\quad\quad\quad \overline{D}_{\dot \alpha}= \partial_{\dot \alpha}+i\theta^{\beta}\sigma_{\beta \dot \alpha}^{\mu}\partial_{\mu}.     \label{eq 4.15}
\end{equation}
Since these covariant derivatives obey Eqs.\ \eqref{eq 4.5} and \eqref{eq 4.6} they can be used to impose constraints on the general superfield to reduce its number of components in a way consistent with the SUSY transformations. Note from Eqs.\ \eqref{eq 4.14} and \eqref{eq 4.15} that the superspace coordinates $\theta$ and $\overline{\theta}$ have mass dimension $M^{-\frac{1}{2}}$.

Because of the nilpotency of the Grassmann numbers $\theta_{\alpha}$, the superfields can be written explicitly as an expansion series in powers of $\theta$ and $\overline{\theta}$ with a finite number of terms. The most general superfield has the following expansion

\begin{eqnarray}
{\bf \Omega}(x^{\mu},\theta,\overline{\theta}) & = & f(x)+\theta\psi(x)+\overline{\theta}\overline{\chi}(x)+\theta^{2} m(x)+\overline{\theta}^{2} n(x)
\nonumber \\
& + & \overline{\theta}\sigma^{\mu}\theta v_{\mu}(x)+ \theta^{2}\overline{\theta} \ \overline{\zeta}(x)+\overline{\theta}^{2}\theta \lambda(x)+\overline{\theta}^{2}\theta^{2} D(x),    \label{eq 4.16}
\end{eqnarray}
where $f(x)$, $m(x)$, $n(x)$ and $D(x)$ are complex scalar fields, while $\psi(x)$, $\lambda(x)$, $\overline{\chi}(x)$ and $\overline{\zeta}(x)$ are Weyl spinor fields and $v_{\mu}$ is a complex four-vector field. Note that in the general superfield there are 16 bosonic degrees of freedom and 16 fermionic degrees of freedom. As a consequence of supersymmetry, all the superfields have the same number of bosonic and fermionic degrees of freedom. However, by imposing SUSY covariant constraints on the general superfield one can obtain other supermultiplets with smaller number of component fields. We use two types of irreducible supermultiplets that are important to build the Lagrangeans of $\mathcal{N}=1$ supersymmetric gauge theories, which are the chiral and vector superfields.

The chiral and the antichiral superfields are defined as superfields that obey the following SUSY invariant constraints
\begin{equation}
\overline{D}_{\dot \alpha}{\bf \Phi}=0 \quad\quad\quad\quad\quad\quad \text{chiral superfield}   
\end{equation}
\begin{equation}\quad\quad
D_{\alpha} \overline{{\bf \Phi}}=0 \quad\quad\quad\quad\quad\quad \text{antichiral superfield}.   
\end{equation}
To obtain the general solutions of the equations above it is useful to define new bosonic coordinates $y^{\mu}$ of the form
\begin{equation}
y^{\mu}=x^{\mu}+i\theta\sigma^{\mu}\overline{\theta} \quad\quad\quad\quad\quad\quad\quad\quad\quad\quad \overline{y}^{\mu}=\left(y^{\mu}\right)^{\ast}=x^{\mu}-i\theta\sigma^{\mu}\overline{\theta}.    
\end{equation} 
Observing that $D_{\alpha}\overline{\theta}_{\dot \beta}=0$ and $\overline{D}_{\dot \alpha}\theta_{\beta}=0$ one can note that $D_{\alpha}\overline{y}^{\mu}=0$ and $\overline{D}_{\dot \alpha}y^{\mu}=0$. This implies that $\Phi$ depends on $y$ and $\theta$, whereas $\overline{\Phi}$ depends on $\overline{y}$ and $\overline{\theta}$. In this basis of coordinates the chiral and the antichiral superfields have the following expansion
\begin{equation}
\Phi(y,\theta)=\phi(y)+\sqrt{2}\theta \psi(y)-\theta^{2} F(y)   
\end{equation}
\begin{equation}
\overline{\Phi}(y^{\ast},\overline{\theta})= \overline{\phi}(y^{\ast})+\sqrt{2}\overline{\theta} \ \overline{\psi}(y^{\ast})-\overline{\theta}^{2}\overline{F}(y^{\ast}).   
\end{equation}
Physically, a chiral superfield describes one complex scalar $\phi$ and one Weyl fermion $\psi$. The field $F$ represents an auxiliary scalar field with mass dimension $M^{2}$. In supersymmetric extensions of the SM, left-handed chiral superfields are used to describe left-handed quarks and left-handed leptons with their corresponding supersymmetric partners. Performing a Taylor expansion of $\Phi$ and $\overline{\Phi}$ around $x^{\mu}$ we see that, as functions of the usual superspace coordinates, the chiral and antichiral superfields assume the forms
\begin{eqnarray}
\Phi(x, \theta, \overline{\theta}) & = & \phi(x)+\sqrt{2}\theta \psi(x)+i\theta \sigma^{\mu}\overline{\theta}\partial_{\mu}\phi(x)-
\theta^{2} F(x) \nonumber \\
 & - & \frac{i}{\sqrt{2}}\theta^{2}
\partial_{\mu}\psi(x)\sigma^{\mu}\overline{\theta}-
\frac{1}{4}\theta^{2}\overline{\theta}^{2} \Box\phi(x)   
\end{eqnarray}
\begin{eqnarray}
\overline{\Phi}(x, \theta, \overline{\theta}) & = & \overline{\phi}(x)+
\sqrt{2}\overline{\theta} \ \overline{\psi}(x)-i\theta \sigma^{\mu}\overline{\theta}\partial_{\mu}\overline{\phi}(x)-\overline{\theta}^{2}\overline{F}(x) \nonumber 
\\
 & + & \frac{i}{\sqrt{2}}\overline{\theta}^{2}\theta\sigma^{\mu}\partial_{\mu}\overline{\psi}(x)-
\frac{1}{4}\theta^{2}\overline{\theta}^{2}
\Box\overline{\phi}(x). 
\end{eqnarray}
  
To construct SUSY invariant Lagrangians using chiral superfields it is necessary to know how their component fields respond to SUSY transformations. Plugging the differential SUSY generators into eq.\ \eqref{eq 4.13} one gets
\begin{eqnarray}
\delta \phi(y) & = & \sqrt{2}\epsilon\psi(y)
\\
\delta \psi_{\alpha}(y) & = & \sqrt{2}i(\sigma^{\mu}\overline{\epsilon})_{\alpha}
    \partial_{\mu}\phi(y)+\sqrt{2}\epsilon_{\alpha}F(y)
\\
\delta F(y) & = & -\sqrt{2}i\partial_{\mu}[\psi(y)\sigma^{\mu}\overline{\epsilon}].
\end{eqnarray}
Note that $\delta F(y)$ is a total derivative, which implies that the integration of the auxiliary field $F$ over spacetime is invariant under SUSY transformations. The result is similar for the SUSY transformations of antichiral superfields. In simple SUSY models with Lagrangians constructed by the product of chiral superfields one only needs to obtain the $F$ component of this product and integrate over spacetime to obtain a SUSY-invariant action. The Wess-Zumino model \cite{Wess:1974tw} is an example of a SUSY model constructed in this way.

The vector superfield is another irreducible representation of the SUSY algebra defined from the general superfield by imposing a condition of reality
\begin{equation}
V(x, \theta, \overline{\theta})=\overline{V}(x, \theta, \overline{\theta}),   
\end{equation}
which is useful to make the vector field $v_{\mu}$ become real as it should be to identify a spin-1 gauge boson. This condition is preserved under SUSY transformations and, after applying it in the general superfield \eqref{eq 4.16}, one obtains the following expansion for the vector superfield
\begin{eqnarray}
V(x, \theta, \overline{\theta}) & = & \phi(x)+ \theta \psi(x)+ \overline{\theta} \ \overline{\psi}(x)+\theta^{2}m(x)+
\overline{\theta}^{2}\overline{m}(x) \nonumber 
\\
& + & \theta \sigma^{\mu}\overline{\theta} v_{\mu}(x)+\theta^{2}\overline{\theta} \ \overline{\lambda}(x)+
\overline{\theta}^{2}\theta\lambda(x)+
\theta^{2}\overline{\theta}^{2}D(x), 
\end{eqnarray}
where $\phi(x)$ and $D(x)$ are real scalar fields, $m(x)$ is a complex scalar field and $v_{\mu}(x)$ is a real vector field, whereas $\psi(x)$ and $\lambda(x)$ are Weyl spinors. Notice that this superfield carries 8 bosonic and 8 fermionic degrees of freedom, which is still a big number. However, it is possible to reduce the number of degrees of freedom via gauge fixing after realizing the supersymmetric version of gauge transformations on the vector superfield.

Considering $\Lambda$ a chiral superfield and $\overline{\Lambda}$ its complex conjugate, the following transformation
\begin{equation}
V \rightarrow V-i(\Lambda-\overline{\Lambda})
\end{equation}
can be seen as a gauge transformation because it reproduces the usual (abelian) gauge transformation of the vector field $v_{\mu}$ and, if this transformation is a symmetry of the theory then, by an appropriate choice of the field components of $\Lambda$, one can transform away the components $\phi(x)$, $\psi(x)$ and $m(x)$ of $V(x, \theta, \overline{\theta})$. As a result of this operation, the final expression of the vector superfield simplifies to
\begin{equation}
V_{\text{WZ}}(x, \theta, \overline{\theta})=\theta \sigma^{\mu}\overline{\theta} v_{\mu}(x)+i\theta^{2}\overline{\theta} \ \overline{\lambda}(x)
-i\overline{\theta}^{2}\theta\lambda(x)
+\frac{1}{2}\theta^{2}\overline{\theta}^{2}D(x),
\end{equation}  
and this choice is called the Wess-Zumino gauge \cite{Wess:1974jb}. The highest $D$ component is another auxiliary and, as the $F$ component of the chiral superfield, it transforms into a total derivative under SUSY transformations. 

\section{Construction of $\mathcal{N}=1$ SYM Lagrangians}
\indent

The construction of general supersymmetric gauge theories becomes easier and more practical with the formalism of superfields in superspace.
The Lagrangians of these theories are constructed from a set of chiral and vector superfields used to write terms invariant under local gauge transformations that belong to a compact Lie group $G$, which can be a direct product of multiple group factors. In order for the action to be completely invariant under supersymmetry, the space-time Lagrangian density can at most change as a total space-time derivative under supersymmetry transformations. We saw in the previous section that the highest components of the chiral and vector superfields transform as total derivatives under SUSY transformations. Hence the supersymmetric Lagrangians are constructed from the F- and D-terms obtained from a set of chiral and vector superfields.

In supersymmetric models the matter interactions is described by the superpotential, a holomorphic function of the chiral superfields $\Phi_{i}$. For a renormalizable theory the most general form of the superpotential is 

\begin{equation}
W= L_{i} \Phi_{i}+ \frac{1}{2} M_{ij}\Phi_{i}\Phi_{j}+\frac{1}{3} Y_{ijk}\Phi_{i}\Phi_{j}\Phi_{k}.
\end{equation}
To preserve the gauge symmetry each term in the superpotential must form a gauge singlet, which implies that $L_{i}$ is zero except for singlet superfields. Terms involving more than three chiral superfields would have mass dimension higher than four and hence are forbidden by renormalizability. 

According to the non-renormalization theorem of supersymmetric theories \cite{Salam:1974jj,Grisaru:1979wc}, the superpotential is not renormalized in perturbation theory. This indicates that any fine-tuning of the potential at tree-level will not receive any higher order loop contributions. Hence, this makes supersymmetry a potential solution to the hierarchy problem.

In the case of non-Abelian gauge theories the components of the vector superfields are expressed in the form of matrices $V=V^{a}T^{a}$, where $a=1,....,$ dimG and $T^{a}$ are the generators of the gauge group G. Generally, the gauge transformations of the vector superfields are of the form
\begin{equation}
e^{2gV} \rightarrow e^{i \overline{\Lambda}} e^{2gV} e^{-i\Lambda},
\end{equation}
where $g$ is the gauge coupling. The kinetic Lagrangian of the gauge fields is constructed using the supersymmetric generalization of the field-strength $F_{\mu\nu}$ which is given by

\begin{equation}
\mathcal{W}_{\alpha}=-\frac{1}{4}\overline{D} \ \overline{D}
\left(e^{-2g_{a}V^{a}T^{a}}D_{\alpha}e^{2g_{a}V^{a}T^{a}}\right).
\end{equation}
The field-strength $\mathcal{W}_{\alpha}$ is a left-chiral superfield that transforms covariantly under the gauge transformation as $\mathcal{W}_{\alpha} \rightarrow e^{i\Lambda}\mathcal{W}_{\alpha} e^{-i\Lambda}$. Consequently, the $\mathcal{N}=1$ supersymmetric Yang-Mills Lagrangian can be obtained from the following F-term \cite{Lykken:1996xt,Bilal:2001nv,Martin:1997ns}

\begin{eqnarray}
\mathcal{L}_{\text{SYM}} & = & \frac{1}{32\pi}\text{Im} \left(\tau
\int d^{2}\theta \ \text{Tr}\mathcal{W}^{\alpha}\mathcal{W}_{\alpha}\right)  \nonumber \\
 & = & \left(-\frac{1}{4} F^{a}_{\mu\nu}F^{a \mu\nu}-
i\lambda^{a}\sigma^{\mu}D_{\mu}\overline{\lambda}^{a}
+\frac{1}{2} D^{a}D^{a}\right)+
\frac{\Theta_{YM}}{32\pi^{2}}g^{2}
F_{\mu\nu}^{a}\widetilde{F}^{a \mu\nu},  
\end{eqnarray}
where $\tau=\frac{\Theta_{YM}}{2\pi}+\frac{4\pi}{g^{2}} i$ is a complex coupling that carries the gauge coupling and the CP-violating parameter. The term proportional to $\Theta_{YM}$ is a total derivative and thus it can be integrated out. 

The chiral and anti-chiral superfields in a representation R have, respectively, the following gauge transformations: $\Phi \rightarrow e^{i \Lambda^{a}T_{R}^{a}} \Phi$ and $\overline{\Phi} \rightarrow \overline{\Phi}
e^{-i\overline{\Lambda^{a}}T_{R}^{a}}$. The combination of these transformations with the gauge transformation of the vector superfields generate a gauge invariant vector superfield of the form $\overline{\Phi}e^{2gV^{a}T^{a}}\Phi$. Therefore the complete Lagrangian for charged matter and interaction reads 

\begin{eqnarray}
\mathcal{L}_{\text{matter}} & = & \int d^{2}\theta d^{2}\overline{\theta} \ \overline{\Phi}e^{2gV^{a}T^{a}}\Phi \ + \ \int d^{2}\theta \ W(\Phi^{i}) + \int d^{2}\overline{\theta} \
\overline{W}(\overline{\Phi}_{i})  \nonumber \\
& = & 
\left(D_{\mu}\phi_{i}^{\ast}\right)
\left(D^{\mu}\phi^{i}\right)
-i \overline{\psi}_{i}\overline{\sigma}^{\mu}D_{\mu}\psi^{i}
+F_{i}^{\ast}F^{i} + i\sqrt{2}g\left(\overline{\phi}\lambda\psi-
\overline{\psi} \ \overline{\lambda}\phi\right)
 \nonumber \\ 
& + &  
g \ \phi_{i}^{\ast} \left(D^{a}T^{a}_{R}\right)^{i}_{\ j} \phi^{j} + \int d^{2}\theta \ W(\Phi^{i}) + \int d^{2}\overline{\theta} \
\overline{W}(\overline{\Phi}_{i}),  \label{eqmatterLagrangian}
\end{eqnarray}
where, for simplicity, we used the Wess-Zumino gauge and introduced the usual gauge-covariant derivative $D_{\mu}=\partial_{\mu}+igA_{\mu}^{a}T^{a}$. This part of the Lagrangian contains the kinetic terms for the matter fermions $\psi^{i}$ and the scalar fields $\phi^{i}$, as well as Yukawa-type interactions between the gauginos $\lambda^{a}$ and the matter fields. 

In order to obtain the full $\mathcal{N}=1$ SYM Lagrangian written in terms of the component fields one shall first expand the superpotential in powers of $\theta$  as

\begin{equation}
\\ W(\Phi^{i})=W(\phi^{i}) + \sqrt{2}\left. \frac{\partial W}{\partial \phi^{i}}\right|_{\Phi^{i}=\phi^{i}}\theta \psi^{i}-
\theta^{2}\left(\left. \frac{\partial W}{\partial \phi^{i}}\right|_{\Phi^{i}=\phi^{i}}F^{i}+\frac{1}{2}\left. \frac{\partial^{2}W}{\partial\phi^{i}\partial\phi^{j}} \right|_{\Phi^{i}=\phi^{i}} \psi^{i}\psi^{j}\right).
\end{equation}
Plugging back this in eq.\ \eqref{eqmatterLagrangian} and integrating out the unphysical auxiliary fields $F^{i}$ and $D^{a}$ one finally obtains the on-shell SYM Lagrangian \cite{Lykken:1996xt,Bilal:2001nv,Martin:1997ns}:

\begin{eqnarray}
\mathcal{L} & = & \mathcal{L}_{\text{SYM}}+\mathcal{L}_{\text{matter}} \nonumber \\
& = & \text{Tr}\left(-\frac{1}{4} F_{\mu\nu}F^{\mu\nu}-
i\lambda\sigma^{\mu}D_{\mu}\overline{\lambda}\right)+
\frac{\Theta}{32\pi^{2}}g^{2} \text{Tr}\left(
F_{\mu\nu}\widetilde{F}^{\mu\nu}\right) \nonumber \\
 & + & \left(D_{\mu}\phi_{i}^{\ast}\right)
\left(D^{\mu}\phi^{i}\right)
-i \overline{\psi}_{i}\overline{\sigma}^{\mu}D_{\mu}\psi^{i}
+i\sqrt{2}g\left(\overline{\phi}\lambda\psi-
\overline{\psi} \ \overline{\lambda}\phi\right) \nonumber \\
 & - &
\frac{1}{2}\left. \frac{\partial^{2}W}{\partial\phi^{i}\partial\phi^{j}} \right|_{\Phi^{i}=\phi^{i}}\psi^{i}\psi^{j}
-\frac{1}{2}\left. \frac{\partial^{2}\overline{W}}{\partial\phi^{\ast}_{i}\partial\phi^{\ast}_{j}} \right|_{\overline{\Phi}_{i}=\phi^{\ast}_{i}}
\overline{\psi}_{i} \ \overline{\psi}_{j}  \nonumber \\
& - &
\left(\sum_{\substack{i}} \bigg| \left. \frac{\partial W}{\partial \phi^{i}}\right|^{2}_{\Phi^{i}=\phi^{i}}+\frac{1}{2} g^{2}\sum_{\substack{a}} \vert \overline{\phi} \ T_{R}^{a} \ \phi \vert^{2} \right),
\end{eqnarray}
where in the last line we have the supersymmetric scalar potential. If there is a U(1) factor in the gauge group, there is one more term that can be added to the Lagrangian known as the Fayet-Iliopoulos term.

\section{Supersymmetry breaking}
\indent

Exact supersymmetry predicts that all particles that appears in the same supermultiplet should be mass degenerated. In other words, exact supersymmetric field theories requires the SM particles to have the same mass as their superpartners, which is completely excluded by experiments. Thus, if supersymmetry occurs in nature it must be broken either spontaneously or explicitly to give higher masses to the supersymmetric partners of the SM particles.

In the following subsections we will first describe how supersymmetry can be softly broken in a consistent way with its theoretical advantages and then we will show which are the main mechanisms used to break SUSY spontaneously and give rise to the Lagrangian terms that break SUSY explicitly.

\subsection{Soft supersymmetry breaking}
\indent

The simplest phenomenologically accepted approach to break supersymmetry is to parametrize the breaking mechanism by introducing additional terms into the Lagrangian that breaks supersymmetry. This is an approach that starts with the SUSY breaking parameters at a low energy scale (like, for example, the electroweak scale), use them to find parameter regions where the theoretical predictions are in agreement with the low energy observables and, finally, derive some theoretical implications at the GUT scale. For this reason, it is called \emph{bottom-up approach}. 

The terms that break supersymmetry should have positive mass dimension so that they do not introduce new divergencies to the SUSY theory in such a way that they do not affect the SUSY relations between the Yukawa and gauge couplings that cancels the quadratic divergencies. For this reason they are known in the literature as \emph{soft breaking terms}. In 1982 L. Girardello and M. T. Grisaru \cite{Girardello:1981wz} showed that the possible soft supersymmetry-breaking terms that can be added to the Lagrangian of a general 4-dimensional SUSY theory are \footnote{In certain cases one can also introduce non-holomorphic trilinear scalar couplings of the form $c^{ijk} \phi_{i}^{\ast}\phi_{j}\phi_{k} + C.C.$ \cite{Martin:1997ns}. In models with gauge singlet scalar fields one can introduce linear terms to the Lagrangean.}

\begin{eqnarray}
\mathcal{L}_{soft} = & - & \frac{1}{2}\left(
M_{a}\lambda^{a}\lambda^{a}+ H.C.\right)
-(m^{2})^{i}_{\ j}\phi^{\ast j}\phi_{i} \nonumber \\
& - & \frac{1}{2}\left(B^{ij}\phi_{i}\phi_{j}+ H.C.\right)
-\frac{1}{6}\left(T^{ijk}\phi_{i}\phi_{j}\phi_{k}+H.C.\right).    
\end{eqnarray}
The first term represents gaugino masses for each gauge group and the second one represents scalar squared-mass terms for all scalar fields, whereas the third and the fourth refers to bilinear and trilinear scalar interaction terms, respectively. It has been shown that a softly broken supersymmetric theory with $\mathcal{L}_{soft}$ as given above is free of quadratic divergencies in quantum corrections to scalar masses to all orders in perturbation theory \cite{Girardello:1981wz}.

\subsection{Mechanisms of supersymmetry breaking}
\indent

The soft SUSY breaking terms can be generated in a more fundamental level through spontaneous supersymmetry breaking in which the Lagrangian remains invariant under SUSY transformations while the vacuum does not respect supersymmetry. This can occur when at least one of the auxiliary fields $F^{i}$ or $D^{a}$ acquires a VEV. There are two mechanisms that can break supersymmetry spontaneously, which are the \emph{O’Raifeartaigh mechanism} \cite{ORaifeartaigh:1975nky} where SUSY is broken by a nonvanishing F-term VEV and the \emph{Fayet-Iliopoulos mechanism} \cite{Fayet:1974jb} where SUSY is broken by a non-zero D-term VEV. These mechanisms are useful to break SUSY in simple models that have small field content, which is not the case of the MSSM and its extensions.

The main scheme used to study the origin and the effects of supersymmetry breaking in low energy is as follows: SUSY should be broken in a \emph{hidden sector} which contains exotic fields that couples very weakly to the \emph{visible sector} which contains the SM particles and their superpartners. The main scenarios of mediating the spontaneous breaking of SUSY studied in the literature are:

\begin{itemize}

\item The {\bf gravity-mediated} scenario \cite{Nilles:1983ge}\cite{Martin:1997ns}\cite{Chamseddine:1982jx,Barbieri:1982eh,Hall:1983iz,Chung:2003fi}, where the breaking of SUSY is communicated to the MSSM via gravitational-strength interactions such as higher order operators suppressed by $M_{P}$, which are naturally addressed within the context of $\mathcal{N}=1$ supergravity. In the hidden sector there is a matter superfield $X$ whose F-term gets a nonvanishing VEV and generates soft mass terms in the low energy effective field theory of the following order
  
\begin{equation}
m_{soft}\sim \frac{\langle F_{X} \rangle}{M_{P}}\sim \frac{M_{SUSY}^{2}}{M_{P}},    
\end{equation}   
where $M_{SUSY}$ is the scale where SUSY is spontaneously broken. Therefore, in order to get mass terms for the supersymmetric particles of the order of TeV scale, SUSY must be broken at $M_{SUSY}\sim 10^{11}$ GeV. 

\item The {\bf Gauge-Mediated Supersymmetry Breaking} (GMSB) scenario, where there is an intermediate sector called \emph{messenger sector} which contains fields that couples to the hidden sector and interacts with the visible sector via gauge interactions \cite{Dine:1981gu,Nappi:1982hm,AlvarezGaume:1981wy,Dine:1993yw,Dine:1994vc,Dine:1995ag}. The mediation of SUSY breaking is communicated to the MSSM through gauge interactions and it provides soft mass terms for the superpartners through loop diagrams involving messenger particles. In this scenario, $M_{SUSY}$ is generally much smaller than it is in gravity-mediated scenarios.

\item The {\bf Anomaly-Mediated Supersymmetry Breaking} (AMSB) scenario \cite{Randall:1998uk,Giudice:1998xp}, where the breaking of SUSY is communicated to the visible sector via a combination of gravity and anomalies. In this scenario the MSSM soft terms are originated from an anomalous violation of the supersymmetric extension of scale invariance called superconformal symmetry. 

\end{itemize}

These scenarios of SUSY breaking mediation produce different mass spectrums and have different predictions in low energy. Although the phenomenological tools that we use in our numerical analysis can accomodate all of them, we will focus on the low energy limit of the UMSSM where all the soft SUSY breaking parameters of the model are free input parameters given at the SUSY breaking scale.

\section{The Minimal Supersymmetric Standard Model}
\indent

The MSSM can be seen as a renormalizable supersymmetric non-Abelian gauge theory constructed with the usage of the superfield formalism. As a consequence of minimal supersymmetry ($\mathcal{N}=1$), each field of the SM has only one superpartner with both fields accomodated in the same supermultiplet. All the quarks, leptons and Higgs matter fields of the SM appear in chiral superfields with their corresponding superpartners, whereas the gauge fields of the SM appear in vector superfields with their corresponding superpartners called gauginos. The MSSM is a model that respects R-parity because it does not allow its superpotential to have gauge-invariant terms that violates lepton or baryon number.

\subsection{Description of the MSSM}
\indent

As its name indicates, the minimal supersymmetric standard model \cite{Nilles:1983ge,Haber:1984rc,Haber:1993wf,Drees:2004jm,Baer:2006rs,Aitchison:2007fn} is the most simplified way to extend the SM with the usage of supersymmetry. It has the same gauge structure as the SM, i.e. the $SU(3)_{C}\times SU(2)_{L}\times U(1)_{Y}$ gauge group, with the smalest particle content that respects SUSY implies in a good low energy phenomenology. With respect to the particle content of the SM, besides providing a supersymmetric partner for all particles of the SM, the MSSM needs 2 Higgs doublets with opposite hypercharges to generate masses for both up- and down-type quarks. These two Higgs doublets are also necessary to cancel the chiral anomalies of the higgsinos because in supersymmetric extensions of the SM the Yukawa interactions arise from the superpotential, which is an analytic function of chiral superfields only. Left-handed fermions are accomodated in left-handed supermultiplets while right-handed fermions in right-handed superfields. As in the SM, there are no right-handed neutrinos in the MSSM but they are added in many extensions of the MSSM known in the literature, such as the UMSSM which we describe in chapter \ref{chapterUMSSM}. 

The supermultiplets of the MSSM with their corresponding particle content and gauge quantum numbers are listed in Table \ref{Chapter4Tabel1}. As in the SM, there are three families of quarks and leptons and, to be consistent with supersymmetry, there are also three families of squarks and sleptons in the MSSM.

The superpotential of the MSSM contains all the renormalizable gauge-invariant operators that respects R-parity. It is given by
\begin{equation}
{\bf \hat{W}}_{MSSM}=(Y_{u})_{ij}{\bf \hat{H}}_{u}{\bf \hat{Q}}_{i}{\bf \hat{U}}^{c}_{j}
-(Y_{d})_{ij}{\bf \hat{H}}_{d}{\bf \hat{Q}}_{i}{\bf \hat{D}}^{c}_{j}
-(Y_{e})_{ij}{\bf \hat{H}}_{d}{\bf \hat{L}}_{i}
{\bf \hat{E}}^{c}_{j}+\mu {\bf \hat{H}}_{u}{\bf \hat{H}}_{d},   \label{eqMSSMsuperpotential}
\end{equation}
where $\mu$ is a supersymmetric mass term for the Higgs sector and ${\bf Y_{u}}$, ${\bf Y_{d}}$, ${\bf Y_{e}}$ are the Yukawa matrices. The $\mu$ term $\mu {\bf \hat{H}}_{u}{\bf \hat{H}}_{d}$ is unique to the MSSM and there are no theoretical predictions from the model for the energy scale on which this parameter should acquire values. Thus, as a consequence, it behaves as a free parameter in the MSSM. However, in order to have acceptable loop corrections to the mass of the SM Higgs boson and to avoid another hierarchy problem, the $\mu$ parameter must not be much higher than the electroweak scale. This means that in the MSSM $\mu$ has to be tuned to a value close to $v_{\text{ew}}$, which is known in the literature as the $\mu$ {\it problem} \cite{Kim:1983dt}.

The superpotential of the MSSM does not include all the possible terms that are compatible with the symmetries of the model. There are other terms that respects SUSY and the gauge symmetry but violates the conservation of baryon (B) and lepton (L) numbers, which constitutes the so-called R-parity violation operators \cite{Sakai:1981pk,Weinberg:1981wj,Baer:2006rs}. Processes that simultaneously violates B and L can lead to rapid proton decay \cite{Hall:1983id,Dreiner:1997uz,Farrar:1978xj,Barbier:2004ez}, unless the coefficients are strongly suppressed. In order to avoid these phenomenologically dangerous processes a discrete symmetry called R-parity is added to the MSSM

\begin{equation}
R_{p}= (-1)^{3(\text{B}-\text{L})+2s}, 
\label{eqRparity}
\end{equation} 
where $s$ stands for the spin of the particle. In a given superfield the R-parity of the component fields are not the same: it can easily be noted that eq.\ \eqref{eqRparity} gives $R_{p}=+1$ for ordinary particles and $R_{p}=-1$ for supersymmetric particles.

The conservation of R-parity has two important consequences: supersymmetric particles can be produced only in pairs and the decay products of a supersymmetric particle other than the LSP must contain an odd number of LSP (usually one). As a consequence, the lightest supersymmetric particle must be a stable particle that does not decay into lighter non-supersymmetric particles and, being electrically neutral and colorless, it can be an interesting candidate for CDM.

The soft SUSY breaking Lagrangian of the MSSM reads

\begin{align}
-\mathcal{L}_{\cancel{SUSY}} &= m^{2}_{H_{d}} \left(\vert H_{d}^{0}\vert^{2}+\vert H_{d}^{-}\vert^{2}\right) +m^{2}_{H_{u}} \left(\vert H_{u}^{0}\vert^{2}+\vert H_{u}^{+}\vert^{2}\right) \nonumber \\
 &+\tilde{Q}^{\dagger}{\bf m^{2}_{\tilde{Q}}}\tilde{Q}+ \tilde{d}_{R}^{\dagger}{\bf m^{2}_{\tilde{D}^{c}}}\tilde{d}_{R} +\tilde{u}_{R}^{\dagger}{\bf m^{2}_{\tilde{U}^{c}}}\tilde{u}_{R}+\tilde{L}^{\dagger}{\bf m^{2}_{\tilde{L}}}\tilde{L}+\tilde{e}_{R}^{\dagger}{\bf m^{2}_{\tilde{E}^{c}}}\tilde{e}_{R} 
\nonumber \\
 &+\frac{1}{2}\Big( M_{1}\lambda_{\tilde{B}} \lambda_{\tilde{B}}+M_{2} \sum_{\substack{k=1}}^{3} \lambda^{k}_{\tilde{W}} \lambda^{k}_{\tilde{W}} +M_{3} \sum_{\substack{a=1}}^{8} \lambda^{a}_{\tilde{g}}\lambda^{a}_{\tilde{g}} + H.C.\Big) + \Big( B\left(H_{u}^{+}H_{d}^{-}-H_{u}^{0}H_{d}^{0}\right) + H.C. \Big)
\nonumber \\
 &+\Big( \tilde{u}_{R}^{c}{\bf T_{u}}\tilde{u}_{L}H_{u}^{0}-\tilde{d}_{R}^{c}{\bf T_{d}}\tilde{d}_{L}H_{d}^{0}-\tilde{e}_{R}^{c}{\bf T_{e}}\tilde{e}_{L}H_{d}^{0} + H.C. \Big), \label{eqMSSMsoftSUSYbreaking}
\end{align}
with ${\bf m_{\varphi}^{2}}$ being a 3 $\times$ 3 matrix in flavor space with $\varphi={\tilde{Q},\tilde{U}^{c},\tilde{D}^{c},\tilde{L},\tilde{E}^{c}}$. This Lagrangian contains soft mass terms for all the scalar particles $\Big( (m_{\varphi}^{2})_{ij},m^{2}_{H_{d}},m^{2}_{H_{u}} \Big)$, mass terms for the gauginos $(M_{1}, M_{2}, M_{3})$, and also bilinear and trilinear couplings between the scalar fields. These terms are invariant under the gauge symmetry of the MSSM and they all have positive mass dimension, thus they do not reintroduce quadratic divergences that affects the solution that SUSY gives to the hierarchy problem.

\begin{table}[h]
\centering
\begin{tabular}{ |p{2.8cm}|p{3.5cm}|p{4cm}|p{1.3cm}|p{1.2cm}|p{1cm}|  }
\hline
\quad\quad Superfields &\centering  Bosonic Fields  &\centering  Fermionic Fields  &  $SU(3)_{C}$  &  $SU(2)_{I}$  &  $U(1)_{Y}$\\
\hline
\multicolumn{1}{|c|}{\centering Gauge Supermultiplets}  & \centering Spin 1  &\centering  Spin $\frac{1}{2}$  & & &\\ 
\hline
\multicolumn{1}{|c|}{${\bf \hat{V}}_{G}^{a}$}  & \centering $G^{a}_{\mu}$  &\centering   $\lambda_{\tilde{g}}^{a}$  & \centering {\bf 8}  & \centering {\bf 1} &\quad 0 \\
\multicolumn{1}{|c|}{${\bf \hat{V}}_{W}^{k}$}  & \centering $W^{k}_{\mu}$  &\centering   $\lambda_{\tilde{W}}^{k}$  & \centering {\bf 1}  & \centering {\bf 3} &\quad 0 \\
\multicolumn{1}{|c|}{${\bf \hat{V}}_{B}$}  & \centering $B_{\mu}$  &\centering   $\lambda_{\tilde{B}}$  & \centering {\bf 1}  & \centering {\bf 1} &\quad 0 \\
\hline
\multicolumn{1}{|c|}{\centering Matter Supermultiplets}  &  \centering Spin 0  & \centering Spin $\frac{1}{2}$  & & &\\
\hline
\multicolumn{1}{|c|}{${\bf \hat{Q}}_{i}$}  & \centering $(\tilde{u}_{L},\tilde{d}_{L})_{i}$  &\centering   $(u_{L}, d_{L})_{i}$  & \centering {\bf 3}  & \centering {\bf 2} &\quad $\frac{1}{6}$ \\
\multicolumn{1}{|c|}{${\bf \hat{U}}^{c}_{i}$}  & \centering $\tilde{u}^{c}_{R i}$  &\centering   $u^{c}_{R i}$  & \centering ${\bf \overline{3}}$  & \centering {\bf 1} & $-\frac{2}{3}$ \\
\multicolumn{1}{|c|}{${\bf \hat{D}}^{c}_{i}$}  & \centering $\tilde{d}^{c}_{R i}$  &\centering   $d^{c}_{R i}$  & \centering ${\bf \overline{3}}$  & \centering {\bf 1} &\quad $\frac{1}{3}$ \\
\hline
\multicolumn{1}{|c|}{${\bf \hat{L}}_{i}$}  & \centering $(\tilde{\nu}_{L},\tilde{e}_{L})_{i}$  &\centering   $(\nu_{L}, e_{L})_{i}$  & \centering {\bf 1}  & \centering {\bf 2} & $-\frac{1}{2}$ \\
\hline
\multicolumn{1}{|c|}{${\bf \hat{E}}^{c}_{i}$}  & \centering $\tilde{e}^{c}_{R i}$  &\centering $e^{c}_{R i}$   & \centering ${\bf \overline{1}}$  & \centering {\bf 1} &\quad 1 \\
\hline
\multicolumn{1}{|c|}{${\bf \hat{H}}_{u}$}  & \centering $(H^{+}_{u}, H^{0}_{u})$  &\centering   $(\tilde{H}^{+}_{u}, \tilde{H}^{0}_{u})$  & \centering {\bf 1}  & \centering {\bf 2} & $\quad \frac{1}{2}$ \\
\hline
\multicolumn{1}{|c|}{${\bf \hat{H}}_{d}$}  & \centering $(H^{0}_{d}, H^{-}_{d})$  &\centering   $(\tilde{H}^{0}_{d}, \tilde{H}^{-}_{d})$  & \centering {\bf 1}  & \centering {\bf 2} & $-\frac{1}{2}$ \\
\hline
\end{tabular}
\caption{MSSM superfields and their gauge properties. As in the SM, there are 3 families of quarks and leptons with their corresponding superpartners ($i={1,2,3}$).}
\label{Chapter4Tabel1}
\end{table}


\subsection{Electroweak symmetry breaking and the the Higgs sector}
\indent

In a general global SUSY model that contains a certain number of scalar fields $\phi_{i}$ and a gauge group formed by direct product of simple Lie groups $G_{A}$ with gauge couplings $g_{A}$ and generators $T_{A}$, the scalar potential has the following general formula \cite{Lykken:1996xt,Bilal:2001nv,Martin:1997ns}

\begin{equation}
V(\phi,\overline{\phi})= \sum_{\substack{i}} \bigg| \left. \frac{\partial W}{\partial \phi^{i}}\right|^{2}_{\Phi^{i}=\phi^{i}}+\frac{1}{2}\sum_{\substack{A}} g_{A}^{2}\sum_{\substack{a}} \vert \phi^{\dagger} \ T_{A}^{a} \ \phi \vert^{2}, 
\end{equation} 
where the first term appears when one integrates out the auxiliary fields $F_{i}$ of the chiral superfields that are used to construct the superpotential $W$ of the model, thus it is referred to as the F-term; while the second one appears when one integrates out the auxiliary fields $D^{a}$ of the vector superfields and thus it is known as the D-term of the scalar potential.

Similar to the SM, the Higgs mechanism is the necessary ingredient to provide masses for fermions and vector bosons. Since the MSSM has two complex Higgs doublets with eight degrees of freedom for scalars on the total, the scalar potential of the MSSM is much bigger than the scalar potential of the SM. The tree level scalar potential of the Higgs fields is given by the
sum of the contributions from the F-term, the D-term and the soft supersymmetry breaking Lagrangian. These contributions are given by

\begin{subequations}
\begin{align}
\begin{split}
  V_{F}(H_{u},H_{d})&= \vert \mu \vert^{2} \Big( \vert H_{u}^{+} \vert^{2}+ \vert H_{u}^{0}\vert^{2}+\vert H_{d}^{0} \vert^{2}+ \vert H_{d}^{-}\vert^{2} \Big)     
\end{split}\\
\begin{split}  
  V_{D}(H_{u},H_{d})&= \frac{1}{8} \left(g_{1}^{2}+g_{2}^{2}\right) \Big( \vert H_{u}^{0}\vert^{2}+\vert H_{u}^{+} \vert^{2}-\vert H_{d}^{0} \vert^{2}-\vert H_{d}^{-} \vert^{2} \Big)^{2}
\\
&+\frac{1}{2} g_{1}^{2}\vert H_{u}^{+}H_{d}^{0 \ast}+H_{u}^{0}H_{d}^{- \ast} \vert^{2}    
\end{split}\\
\begin{split}  
  V_{soft}(H_{u},H_{d})&= m^{2}_{H_{u}}\Big(\vert H_{u}^{+} \vert^{2}+ \vert H_{u}^{0}\vert^{2}\Big)+m^{2}_{H_{d}}\Big(\vert H_{d}^{0} \vert^{2}+ \vert H_{d}^{-}\vert^{2}\Big)
\\
&+ \Big( B\left(H_{u}^{+}H_{d}^{-}-H_{u}^{0}H_{d}^{0}\right) + H.C. \Big)              
\end{split}  \\
\begin{split}
  V(H_{u},H_{d})&= V_{F}(H_{u},H_{d})+V_{D}(H_{u},H_{d})+V_{soft}(H_{u},H_{d}),     
\end{split}
\end{align}
\end{subequations}
where $g_{1}$ and $g_{2}$ are the $U(1)_{Y}$ and $SU(2)_{I}$ gauge couplings, respectively. As in the SM, in order to preserve the electromagnetic gauge symmetry, one can set $H_{u}^{+}=H_{d}^{-}=0$ and then the scalar potential becomes

\begin{align}
V(H_{u}^{0},H_{d}^{0}) &= \Big(\vert \mu \vert^{2}+m^{2}_{H_{u}} \Big)\vert H_{u}^{0}\vert^{2}+\Big(\vert \mu \vert^{2}+m^{2}_{H_{d}} \Big)\vert H_{d}^{0}\vert^{2}+\Big(-B H_{u}^{0}H_{d}^{0}+H.C. \Big)
\nonumber \\
&+\frac{1}{8} \left(g_{1}^{2}+g_{2}^{2}\right)\Big(\vert H_{u}^{0}\vert^{2}-\vert H_{d}^{0}\vert^{2}\Big)^{2}.   \label{Higgs scalar potential}
\end{align}
Note that the quartic couplings of the Higgses are completely determined by the gauge couplings as a consequence of supersymmetry and, therefore, they cannot be seen as free parameters  of the MSSM.

In the minimum of the scalar potential, the real part of the neutral Higgs fields acquire non-zero vacuum expectation values that breaks spontaneously the electroweak symmetry into the electromagnetic gauge group $U(1)_{em}$. The neutral components of the Higgs doublets are expanded as

\begin{subequations}
\begin{align}
\begin{split}
  H_{d}^{0}&=\frac{1}{\sqrt{2}}\left(v_{d}+\phi_{d}+i\sigma_{d}\right)     
\end{split}\\
\begin{split}  
  H_{u}^{0}&=\frac{1}{\sqrt{2}}\left(v_{u}+\phi_{u}+i\sigma_{u}\right)     
\end{split},
\end{align}
\end{subequations}
in which $v=\sqrt{v_{u}^{2}+v_{d}^{2}}=246$ GeV and $\tan{\beta}=v_{u}/v_{d}$. The Higgs scalar potential is minimized by solving the equations $\partial V/\partial \phi_{u,d}=\partial V/\partial \sigma_{u,d}=0$, the so-called tadpole equations. The minimization conditions of the scalar potential can be written as

\begin{equation}
m^{2}_{H_{u}}+\vert \mu \vert^{2}-B \text{cot}\ \beta-\frac{1}{2} m_{Z}^{2}\text{cos} (2\beta)=0
\end{equation}
\begin{equation}
m^{2}_{H_{d}}+\vert \mu \vert^{2}-B \text{tan}\ \beta+\frac{1}{2} m_{Z}^{2}\text{cos} (2\beta)=0,
\end{equation}
where $m_{Z}$ is the tree-level mass term of the neutral vector boson of the SM. These conditions show that the soft SUSY breaking parameters $m^{2}_{H_{u}}$, $m^{2}_{H_{d}}$, $B$ and the supersymmetric parameter $\mu$ must all be of approximately the same order of magnitude as $M_{Z}$ in such a way that the EWSB occurs in a natural manner without requiring large fine-tuning of these independent parameters. The mass matrices of the Higgs states are computed from the second derivatives of the Higgs potential \eqref{Higgs scalar potential} taken at its absolute minimum. In the gauge basis the mass matrices are not diagonal, but they can be diagonalized by unitary matrices that change the basis to the basis of the physical states, which can be seen as the eigenstates of the mass matrices.

After the electroweak symmetry breaking, three of the eight degrees of freedom of the two Higgs doublets are eaten by the weak gauge bosons and become their longitudinal components, resulting in five physical Higgs bosons which are classified as: two CP-even $h^{0}$ and $H^{0}$, one CP-odd denoted as $A^{0}$ and two charged $H^{\pm}$. The physical CP-odd Higgs boson $A^{0}$ is obtained from the imaginary parts of $H_{u}^{0}$ and $H_{d}^{0}$ by the relations

\begin{equation} 
\left( \begin{array}{c} G^{0} \\ A^{0} 
\end{array} \right) = \left( 
\begin{array}{cc} 
\sin{\beta} & -\cos{\beta} \\ 
 \cos{\beta}  & \sin{\beta} \end{array} 
\right) 
\left( \begin{array}{c} \sigma_{u} \\  \sigma_{d}
\end{array} \right),
\end{equation}
where $G^{0}$ is the neutral Goldstone boson that, after a gauge transformation to the unitary gauge, becomes the longitudinal part of the massive $Z^{0}$ boson. The tree-level mass of the pseudoscalar Higgs is given by

\begin{equation}
m_{A^{0}}^{2}=\frac{B}{\sin{\beta}\cos{\beta}}.
\end{equation}
The physical charged Higgs bosons come from the charged components of the Higgs doublets

\begin{equation} 
\left( \begin{array}{c} G^{+} \\ H^{+} 
\end{array} \right) = \left( 
\begin{array}{cc} 
\sin{\beta} & -\cos{\beta} \\ 
 \cos{\beta}  & \sin{\beta} \end{array} 
\right) 
\left( \begin{array}{c} H_{u}^{+} \\  H_{d}^{- \ast}
\end{array} \right)
\end{equation}
with $G^{-}=G^{+ \ast}$ and $H^{-}=H^{+ \ast}$. In the unitary gauge the charged Goldstone bosons $G^{\pm}$ are eaten by the weak charged vector bosons $W^{\pm}$ and become their longitudinal components. The mass of the charged Higgs bosons is related to the mass of the pseudoscalar Higgs via 

\begin{equation}
m_{H^{\pm}}^{2}=m_{A^{0}}^{2}+M_{W}^{2}.
\end{equation}

Moving on to the real components of the neutral Higgs fields, the rotation to the eigenstates $h^{0}$ and $H^{0}$ is given by

\begin{equation} 
\left( \begin{array}{c} h^{0} \\ H^{0} 
\end{array} \right) = \left( 
\begin{array}{cc} 
\cos{\alpha} & -\sin{\alpha} \\ 
 \sin{\alpha}  & \cos{\alpha} \end{array} 
\right) 
\left( \begin{array}{c} \phi_{u} \\  \phi_{d}
\end{array} \right),
\end{equation}
where the mixing angle $\alpha$ is determined by \cite{Martin:1997ns}

\begin{equation} 
\frac{\sin{2\alpha}}{\sin{2\beta}}=-\left(\frac{m^{2}_{H^{0}}+m^{2}_{h^{0}}}{m^{2}_{H^{0}}-m^{2}_{h^{0}}}\right) \quad\quad\quad\quad\quad\quad\quad\quad \frac{\tan{2\alpha}}{\tan{2\beta}}= \left(\frac{m^{2}_{A^{0}}+m^{2}_{Z}}{m^{2}_{A^{0}}-m^{2}_{Z}}\right),
\end{equation}
and it is usually chosen to be negative $(-\pi <2\alpha<0)$ because, by convention, $m_{H^{0}}>m_{h^{0}}$. The tree-level masses of the CP-even Higgs bosons are given by

\begin{equation}
m^{2}_{h^{0},H^{0}}=\frac{1}{2}\left(m_{A^{0}}^{2}+m_{Z}^{2}\mp \sqrt{\left(m_{A^{0}}^{2}-m_{Z}^{2}\right)^{2}+4m_{Z}^{2}m_{A^{0}}^{2}\sin^{2}{2\beta}} \right).   \label{Higgs eigenvalues}
\end{equation}

The light state $h^{0}$ is assumed to be the 125 GeV Higgs boson of the SM. In principle, since $m_{A^{0}}$, $m_{H^{+}}$ and $m_{H^{0}}$ can be arbitrarily large because they all increase with $\vert B\vert/\sin{2\beta}$, $m_{h^{0}}$ is bounded from above. From eq. \eqref{Higgs eigenvalues} one obtains the following upper limit for the tree-level mass of the lightest Higgs \cite{Inoue:1982ej}
\begin{equation}
m^{2}_{h^{0}}\leq \vert m_{Z}\cos{2\beta}\vert,   
\end{equation}  
which is a consequence of the fact that the Higgs self couplings are obtained by the electroweak gauge couplings in the MSSM. This upper bound is ruled out by experiments, but this does not rule out the MSSM because the mass of the SM Higgs can receive radiative corrections already at the 1-loop level that are large enough to push $m_{h^{0}}$ towards the measured value of 125 GeV. Because of the size of the Yukawa couplings, the dominant contribution comes from top-stop loops. These corrections become large for large $\tan \beta$ and when stop masses are much bigger than the mass of the top quark.

\subsection{Gluinos}
\indent

As the superpartner of the gluon, gluino is a color octet Majorana fermion with mass given by $\vert M_{3} \vert$ and therefore its role and interactions are directly related to the properties of the supersymmetric version of QCD (SQCD). Since $SU(3)_{C}$ gauge symmetry is preserved, the gluino cannot mix with any other fermion, and it must be interpreted as a physical state. Gluino is being deeply searched by the ATLAS and CMS groups of the LHC, and because of its strong interactions with quarks and gluons, gluino is naturally expected to be one of the heaviest supersymmetric particles in many scenarios of the MSSM.

\subsection{Neutralinos and charginos}
\indent

The mixing between the neutral Higgsinos $(\tilde{H}^{0}_{u},\tilde{H}^{0}_{d})$ and the neutral gauginos $\lambda_{\tilde{B}}$ (bino) and $\lambda_{\tilde{W}^{3}}$ (wino) gives rise to four Majorana physical states called neutralinos $\tilde{\chi}^{0}_{i}$ with $i \in \lbrace 1,2,3,4 \rbrace$. The contribution to the neutralino mass matrix comes from the soft SUSY breaking gaugino mass terms, the $\mu$ term and the couplings of Higgs fields to gauginos and Higgsinos when the neutral Higgses acquire non-zero VEVs. In the basis $\left(\lambda_{\tilde{B}}, \lambda_{\tilde{W}^{3}}, \tilde{H}_d^0, \tilde{H}_u^0\right)$, the neutralino mass matrix is given by

\begin{equation} 
\mathcal{M}_{\tilde{\chi}^0} = \left( 
\begin{array}{cccc}
M_1 &0 &-\frac{1}{2} g_1 v_d  &\frac{1}{2} g_1 v_u \\ 
0 &M_2 &\frac{1}{2} g_2 v_d  &-\frac{1}{2} g_2 v_u \\ 
-\frac{1}{2} g_1 v_d  &\frac{1}{2} g_2 v_d  &0 &- \mu \\ 
\frac{1}{2} g_1 v_u  &-\frac{1}{2} g_2 v_u  &- \mu  &0  
\end{array} 
\right).    
\end{equation} 
This matrix is diagonalized by a unitary 4 $\times$ 4 matrix $N$ which gives the mass eigenstates $\left(\tilde{\chi}^{0}_{1},\tilde{\chi}^{0}_{2},\tilde{\chi}^{0}_{3},\tilde{\chi}^{0}_{4}\right)$ ordered in mass as a linear combination of $\left(\lambda_{\tilde{B}}, \lambda_{\tilde{W}^{3}}, \tilde{H}_d^0, \tilde{H}_u^0\right)$.

In supersymmetric models that respects R parity, the Lightest Supersymmetric Particle (LSP) is stable and can be eventually produced in the decay chain of all other heavier superparticles \cite{Martin:1997ns}. Usually the lightest neutralino $\tilde{\chi}^{0}_{1}$ is the LSP of the MSSM and it behaves as a neutral massive particle that interacts weakly with the other particles. As a consequence, $\tilde{\chi}^{0}_{1}$ is the most commonly known supersymmetric WIMP DM candidate studied in the literature, although SUSY also can provide other viable WIMP DM candidates. 

The two charged higgsinos $(\tilde{H}^{+}_{u},\tilde{H}^{-}_{d})$ can mix with the two charged gauginos $(\tilde{W}^{+}, \tilde{W}^{-})$ and form two charged fermions known as charginos $\chi_{1,2}^{\pm}$. If we use the following gauge-eigenstate basis for the positively and negatively charged states

\begin{equation} 
\psi^{+} = \left( \begin{array}{c} \tilde{W}^{+} \\  \tilde{H}^{+}_{u}
\end{array} \right)  \quad\quad\quad\quad\quad\quad\quad\quad\quad\quad\quad  
\psi^{-} = \left( \begin{array}{c} \tilde{W}^{-} \\  \tilde{H}^{-}_{d}
\end{array} \right),
\end{equation}
the chargino mass terms can be written as
\begin{equation}
-\frac{1}{2}\left[(\psi^{+})^{\text{T}} {\bf M}_{c}^{\text{T}} \psi^{-} + (\psi^{-})^{\text{T}} {\bf M}_{c} \psi^{+} \right] + \text{h.c.},    
\end{equation}
where
\begin{equation} 
{\bf M}_{c} = \left( 
\begin{array}{cc} 
M_{2} & \sqrt{2} \sin{\beta} M_{W^{+}} \\ 
 \sqrt{2} \cos{\beta} M_{W^{+}}  & \mu  \end{array} 
\right). 
\end{equation}
There are two independent mixings that occur to diagonalize the chargino mass matrix and, as a consequence, the mass eigenstates are obtained by two unitary rotation matrices. They can be defined as
\begin{equation}
\tilde{\chi}^{+}={\bf V} \psi^{+}= \left( \begin{array}{c} \chi^{+}_{1} \\  \chi^{+}_{2}
\end{array} \right),      
\end{equation}

\begin{equation}
\tilde{\chi}^{-}={\bf U} \psi^{-}= \left( \begin{array}{c} \chi^{-}_{1} \\  \chi^{-}_{2}
\end{array} \right).      
\end{equation}

\subsection{Sfermions}
\indent

The sfermion sector includes all the superpartners of the leptons and quarks,
namely to each SM fermionic field $\psi$ there is a complex scalar field $\tilde{\psi}$. The mass matrices of the sfermions receive contributions from the soft SUSY breaking Lagrangian \eqref{eqMSSMsoftSUSYbreaking} and from EWSB. Mixing can occur between the left-handed and right-handed components of the squarks and sleptons.



\section{Phenomenological tools to explore SUSY models} \label{CHAPSUSYlastsection}
\indent

Supersymmetric extensions of the SM are usually models that possesses a big particle content with many interaction terms in the Lagrangian and they have a large number of free parameters. For example, in the simplest supersymmetric extension of the SM, the MSSM, the particle content is increased by a factor bigger than two and there are a lot of new interactions between all the particle fields required by SUSY. As a consequence, the calculations of the relevant masses, vertices, tadpole equations and renormalization group equations are tedious.

To realize a comprehensive analysis of  a new SUSY model one has to first deal with a long list of tasks, which are: choose a gauge group, make sure it is free of gauge anomalies, calculate the Lagrangian, break some symmetries when necessary, solve the tadpole equations to find the minimization conditions of the scalar potential, calculate masses and vertices and finally calculate the renormalization group equations that are needed to connect the values of parameters of the model at different energy scales. Since most of the public tools used to study SUSY models are restricted to the MSSM or small extensions of it, more generic and sophisticated tools are needed to explore non-minimal SUSY models with the same precision as the MSSM. For this purpose the {\tt Mathematica} package {\tt SARAH} \cite{Staub:2008uz,Staub:2009bi,Staub:2010jh,Staub:2012pb,Staub:2013tta,Staub:2015kfa} has been created to raise up the possibility to explore non-minimal SUSY models in a faster way. In the next subsections we give a brief resume about the {\tt SARAH} framework and its connection with other useful tools.

\subsection{{\tt SARAH}} 
\indent

{\tt SARAH} is optimized for handling a wide range of BSM models. Originally it was created to study only SUSY models, but now recent versions of {\tt SARAH} can also explore non-supersymmetric extensions of the SM. The basic idea of {\tt SARAH} is to give the user the opportunity to build a model in an easier way, which is summarized as: choose a local gauge group, choose global symmetries, define the particle content and representations and finally write explicit non-gauge interactions for the superpotential. All these aspects must be stored in the model file that {\tt SARAH} reads. Additionally, the user has to specify which fields acquire VEVs and which fields mix after symmetry breaking. After the initialisation of a specified model, {\tt SARAH} realizes some operations to check its (self-)consistency:

\begin{itemize}

\item Check for gauge and mixed gauge/gravity anomalies;

\item Check for Witten anomalies \cite{Witten:1982fp};

\item Check if all terms in the (super)potential respect the global and gauge symmetries;

\item Check if terms allowed by the symmetries are missing in the (super)potential;

\item Check if additional mass eigenstates can mix in principle;

\item Check if all mass matrices are irreducible.

\end{itemize}

As a {\tt Mathematica} package used to build and analyze SUSY and non-SUSY models, {\tt SARAH} performs a lot of analytical calculations for a given model beyond the derivation of the Lagrangian. Let's give a summary of them. 

\begin{enumerate}
\item[{\bf a)}] {\bf Tadpole Equations:} During the evaluation of a model {\tt SARAH} calculates all the minimization conditions of the tree-level scalar potential, which are known in the Literature as the {\it tadpole equations}. {\tt SARAH} also calculates one- and two-loop corrections to tadpoles and self-energies for all particles; 

\item[{\bf b)}] {\bf Masses:} {\tt SARAH} calculates the mass matrices of the states which are rotated from gauge eigenstates to mass eigenstates. Additionally, it also calculates the masses of states that do not mix with other fields;

\item[{\bf c)}] {\bf Vertices:} {\tt SARAH} calculates in an efficient way all tree-level interaction vertices from the Lagrangian;

\item[{\bf d)}] {\bf Renormalization Group Equations:} In order to take into account the variation of the parameters of a model with respect to the energy scale, {\tt SARAH} calculates the full two-loop RGEs for SUSY and non-SUSY models including the full CP and flavour structure. 
\end{enumerate}
Finally, {\tt SARAH} can export all of these information to {\LaTeX} files so that the user can obtain a pdf file with all the analytical results calculated by {\tt SARAH}. Further details about the usage of {\tt SARAH} for analytical and numerical analyses of a model can be found in Refs.\cite{Staub:2015kfa,Vicente:2015zba,Staub:2016dxq}.

Since {\tt SARAH} is a {\tt Mathematica} package, it is not suited for deep numerical studies. However, {\tt SARAH} can use the analytical information derived about a model and pass it to other phenomenological tools. Once the model implementation is successfully done, {\tt SARAH} can generate the required input files for many other popular tools so that the user can realize a deeper phenomenological study of the model. A schematic illustration of the workflow is shown in Figure~\ref{FigSARAH}. In the next subsections we give a summary of {\tt SPheno} and {\tt MicrOMEGAs}, which were the tools mostly used in the project contained in this part of this thesis, together with a brief resume of some other phenomenological tools.

\begin{figure}[h!]
\centering
\includegraphics[width=0.85\textwidth]{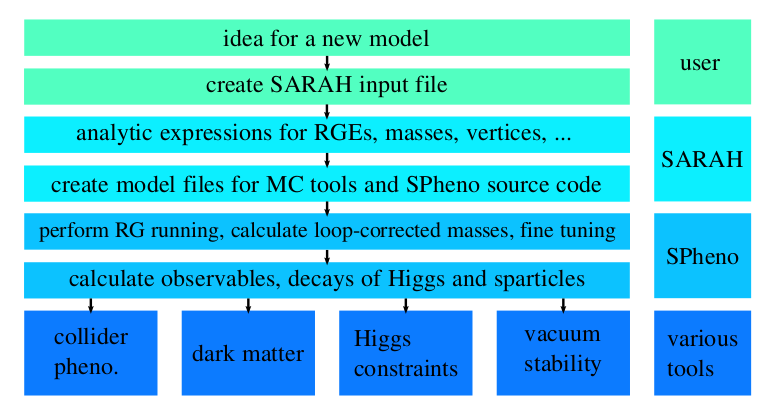}
\caption{\label{fig:overleaf} The workflow within the framework of SARAH/SPheno.} \label{FigSARAH}
\end{figure}

\subsection{{\tt SPheno}} 
\indent

To calculate certain observables of a parameter region of a model one needs to know the masses of the particles and the interaction couplings that are necessary to take into account in the calculation of the observable. {\tt SARAH} criates a set of {\tt Fortran} files for the spectrum generator {\tt SPheno} \cite{Porod:2003um, Porod:2011nf} that contains all the derived information about the mass matrices, vertices, tadpole equations, loop corrections and RGEs of the given model. {\tt SPheno} is a spectrum calculator that calculates, for given values of the input parameters, all masses and mixing matrices of a model, including full corrections of one loop order for all the masses and the main corrections of two loop order for the scalar particles of the model. Furthermore, after the calculation of the masses, it also calculates some low-energy observables (such as the anomalous magnetic moment of the muon and other lepton dipole moments), many flavor observables, as well as two- and three-body decay widths. {\tt SPheno} stands for S(upersymmetric) Pheno(menology) and, although the code was originally developed to cover only the MSSM and its first simple extensions, now it can cover many other models (including the non-supersymmetric ones) thanks to {\tt SARAH}.

\subsection{{\tt MicrOMEGAs}} 
\indent

{\tt MicrOMEGAs} \cite{Belanger:2001fz,Belanger:2014hqa,Belanger:2014vza} is a very popular tool used by many phenomenologists and dark matter model builders to compute the properties and signatures of a dark matter candidate in a generic model of new physics. This tool calculates not only the relic density for one or more stable massive particle in a model, but it also gives the cross sections for direct and indirect DM searches. {\tt MicrOMEGAs} uses the {\tt CalcHEP} package \cite{Pukhov:2004ca, Belyaev:2012qa} to evaluate Feynman diagrams and calculate all tree-level matrix elements of all subprocesses used for the relic density calculation and for the direct and indirect DM searches.

In order to implement the user's model in {\tt MicrOMEGAs}, {\tt SARAH} generates model files for {\tt CalcHEP} and also writes main files which can be used to run {\tt MicrOmegas}. Finally, to obtain the dark matter properties of the dark matter candidate of the model in a certain point of the parameter space, the mass spectrum file produced by {\tt SPheno} is used as an input file for the main file generated by {\tt SARAH} that is used to run {\tt MicrOmegas}.


\subsection{Other tools} 
\indent

{\tt SARAH} writes model files for many other tools that can be used to perform other useful phenomenological analysis for the model, such as: collider studies, check Higgs constraints and check the vacuum stability. Given the wealth of LHC results, one needs Monte-Carlo (MC) event generators to perform an efficient collider study with the recent data obtained by LHC. {\tt SARAH} provides interfaces to the two event generators {\tt CalcHEP} and {\tt WHIZARD} \cite{Kilian:2007gr,Moretti:2001zz} and, additionally, it writes a universal {\tt FeynRules} output file (UFO) \cite{Degrande:2011ua}, which is used to implement new models into several other MC tools such as {\tt MadGraph} \cite{Alwall:2011uj,Alwall:2014hca}, {\tt GoSam} \cite{Cullen:2011ac}, {\tt Herwig++} \cite{Gieseke:2003hm,Gieseke:2006ga,Bellm:2013hwb} and {\tt Sherpa} \cite{Gleisberg:2003xi,Gleisberg:2008ta,Hoche:2014kca}. The consistency of the Higgs sector with experimental data of a given parameter point is checked by {\tt HiggsBounds} \cite{Bechtle:2008jh,Bechtle:2011sb,Bechtle:2013wla} and {\tt HiggsSignals} \cite{Bechtle:2013xfa}. Vacuum stability of the scalar potential of the model can be tested with {\tt Vevacious} \cite{Camargo-Molina:2013qva}, a tool that operates to find the global minimum of the one-loop effective potential. With all these improved numerical tools available, the barriers to performing studies of new physics beyond the SM have been substantially reduced.

\chapter{$U(1)'$ Extensions of the MSSM}   \label{chapterUMSSM}
\indent

\section{Introduction}
\indent

The possibility of adding an extra $U(1)'$ gauge symmetry to the SM is well motivaded by many studies beyond the SM, including superstring constructions \cite{Cvetic:1995rj}, grand unified theories \cite{Langacker:1980js,London:1986dk}\cite{Langacker:1998tc}, models of dynamical symmetry breaking \cite{Hill:2002ap}, little Higgs models \cite{ArkaniHamed:2001nc}, large extra dimensions \cite{Masip:1999mk} and Stueckelberg mechanism \cite{Kors:2004dx}. Models with an extended Abelian gauge group generally are expected to arise from the breaking of an $SO(10)$ or $E_{6}$ symmetry to the SM gauge symmetry. They can be the low energy limit of some superstring models and have interesting consequences both at the theoretical and low energy phenomenological level. In the supersymmetric version, $U(1)'$ extensions of the MSSM can provide a natural solution to the $\mu$ problem of the MSSM \cite{Kim:1983dt,Cohen:2008ni} where the $\mu$ term is generated dynamically by the vacuum expectation value (VEV) of the SM singlet field $S$ which breaks the $U(1)'$ symmetry \cite{Cvetic:1997ky}. Although this solution is similar to the one provided by the next-to-minimal supersymmetric standard model (NMSSM) \cite{Ellwanger:2009dp}, the UMSSM is free of the cosmological domain wall problem because the $U(1)'$ symmetry forbids the appearance of domain walls which are created by the $Z_{3}$ discrete symmetry of the NMSSM \cite{Han:2004ju}. 

One of the main motivations to study the UMSSM is that, besides the well-known neutralino of the MSSM, it provides another good WIMP candidate to describe the dark matter properties: the right-handed sneutrino \cite{Belanger:2011rs,Belanger:2015cra,Belanger:2017vpq}. This is in contrast with the left--handed sneutrinos of the MSSM, which have been ruled out as DM candidates by direct WIMP searches because their scattering cross sections on nuclei are too large \cite{Falk:1994es}. Right--handed sneutrinos have small scattering cross sections on nuclei. Moreover, being scalar $SU(2)$ singlets, a right--handed sneutrino only has two degrees of freedom; in contrast, a higgsino--like neutralino, which also has unsuppressed couplings to the $Z'$ boson in many cases, effectively has eight (an $SU(2)$ doublet of Dirac fermions, once co--annihilation has been included). Another nice feature of the UMSSM is that it can also be used to study electroweak baryogenesis by the fact that the interactions of the singlet Higgs with the Higgs doublets can give rise to a strongly first order phase transition \cite{Ahriche:2010ny}.
 
The confirmation that neutrinos should have tiny masses to explain their oscilations is viewed as a natural motivation to add right-handed (RH) neutrinos to the SM field content. UMSSM models can be used to obtain neutrino masses that are consistent with neutrino oscillation data in such a way that the exact details depend on the form of the extra $U(1)'$ gauge symmetry \cite{Kang:2004ix}. If the neutrinos are Majorana particles, the smallness of their masses is usually explained through a see-saw mechanism \cite{Minkowski:1977sc,Mohapatra:1979ia,Schechter:1980gr} which requires the existence of a heavy RH neutrino whose natural mass scale is generally around $10^{12}$ GeV.

\section{Description of the UMSSM} \label{section5.2}
\indent

The UMSSM is well known as an Abelian extention of the MSSM with gauge group $SU(3)_{C}\times SU(2)_{L}\times U(1)_{Y}\times U(1)'$, which can
result from the breaking of the $E_6$ gauge symmetry \cite{London:1986dk, Hewett:1988xc}. In other words, it can be seen as
the low energy limit of a -- possibly string-inspired -- $E_6$ grand unified gauge theory. $E_{6}$ contains $SO(10)\times U(1)_{\psi}$ and, since $SO(10)$ can be decomposed into $SU(5)\times U(1)_{\chi}$, after applying the Hosotani mechanism \cite{Hosotani:1983xw,Hosotani:1983vn} and noting that $SU(5)$ contains the SM gauge group, one can break $E_{6}$ directly into $SU(3)_{C}\times SU(2)_{L}\times U(1)_{Y}\times U(1)_{\psi}\times U(1)_{\chi}$. Here we assume that only one extra $U(1)$ factor survives
at the relevant energy scale, which in general is a linear combination of $U(1)_\psi$ and $U(1)_\chi$, parameterized by a mixing angle $\theta_{E_{6}}$ \cite{London:1986dk}

\begin{equation}
U(1)'=\text{sin} \theta_{E_{6}} U(1)'_{\psi}+ \text{cos} \theta_{E_{6}} U(1)'_{\chi},  \label{eq 1.1}
\end{equation}
with $\theta_{E_{6}}\in [-\frac{\pi}{2},\frac{\pi}{2}]$. The $U(1)'$
charges of all the fields contained in the model are then given by

\begin{equation}
Q'(\theta_{E_{6}})=\text{sin} \theta_{E_{6}} Q'_{\psi}+ \text{cos} \theta_{E_{6}} Q'_{\chi}  \label{eq 1.2}
\end{equation}
where $Q'_{\psi}$ and $Q'_{\chi}$ are the charges associated to the gauge groups $U(1)'_{\psi}$ and $U(1)'_{\chi}$, respectively. 
In Table \ref{T1} we give the $U(1)'$ charge of all relevant matter and Higgs fields in the UMSSM for certain values of the mixing angle $\theta_{E_6}$.

\begin{table}
\begin{center}
\begin{tabular}{ |c|c|c|c|c|c|c| }  
\hline
    {$ $} & {$2\sqrt{6} Q'_{\psi}$} & {$2\sqrt{10} Q'_{\chi}$} & {$2\sqrt{10} Q'_{N}$} & {$2\sqrt{15} Q'_{\eta}$} & {$2\sqrt{15} Q'_{S}$} & {$2 Q'_{I}$}  \\ 
    \hline
    $\theta_{E_{6}}$  & $\frac{\pi}{2}$ & 0 & arctan$\sqrt{15}$ & -arctan$\sqrt{5/3}$ & arctan($\sqrt{15}/9$) & arctan$\sqrt{3/5}$  \\ \hline
    $Q'_{Q}$  & 1  & -1 & 1  & -2 & -1/2 & 0    \\
    $Q'_{U^{c}}$  & 1  & -1 & 1  & -2 & -1/2 & 0   \\
    $Q'_{D^{c}}$  & 1  & 3 & 2  & 1 & 4  & -1      \\ 
    $Q'_{L}$  & 1   & 3 & 2   & 1 & 4 & -1  \\
    $Q'_{N^{c}}$  & 1  & -5 & 0  & -5 & -5 & 1     \\
    $Q'_{E^{c}}$  & 1  & -1 & 1  & -2 & -1/2 & 0    \\
    $Q'_{H_{u}}$  & -2  & 2 & -2  & 4 & 1  & 0      \\ 
    
    $Q'_{H_{d}}$  & -2  & -2 & -3  & 1 & -7/2 & 1   \\
    $Q'_{S}$ & 4   & 0 & 5   & -5 & 5/2 & -1    \\
\hline
\end{tabular}
\caption{$U(1)'$ charges of the UMSSM chiral superfields for certain values of $\theta_{E_{6}}$.}
\label{T1}
\end{center}
\end{table}

\begin{figure}[h!]
\centering
\includegraphics[width=0.8\textwidth]{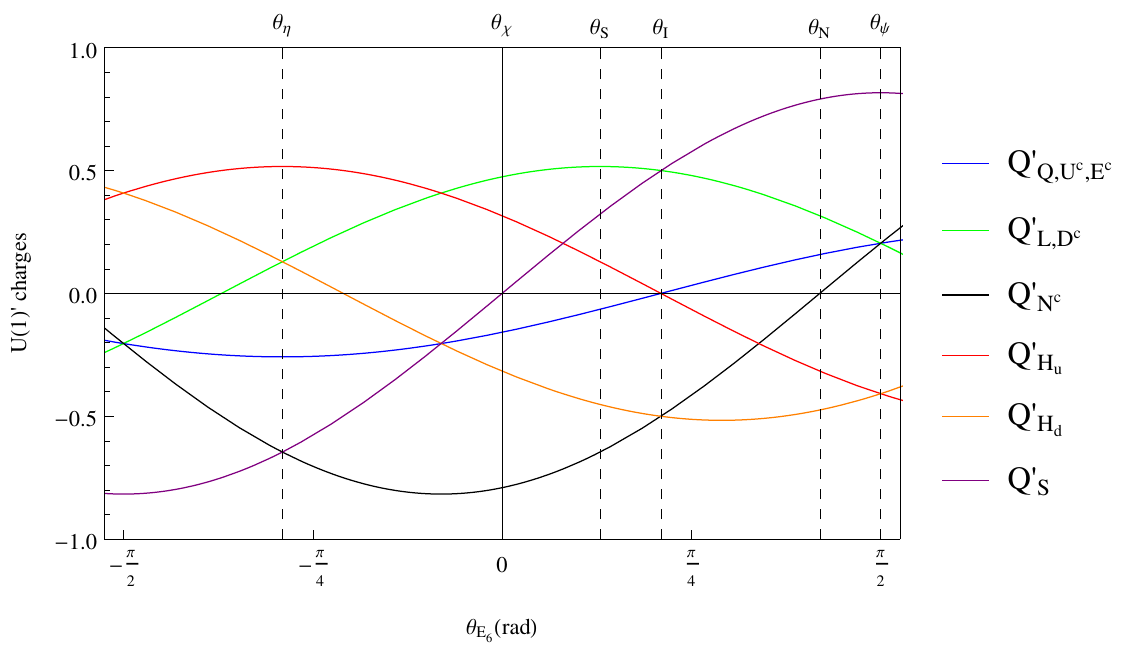}
\caption{\label{fig:overleaf}$U(1)'$ charges of all the matter fields of the UMSSM as a function of $\theta_{E_{6}}$.} \label{Fig1}
\end{figure}

In addition to the new vector superfield $\hat{B'}$ and the MSSM superfields, the UMSSM
contains one electroweak singlet supermultiplet
$\hat{S}\equiv (s,\tilde{s})$, with a scalar field $s$ that breaks the $U(1)'$ gauge symmetry, and three RH neutrino supermultiplets
$\hat{N}^{c}_{i}\equiv (\tilde{\nu}_{R}^{c},\nu_{R}^{c})_{i}$ whose fermionic components are needed to cancel anomalies related to the $U(1)'$ gauge symmetry. It should be noted that $U(1)_\psi$ and $U(1)_\chi$
are both anomaly--free over complete (fermionic) representations of
$E_6$. Since $U(1)_\chi$ is a subgroup of $SO(10)$, which is also
anomaly--free over complete representations of $SO(10)$, and the SM
fermions plus the right--handed neutrino complete the ${\bf 16}-$dimensional representation of $SO(10)$, $U(1)_\chi$ is
anomaly--free within the fermion content we show in Table \ref{T1}. However, $U(1)_\psi$ will be anomaly--free only after we
include the ``exotic'' fermions that are contained in the ${\bf 27}-$dimensional representation of $E_6$, but are not contained
in the ${\bf 16}$ of $SO(10)$. Here we assume that these exotic superfields are too heavy to affect the calculation of the $\tilde \nu_{R,1}$ relic density. We will see that this assumption is not essential for our result. 

Figure~\ref{Fig1} shows these charges as functions of the mixing angle
$\theta_{E_6}$. We identify by vertical lines values of $\theta_{E_6}$
that generate the well--known $U(1)'$ groups denoted by
$U(1)'_{\psi}$, $U(1)'_N$, $U(1)'_I$, $U(1)'_S$, $U(1)'_\chi$ and
$U(1)'_\eta$. The black curve in Fig.~\ref{Fig1} shows that for
$\theta_{E_6} = \arctan{\sqrt{15}}$ the $U(1)'$ charge of the RH
(s)neutrinos vanishes; this corresponds to the $U(1)'_N$ model of
Table~\ref{T1}. This model is not of interest to us, since the
$\tilde \nu_{R,i}$ are then complete gauge singlets, and do not couple
to any potential $s-$channel resonance. Similarly, for
$\theta_{E_6} = 0$, i.e.  $U(1)' = U(1)'_\chi$, the charge of $\hat S$
vanishes; in that case $s$ cannot be used to break the gauge symmetry,
i.e. the field content we have chosen is not sufficient to achieve the
complete breaking of the (extended) electroweak gauge symmetry down to
$U(1)_{\rm QED}$. All other values of $\theta_{E_6}$ are acceptable
for us.

\begin{table}[h]
\begin{center} 
\begin{tabular}{|c|c|c|c|c|c|} 
\hline  
SF & Spin 0 & Spin \(\frac{1}{2}\) & Generations & \((SU(3)_{C}\times\,SU(2)_{L}\times\,U(1)_{Y}\times\, U(1)')\) \\ 
\hline 
\({\bf \hat{Q}}\) & \((\tilde{u}_{L} \quad \tilde{d}_{L})\) & \((u_{L} \quad d_{L})\) & 3 & \(({\bf 3},{\bf 2},\frac{1}{6},Q'_{Q}) \) \\ 
\({\bf \hat{L}}\) & \((\tilde{\nu}_{L} \quad \tilde{e}_{L})\) & \((\nu_{L} \quad e_{L})\) & 3 & \(({\bf 1},{\bf 2},-\frac{1}{2},Q'_{L}) \) \\ 
\({\bf \hat{H}}_d\) & \((H_{d}^{0} \quad H_{d}^{-})\) & \((\tilde{H}_{d}^{0} \quad \tilde{H}_{d}^{-})\) & 1 & \(({\bf 1},{\bf 2},-\frac{1}{2},Q'_{H_d}) \) \\ 
\({\bf \hat{H}}_u\) & \((H_{u}^{+} \quad H_{u}^{0})\) & \((\tilde{H}_{u}^{+} \quad \tilde{H}_{u}^{0})\) & 1 & \(({\bf 1},{\bf 2},\frac{1}{2},Q'_{H_u}) \) \\ 
\({\bf \hat{D}}^{c}\) & \(\tilde{d}_{R}^{c}\) & \(d_{R}^{c}\) & 3 & \(({\bf \overline{3}},{\bf 1},\frac{1}{3},Q'_{D^{c}}) \) \\ 
\({\bf \hat{U}}^{c}\) & \(\tilde{u}_{R}^{c}\) & \(u_{R}^{c}\) & 3 & \(({\bf \overline{3}},{\bf 1},-\frac{2}{3},Q'_{U^{c}}) \) \\ 
\({\bf \hat{E}}^{c}\) & \(\tilde{e}_{R}^{c}\) & \(e_{R}^{c}\) & 3 & \(({\bf \overline{1}},{\bf 1},1,Q'_{E^{c}}) \) \\ 
\({\bf \hat{N}}^{c}\) & \(\tilde{\nu}_{R}^{c}\) & \(\nu_{R}^{c}\) & 3 & \(({\bf \overline{1}},{\bf 1},0,Q'_{N^{c}}) \) \\ 
\({\bf \hat{S}}\) & \(s\) & \(\tilde{s}\) & 1 & \(({\bf 1},{\bf 1},0,Q'_{s}) \) \\ 
\hline 
\end{tabular} 
\caption{Chiral superfields used in the UMSSM and their corresponding quantum numbers.}
\label{T2}
\end{center}
\end{table}

The superpotential of the UMSSM contains, besides the MSSM
superpotential without $\mu$ term, a term that couples the extra
singlet superfield to the two doublet Higgs superfields; this term is
always allowed, since it is part of the gauge invariant ${\bf 27}^3$
of $E_6$. The superpotential also contains Yukawa couplings for
the neutrinos. We thus have:
\begin{equation} \label{eq 1.3}
\hat W = \hat{W}_{MSSM}|_{\mu=0} + \lambda \hat{S} \hat{H}_u \cdot \hat{H}_d 
+ \hat{N}^{C} {\bf Y_{\nu}} \hat{L} \cdot \hat{H}_u\,, 
\end{equation}
where $\cdot$ stands for the antisymmetric $SU(2)$ invariant product
of two doublets. The neutrino Yukawa coupling ${\bf Y_{\nu}}$ is a
$3 \times 3$ matrix in generation space and $\lambda$ is a
dimensionless coupling. Note that for $\theta_{E_6} \neq 0$ the
$U(1)'$ symmetry forbids both bilinear $\hat N^C_i \hat N^C_j$ and
trilinear $\hat S \hat N^C_i \hat N^C_j$ terms in the
superpotential. In this model the neutrinos therefore obtain pure
Dirac masses, which means that the Yukawa couplings $Y_{\nu,ij}$ must
be of order $10^{-11}$ or less; in our numerical analysis we therefore
set ${\bf Y_\nu} = 0$. 

The electroweak and the $U(1)'$ gauge symmetries are spontaneously
broken when, in the minimum of the scalar potential, the real parts of
the doublet and singlet Higgs fields acquire non--zero vacuum
expectation values. These fields are expanded as
\begin{subequations} \label{vevs}
\begin{align}
\begin{split}
  H_d^0 &= \frac{1}{\sqrt{2}} \left( v_d + \phi_d + i\sigma_d \right)\,;
\end{split}\\
\begin{split}  
  H_u^0 &= \frac{1}{\sqrt{2}} \left( v_u + \phi_u + i\sigma_u \right)\,;  
\end{split}\\
\begin{split}  
  s &= \frac{1}{\sqrt{2}} \left( v_s + \phi_s + i\sigma_s \right)\,.            
\end{split}
\end{align}
\end{subequations}
We define $\tan{\beta} = \frac{v_u} {v_d}$ and
$v = \sqrt{v_d^2 + v_u^2}$ exactly as in the MSSM; this describes the
breaking of the $SU(2)_L \times U(1)_Y$ symmetry, and makes subleading
contributions to the breaking of $U(1)'$. The latter is mostly accomplished
by the VEV of $s$. The coupling $\lambda$ in
eq.(\ref{eq 1.3}) then generates an effective $\mu-$term:
\begin{equation} \label{eq 1.4}
\mu_{\rm eff} = \lambda \frac{v_{s}}{\sqrt{2}}\,. 
\end{equation}

As well known, supersymmetry needs to be broken. We parameterize this by
soft breaking terms:
\begin{align} \label{eq 1.66}
-\mathcal{L}_{SB} &= m^2_{H_d} |H_d|^2 + m^2_{H_u} |H_{u}|^2 + m_S^2 |s|^2
+ \tilde{Q}^\dagger {\bf m^2_{\tilde Q}} \tilde{Q} + 
\tilde{d}_R^{\dagger} {\bf m^2_{\tilde D^C}} \tilde{d}_R \nonumber \\
&+\tilde{u}_{R}^\dagger {\bf m^2_{\tilde U^C}} \tilde{u}_R
+\tilde{L}^\dagger {\bf m^2_{\tilde L}} \tilde{L} 
+\tilde{e}_R^\dagger {\bf m^2_{\tilde{E}^{C}}} \tilde{e}_R
+\tilde{\nu}_{R}^\dagger {\bf m^2_{\tilde N^C}} \tilde{\nu}_{R} \nonumber \\
&+\frac{1}{2}\Big( M_1 \lambda_{\tilde B} \lambda_{\tilde B} + 
M_2 \lambda_{\tilde W} \lambda_{\tilde W} + M_3 \lambda_{\tilde g} \lambda_{\tilde g}
+ M_4\lambda_{\tilde B'} \lambda_{\tilde B'} + h.c.\Big) \\
&+ \Big( \tilde{u}_R^C {\bf T_u} \tilde{Q}_L \cdot H_u 
- \tilde{d}_R^C {\bf T_d} \tilde{Q}_L \cdot H_d
- \tilde{e}_R^C{\bf T_e} \tilde{L}_L \cdot H_d + T_\lambda s H_u \cdot H_d
+ \tilde{\nu}_R^C {\bf T_\nu} \tilde{L}_L \cdot H_u + h.c. \Big)\,. \nonumber
\end{align}
Here we have used the notation of {\tt SPheno} \cite{Porod:2003um,
  Porod:2011nf}. The soft scalar masses and the soft trilinear
parameters of the sfermions are again  $3 \times 3$ matrices in generation
space. In the UMSSM, the $B\mu$ term of the MSSM is induced by the
$T_\lambda$ term after the breaking of the $U(1)'$ gauge symmetry.

In the following subsections we discuss those parts of the spectrum in
a bit more detail that are important for our calculation. These are
the sfermions, in particular sneutrinos; the massive gauge bosons; the
Higgs bosons; and the neutralinos. The lightest right--handed
sneutrino is assumed to be the LSP, which annihilates chiefly through
the exchange of massive gauge and Higgs bosons in the
$s-$channel. Requiring the lightest neutralino to be sufficiently
heavier than the lightest right--handed sneutrino gives important
constraints on the parameter space. The mass matrices in these
subsections have been obtained with the help of the computer code {\tt
  SARAH} \cite{Staub:2008uz, Staub:2013tta, Staub:2015kfa}; many of
these results can also be found in refs.~\cite{Belanger:2011rs,
  Belanger:2015cra, Belanger:2017vpq}.

\section{Gauge bosons}
\label{subsectionGaugeBosonsU1}

\indent

The UMSSM contains three neutral gauge bosons, from the $SU(2)_L, \, U(1)_Y$
and $U(1)'$, respectively. As in the SM and MSSM, after symmetry breaking
one linear combination of the neutral $SU(2)_L$ and $U(1)_Y$ gauge
bosons remains massless; this is the photon. The orthogonal state $Z_0$ mixes
with the $U(1)'$ gauge boson $Z_0'$ via a $2 \times 2$ mass matrix:
\begin{equation} \label{MM}
\mathcal{M}^2_{ZZ'}= \, \left( 
\begin{array}{cc} 
M_{Z_0}^2 & \Delta_Z \\  \Delta_Z  & M_{Z_0^{\prime}}^2 \end{array} 
\right)\,,
\end{equation}
with
\begin{eqnarray}
M_{Z_0}^2 & = & \frac{1}{4} (g_1^2+g_2^2) v^2\,;       \label{37}
\\
\Delta_Z & = & \frac{1}{2} g' \sqrt{g_1^2+g_2^2} \left( 
Q'_{H_d} v_d^2 - Q'_{H_u} v_u^2\right)\,;  \label{42}
\\
M_{Z'_0}^2 & = & g'^2 \Big( Q'^2_{H_d} v_d^2  + Q'^2_{H_u} v_u^{2}  
+ Q'^2_S v_s^2 \Big)\,.           \label{eq 2.10}
\end{eqnarray}
Recall that $g_2, \, g_1$ and $g'$ are the gauge couplings associated
to $SU(2)_L, \, U(1)_Y$ and $U(1)'$, respectively. The eigenstates
$Z$ and $Z'$ of this mass matrix can be written as:
\begin{eqnarray}
Z & = & \cos{\alpha_{ZZ'}} Z_0 + \sin{\alpha_{ZZ'}} Z_0'\,; \nonumber \label{44}
\\
Z' & = & -\sin{\alpha_{ZZ'}} Z_0 + \cos{\alpha_{ZZ'}}Z_0'\,.     \label{45}
\end{eqnarray}
The mixing angle $\alpha_{ZZ'}$ is given by
\begin{equation}
\sin{2\alpha_{ZZ'}} = \frac{2\Delta_{Z}} { M_Z^2 - M_{Z'}^2 }\,.
\end{equation}
The masses of the physical states are
\begin{equation}
M_{Z,Z'}^2 = \frac{1}{2} \left[ M_{Z_0}^2 + M_{Z'_0}^2 \mp 
\sqrt{\left( M_{Z'_0}^2 - M_{Z_0}^2 \right)^2 + 4\Delta_Z^2} \ \right]\,.
\end{equation}

Note that the off--diagonal entry $\Delta_Z$ in eq.(\ref{MM}) is of
order $v^2$. We are interested in $Z'$ masses in excess of $10$ TeV,
which implies $v_s^2 \gg v^2$. The mixing angle $\alpha_{ZZ'}$ is
${\cal O}(M_Z^2 / M^2_{Z'})$, which is automatically below current
limits \cite{Tanabashi:2018oca} if $M_{Z'} \geq 10$ TeV. Moreover, mass mixing
increases the mass of the physical $Z'$ boson only by a term of order
$M^4_Z/M^3_{Z'}$, which is less than $0.1$ MeV for $M_{Z'} \geq 10$
TeV. To excellent approximation we can therefore identify the physical
$Z'$ mass with $M_{Z'_0}$ given in eq.(\ref{eq 2.10}), with the last
term $\propto v_s^2$ being the by far dominant one.

Recall from the discussion of the previous subsection that the mass of
the right--handed sneutrinos can get a large positive contribution
from the $U(1)'$ $D-$term for some range of $\theta_{E_6}$. In fact,
from eqs.(\ref{eq 2.18}) and (\ref{eq 2.10}) together with the charges
listed in Table~\ref{T1} we find that this $D-$term contribution
exceeds $(M_{Z'}/2)^2$ if
\begin{equation} \label{th_range}
-\sqrt{15} < \tan \theta_{E_6} < 0\,.
\end{equation}
For this range of $\theta_{E_6}$ one therefore needs a negative
squared soft breaking contribution $m^2_{\tilde N^C_1}$ in order to
obtain $M_{\tilde \nu_{R,1}} \simeq M_{Z'}/2$.

Note that we neglect kinetic $Z - Z'$ mixing \cite{Kalinowski:2008iq,
  Belanger:2017vpq}. In the present context this is a loop effect
caused by the mass splitting of members of the ${\bf 27}$ of
$E_6$. This induces small changes of the couplings of the physical
$Z'$ boson, which have little effect on our results; besides, this loop
effect should be treated on the same footing as other one--loop
corrections.

\section{The Higgs Sector}
\indent

The Higgs sector of the UMSSM contains two complex $SU(2)_{L}$ doublets
$H_{u,d}$ and the complex singlet $s$. Four degrees of freedom get
``eaten'' by the longitudinal components of $W^\pm, \, Z$ and
$Z'$. This leaves three neutral CP--even Higgs bosons
$h_{i},\, i\in \{1,2,3 \}$, one CP-odd Higgs boson $A$ and two charged
Higgs bosons $H^{\pm}$ as physical states. After solving the
minimization conditions of the scalar potential for the soft breaking
masses of the Higgs fields, the symmetric $3 \times 3$ mass matrix for
the neutral CP--even states in the basis $(\phi_d,\phi_u,\phi_s)$ has
the following tree--level elements:

\begin{subequations}\label{eq:subeqns}
\begin{align}
\left(\mathcal{M}_{+}^{0}\right)_{\phi_{d}\phi_{d}}&= \Big[\frac{g_{1}^{2}+g_{2}^{2}}{4}+ (Q'_{H_d})^{2}g'^{2}\Big]v_{d}^{2} + \frac{T_{\lambda}v_{s}v_{u}}{\sqrt{2}v_{d}}    \label{eq 2.14a} \\  
\left(\mathcal{M}_{+}^{0}\right)_{\phi_{d}\phi_{u}}&= -\Big[\frac{g_{1}^{2}+g_{2}^{2}}{4}-g'^{2}Q'_{H_d}Q'_{H_u}-\lambda^{2}\Big]v_{d}v_{u}-\frac{T_{\lambda}v_{s}}{\sqrt{2}}
\label{eq 2.14b} \\
\left(\mathcal{M}_{+}^{0}\right)_{\phi_{d}\phi_{s}}&=\Big[g'^{2}Q'_{H_d}Q'_{S}+\lambda^{2}\Big]v_{d}v_{s}-\frac{T_{\lambda}v_{u}}{\sqrt{2}}\label{eq 2.14c} \\
\left(\mathcal{M}_{+}^{0}\right)_{\phi_{u}\phi_{u}}&= \Big[\frac{g_{1}^{2}+g_{2}^{2}}{4}+ (Q'_{H_u})^{2}g'^{2}\Big]v_{u}^{2} + \frac{T_{\lambda}v_{s}v_{d}}{\sqrt{2}v_{u}}
\label{eq 2.14d}   \\
\left(\mathcal{M}_{+}^{0}\right)_{\phi_{u}\phi_{s}}&= \Big[g'^{2}Q'_{H_u}Q'_{S}+\lambda^{2}\Big]v_{u}v_{s}-\frac{T_{\lambda}v_{d}}{\sqrt{2}}\label{eq 2.14e}   \\
\left(\mathcal{M}_{+}^{0}\right)_{\phi_{s}\phi_{s}}&= g'^{2}(Q'_{S})^{2}v_{s}^{2}+\frac{T_{\lambda}v_{d}v_{u}}{\sqrt{2}v_{s}}.  \label{eq 2.14f}
\end{align}
\end{subequations}
This matrix is diagonalized by a unitary 3 $\times$ 3 matrix $Z^{H}$ which gives the mass eigenstates $\left(h_{1},h_{2},h_{3}\right)$ ordered in mass as a linear combination of $(\phi_{d},\phi_{u},\phi_{s})$. In general the eigenstates and eigenvalues of the mass matrix above have to be obtained numerically.
 
At tree level, the lightest Higgs mass can be approximately written as \cite{King:2005jy}\cite{Belanger:2017vpq}

\begin{eqnarray}
M_{h_{1}}^{2}|_{\text{tree}}&\simeq& \frac{1}{4}(g_{1}^{2}+g_{2}^{2})v^{2}\text{cos}^{2}2\beta+\frac{1}{2}\lambda^{2}v^{2}\text{sin}^{2}2\beta+g'^{2}v^{2}\left(Q'_{H_d}\text{cos}^{2}\beta+Q'_{H_u}\text{sin}^{2}\beta\right)^{2} \nonumber \\
& - & \frac{v^{2}}{g'^{2}(Q'_{S})^{2}}\Big[\lambda^{2}-\frac{T_{\lambda}\text{sin}^{2}2\beta}{\sqrt2 v_{s}}+g'^{2}Q'_{S}\left(Q'_{H_d}\text{cos}^{2}\beta+Q'_{H_u}\text{sin}^{2}\beta\right)\Big]^{2} \label{eq 1.65}
\end{eqnarray}
where the first term on the right-hand side is the MSSM tree-level upper bound, the second term is an F-term contribution that also appears in the NMSSM, while the third and the last term are respectively a D-term and a combination of an F- and a D-term that comes from the UMSSM. It is also possible to have a Higgs boson lighter than the SM Higgs if the fourth term dominates. The clear dependence of the last two terms on the $U(1)'$ charges shows that the magnitude of the tree-level contribution to the lightest Higgs mass will strongly depend on the value of the $U(1)$ mixing angle $\theta_{E_{6}}$. 

The interaction between the SM singlet superfield and the two Higgs doublets brings a new contribution to the tree level mass of the lightest Higgs, which is the same as in the NMSSM. In addition, the UMSSM has also other contributions to enhance the SM-like Higgs mass that comes from the $U(1)'$ D-term \cite{Cvetic:1997ky}\cite{Barger:2006dh} of the scalar potential and, as a consequence of these contributions, the loop-induced corrections that comes from the stop sector does not need to be large as in the MSSM. As well known, the
mass of $h_1$ also receives sizable loop corrections, in particular
from the top--stop sector \cite{Okada:1990vk, Haber:1990aw}.

The CP-odd sector contains one pseudoscalar Higgs boson $A^{0}$ and two Goldstone bosons $G_{Z}^{0}$ and $G_{Z'}^{0}$ which are absorbed respectively to the physical massive neutral vector bosons $Z^{0}$ and $Z_{2}$ after the breaking of the electroweak gauge symmetry. The Higgs mass-squared matrix of this sector $\mathcal{M}_{-}^{0}$ is diagonalized by $Z^{A}$ which gives the mass eigenstates $\left(G_{Z}^{0},G_{Z'}^{0},A^{0}\right)$ as a linear combination of $(\sigma_{d},\sigma_{u},\sigma_{s})$. The tree level mass of the single physical neutral CP--odd state is
\begin{equation}
M_A^2|_{\text{tree}} = \frac {\sqrt{2} T_\lambda} {\sin{2\beta}} v_s 
\left( 1 + \frac {v^2} {4v_s^2} \sin^2{2\beta} \right)\,.    \label{eq 2.17}
\end{equation}
In our sign convention, $\tan\beta$ and $v_s$ are positive in the minimum of the potential; eq.(\ref{eq 2.17}) then implies that $T_\lambda$ must also be positive.

The charged components of the Higgs doublets do not mix with the neutral Higgs fields because of electric charge conservation, so the charged Higgs sector of the UMSSM is the same as the one in the MSSM. It consists of one charged Higgs boson $H^{+}$ and one charged Goldstone boson $G^{+}_{W}$ that enters as the longitudinal polarization of the $SU(2)_{L}$ charged vector boson $W^{+}$ after the breaking of the electroweak gauge symmetry. The mass-squared matrix of the charged Higgs is diagonalized by $Z^{+}$ which writes the mass eigenstates $\left(G_{W}^{+},H^{+}\right)$ as a linear combination of $(H_{d}^{+},H_{u}^{+})$. The mass of the physical charged Higgs boson at tree level reads
\begin{equation}
M_{H^{+}}^{2}|_{\text{tree}}= M_{W^{+}}^{2}+ \frac{\sqrt{2}  T_{\lambda}}{\sin{2\beta}} v_{s}-\frac{\lambda^{2}}{2}v^{2}. \label{eq 2.175}
\end{equation}
As in the MSSM $M_A^2$ differs from the squared mass of the physical charged Higgs boson only by terms of order $v^2$. Both $A$ and $H^\pm$ are constructed from the components of $H_u$ and $H_d$, without any admixture of $s$.

As noted above, in the limit $v_s \gg v$ the mixing between singlet and
doublet states can to first approximation be neglected. Here we chose
the heaviest state to be (mostly) singlet. From the last eq.(\ref{eq:subeqns})
and eq.(\ref{eq 2.10}) we derive the important result
\begin{equation} \label{ms}
M^2_{Z'}|_{\text{tree}} \simeq M_{h_3}^2|_{\text{tree}} + {\cal O}(v^2)\,.
\end{equation}
Here we have assumed $|T_\lambda| \leq v_s$ because for larger values
of $|T_\lambda|$ the mass of the heavy doublet Higgs can exceed the
mass of the singlet state. As we will see, in the region of parameter
space that minimizes the $\tilde \nu_{R,1}$ relic density we need
$M_A < M_{\tilde \nu_{R,1}}$.

Eq.(\ref{ms}) leads to an $h_3 - Z'$ mass splitting of order
$M^2_Z/M_{Z'}$, which is below $1$ GeV for $M_{Z'} > 10$ TeV. Loop
corrections induce significantly larger mass splittings, with
$M_{Z'} > M_{h_3}$; however, the splitting still amounts to less than
$1\%$ in the relevant region of parameter space, which is well below
the typical kinetic energy of WIMPs in the epoch around their
decoupling from the thermal bath. We thus arrive at the important
result that $M_{\tilde \nu_{R,1}} \simeq M_{Z'}$ {\em automatically}
implies $M_{\tilde \nu_{R,1}} \simeq M_{h_3}$ in our set--up, so that
$\tilde \nu_{R,1}$ annihilation is enhanced by {\em two} nearby
resonances.

\section{Sfermions}

\indent

In the UMSSM, the $U(1)'$ gauge symmetry induces some new D-term contributions to all the sfermion masses, which modify the diagonal components of the MSSM sfermion mass matrices as

\begin{equation} 
\Delta_{F}=\frac{1}{2} g'^{2} Q'_{F} \Big(Q'_{H_d} v_{d}^{2}  + Q'_{H_u} v_{u}^{2}  + Q'_{S} v_{s}^{2} \Big), \label{eq 2.18}
\end{equation}
where $F\in \{Q,L,D^{c},U^{c},E^{c},N^{c}\}$. This D-term contribution can dominate the sfermion mass for large values of $v_{S}$. Moreover, depending on the value of $\theta_{E_{6}}$ this term can induce positive or negative corrections to the sfermion masses. As can be seen in Figure \ref{Fig1}, for $\text{arctan}\sqrt{15}<\theta_{E_{6}}<\frac{\pi}{2}$ all the sfermion masses receive positive corrections in such a way that the corrections to the RH sneutrino masses are the smallest ones, while for $0<\theta_{E_{6}}<\text{arctan}\sqrt{15}$ the RH sneutrino masses receive large negative corrections compared to the corrections that the other sfermion masses receive in the same range. On the other hand, for $\theta_{E_{6}}<0$ the RH sneutrino masses receive positive corrections larger than the corrections that the other sfermion masses acquire from the D-term. As a consequence, the lightest RH sneutrino can easily be the lightest sfermion of the model when $\theta_{E_{6}}>0$, but when $\theta_{E_{6}}<0$ it needs to have a negative soft mass-squared parameter in order to (possibly) become the LSP and behave as a dark matter particle.

The tree--level sneutrino mass matrix written in the basis 
$\left(\tilde{\nu}_L, \tilde{\nu}_R\right)$ is

\begin{equation} 
\mathcal{M}^2_{\tilde{\nu}} = \left( 
\begin{array}{cc}
{\bf m}_{\tilde{\nu}_L\tilde{\nu}_L^*}^{2} &-\frac{1}{2} v_d v_s \lambda {\bf Y_{\nu}^{*}}  + \frac{1}{\sqrt{2}} v_u {\bf T_{\nu}^{*}} \\ 
-\frac{1}{2} v_d v_s \lambda {\bf Y}_{\nu}^{T}  + \frac{1}{\sqrt{2}} v_u {\bf T}_{\nu}^{T}  &{\bf m}_{\tilde{\nu}_R\tilde{\nu}_R^*}^{2}\end{array} 
\right).  \label{eq 2.19}
\end{equation} 
The $3 \times 3$ sub--matrices along the diagonal are given by: 
\begin{subequations} 
\begin{align} 
{\bf m}_{\tilde{\nu}_L\tilde{\nu}_L^*}^{2} &= \Big[\Delta_{L} + \frac{1}{8}\Big(g_{1}^{2} + g_{2}^{2}\Big)\Big(v_{d}^{2}-v_{u}^{2}\Big)\Big]{\bf 1}+\frac{1}{2} v_{u}^{2} {{\bf Y}_{\nu}^{*} {\bf Y}_{\nu}^{T}}+ {\bf m^{2}_{\tilde{L}}}\\ 
{\bf m}_{\tilde{\nu}_R\tilde{\nu}_R^*}^{2} &= \Delta_{N^{c}}{\bf 1} + \frac{1}{2} v_{u}^{2} {{\bf Y}_{\nu}^{T} {\bf Y}_{\nu}^{*}}  + {\bf m^{2}_{\tilde{N}^{c}}},
\end{align}                  
\end{subequations}
where $g_1$ and $g_2$ are the $U(1)_Y$ and $SU(2)_L$ gauge couplings,
respectively. As noted earlier, the neutrino Yukawa couplings have to
be very small. We therefore set ${\bf Y_\nu} = {\bf T_\nu} = 0$, so
that the $6 \times 6$ matrix (\ref{eq 2.19}) decomposes into two
$3 \times 3$ matrices.\footnote{Strictly speaking some neutrino Yukawa
  couplings have to be nonzero in order to generate the required
  sub--eV neutrino masses. However, the $\tilde \nu_L - \tilde \nu_R$
  mixing induced by these tiny couplings is completely negligible for
  our purposes.} Since all interactions of the $\tilde \nu_R$ fields
are due to $U(1)'$ gauge interactions which are the same for all
generations, we can without loss of generality assume that the matrix
${\bf m^2_{\tilde{N}^C}}$ of soft breaking masses is diagonal. The
physical masses of the RH sneutrinos are then simply given by
$m^2_{\tilde \nu_{R,i}} = m^2_{\tilde N^C_i} + \Delta_{N^C}$. Our LSP
candidate is the lightest of the three $\tilde \nu_R$ states, which we
call $\tilde \nu_{R,1}$.


\section{Neutralinos}

\indent

The neutralino sector is formed by the fermionic components of the
neutral vector and Higgs
supermultiplets. So, in addition to the neutralino sector of the MSSM,
the UMSSM has another gaugino state associated with the $U(1)'$
gauge symmetry and a singlino state that comes from the extra
scalar supermultiplet $\hat{S}$. The neutralino mass matrix written in
the basis
$\left(\lambda_{\tilde{B}}, \tilde{W}^0, \tilde{H}_d^0, \tilde{H}_u^0,
  \tilde{S}, \lambda_{\tilde{B}'}\right)$ is:
\begin{equation}  \label{eq 2.21}
\mathcal{M}_{\tilde{\chi}^0} = \left( 
\begin{array}{cccccc}
M_1 &0 &-\frac{1}{2} g_1 v_d  &\frac{1}{2} g_1 v_u  &0 &0\\ 
0 &M_2 &\frac{1}{2} g_2 v_d  &-\frac{1}{2} g_2 v_u  &0 &0\\ 
-\frac{1}{2} g_1 v_d  &\frac{1}{2} g_2 v_d  &0 &- \mu_{\rm eff}  &
- \frac{1}{\sqrt{2}} v_u \lambda  &g' Q'_{H_d} v_d \\ 
\frac{1}{2} g_1 v_u  &-\frac{1}{2} g_2 v_u  &- \mu_{\rm eff}  &0 &
- \frac{1}{\sqrt{2}} v_d \lambda  &g' Q'_{H_u} v_u \\ 
0 &0 &- \frac{1}{\sqrt{2}} v_u \lambda  &- \frac{1}{\sqrt{2}} v_d \lambda  
&0 &g' Q'_{S} v_s \\ 
0 &0 &g' Q'_{H_d} v_d  &g' Q'_{H_u} v_u  &g' Q'_{S} v_s  &M_4\end{array} 
\right)\,.   
\end{equation} 
This matrix is diagonalized by a unitary 6 $\times$ 6 matrix $N$ which
gives the mass eigenstates (in order of increasing mass)
$\tilde{\chi}^0_1, \, \tilde{\chi}^0_2, \, \tilde{\chi}^0_3,
\tilde{\chi}^0_4,\, \tilde{\chi}^0_5,\, \tilde{\chi}^0_6$
as a linear combinations of the current eigenstates. We have ignored a
possible (gauge invariant) mixed $\tilde B \tilde B'$ mass term
\cite{Suematsu:1997qt, Suematsu:1997au}. The properties of the neutralino sector in $E_{6}$ inspired SUSY models have been analysed in \cite{Cvetic:1997ky}\cite{Keith:1996fv,Keith:1997zb,Gherghetta:1996yr,Hesselbach:2001ri,Choi:2006fz}, in particular as considering the neutralino LSP of the UMSSM as a viable dark matter candidate \cite{Kalinowski:2008iq}\cite{deCarlos:1997yv,Barger:2004bz}.

Note that the singlet higgsino (singlino for short) $\tilde S$ and the
$U(1)'$ gaugino $\tilde B'$ mix strongly, through an entry of order
$v_s$. On the other hand, these two new states mix with the MSSM only
through entries of order $v$. Therefore the eigenvalues of the
lower--right $2 \times 2$ submatrix in eq.(\ref{eq 2.21}) are to good
approximation also eigenvalues of the entire neutralino mass
matrix. Note that the smaller of these two eigenvalues decreases with
increasing $M_4$. Requiring this eigenvalue to be larger than
$M_{\tilde \nu_{R,1}} \simeq M_{Z'}/2$ therefore implies
\begin{equation} \label{neut1}
| M_4 | < \frac{3}{2} M_{Z'}\,.
\end{equation}
Moreover, the smallest mass of the MSSM--like states should also be larger
than $M_{Z'}/2$, which implies
\begin{equation} \label{neut2}
|M_1| > \frac{1}{2} M_{Z'}; \ |M_2| > \frac{1}{2} M_{Z'}; \
|\lambda| > \frac{1}{\sqrt{2}} |Q_S g'| \,.
\end{equation}
We have used eqs.(\ref{eq 1.4}) and (\ref{eq 2.10}) in the derivation of
the last inequality.

Since the $B'$ boson and the singlet Higgs ${\bf \hat{S}}$ supermultiplets are electromagnetically neutral they do not contribute to form the chargino spectrum. Hence, the chargino sector of the UMSSM is identical to that of the MSSM.





\section{Conclusions}  
\indent

In this chapter we analyzed the UMSSM, i.e. extensions of the minimal
supersymmetrized Standard Model that contain an additional $U(1)'$
gauge group as well as three additional right--handed (RH) neutrino
superfields which are singlets under the SM gauge group but carry
$U(1)'$ charge. We assume that $U(1)'$ is a subgroup of $E_6$, which
has been suggested as an (effective) GUT group, e.g. in the context of
early superstring phenomenology. In comparison to the MSSM, the field content of the UMSSM contains additionally three RH neutrino superfields, one gauge boson associated to the extra Abelian gauge symmetry and an SM singlet scalar field whose VEV is responsable for the breaking of the $U(1)'$ symmetry.

In the UMSSM the lightest RH sneutrino $\tilde \nu_{R,1}$ can be a good dark matter candidate. It can annihilate into lighter particles through some s-channel processes that are mediated by the Higgs bosons and by the new massive gauge boson. If the RH sneutrino has a mass close to half the mass of the mediator its annihilation cross section can be largely increased and this allows the WIMP candidate to produce the observed dark matter relic density with much higher masses. We will see in the next chapter how heavy can the lightest RH sneutrino be in such a way that it still behaves as a viable thermal dark matter candidate.

\chapter{A Very Heavy Sneutrino as Viable Thermal Dark Matter Candidate in $U(1)'$ Extensions of the MSSM}    
\label{ChapterUPPERLIMIT}
\indent

\section{Introduction}
\indent


It is interesting to ask how heavy a WIMP can be. As long as
no positive WIMP signal has been found, an upper bound on the WIMP
mass can only be obtained within a specific production mechanism,
i.e. within a specific cosmological model. In particular, nonthermal
production from the decay of an even heavier, long--lived particle can
reproduce the correct relic density for any WIMP mass, if the mass,
lifetime and decay properties of the long--lived particle are chosen
appropriately \cite{Gelmini:2006pw}. Here we stick to standard
cosmology, where the WIMP is produced thermally from the hot gas of SM
particles. The crucial observation is that the resulting relic density
is inversely proportional to the annihilation cross section of the WIMP
\cite{kt}. It has been known for nearly thirty years that the unitarity
limit on the WIMP annihilation cross section leads to an upper bound on
its mass \cite{Griest:1989wd}. Using the modern determination of the
DM density \cite{Ade:2015xua},
\begin{equation} \label{relden}
\Omega_{\rm DM} h^2 = 0.1188\pm 0.0010\,,
\end{equation}
the result of \cite{Griest:1989wd} translates into the upper bound
\begin{equation} \label{unitarity}
m_\chi \leq 120 \ {\rm TeV}\,.
\end{equation}

While any elementary WIMP $\chi$ has to satisfy this bound, it is not very
satisfying. Not only is the numerical value of the bound well above the
range that can be probed even by planned colliders; a particle that 
interacts so strongly that the annihilation cross section saturates the
unitarity limit can hardly be said to qualify as a WIMP. In order to put
this into perspective, let us have a look at the upper bound on the WIMP
mass in specific models. Since WIMPs have non--negligible
interactions with SM particles, they can be searched for in a variety
of ways. Direct WIMP search experiments look for the recoil of a
nucleus after elastic WIMP scattering. These experiments have now
begun to probe quite deeply into the parameter space of many WIMP
models \cite{Tanabashi:2018oca, Aprile:2018dbl}. The limits from these experiments
are strongest for WIMP masses around 30 to 50 GeV. For lighter WIMPs
the recoil energy of the struck nucleus might be below the
experimental threshold, whereas the sensitivity to heavier WIMPs
suffers because their flux decreases inversely to the mass.

An $SU(2)$ non--singlet WIMP can annihilate into $SU(2)$ gauge bosons
with full $SU(2)$ gauge strength. For a spin$-1/2$ fermion and using
tree--level expressions for the cross section, this will reproduce the
desired relic density (\ref{relden}) for $m_\chi \simeq 1.1$ TeV for a
doublet (e.g., a higgsino--like neutralino in the MSSM
\cite{Edsjo:1997bg}); about $2.5$ TeV for a triplet (e.g., a
wino--like neutralino in the MSSM \cite{Edsjo:1997bg}); and $4.4$ TeV
for a quintuplet \cite{Cirelli:2005uq}. Including large one--loop
(``Sommerfeld'') corrections increases the desired value of the quintuplet
mass to about $9.6$ TeV \cite{Cirelli:2009uv}.

One way to increase the effective WIMP annihilation cross section is
to allow for co--annihilation with strongly interacting particles
\cite{Griest:1990kh}. Co--annihilation happens if the WIMP is close in
mass to another particle $\chi'$, and reactions of the kind
$\chi + f \leftrightarrow \chi' + f'$, where $f, f'$ are SM
particles, are not suppressed. In this case $\chi \chi'$ and
$\chi'\chi'$ annihilation reactions effectively contribute to the
$\chi$ annihilation cross section. If $\chi'$ transforms
non--trivially under $SU(3)_C$, the $\chi' \chi'$ annihilation cross
section can be much larger than that for $\chi \chi$ initial
states. On the other hand, $\chi'$ then effectively also counts as
Dark Matter, increasing the effective number of internal degrees of
freedom of $\chi$. For example, in the context of the MSSM,
co--annihilation with a stop squark \cite{Boehm:1999bj} can allow even
$SU(2)$ singlet (bino--like) DM up to about $3.3$ TeV
\cite{Harz:2018csl}, or even up to $\sim 6$ TeV if the mass splitting
is so small that the lowest stoponium bound state has a mass below
twice that of the bino \cite{Biondini:2018pwp}. Co--annihilation with
the gluino \cite{Profumo:2004wk} can put this bound up to $\sim 8$ TeV
\cite{Ellis:2015vaa}. Very recently it has been pointed out that
nonperturbative co--annihilation effects after the QCD transition
might allow neutralino masses as large as $100$ TeV if the mass splitting
is below the hadronic scale \cite{Fukuda:2018ufg}; the exact value of the
bound depends on non--perturbative physics which is not well under
control.

The WIMP annihilation cross section can also be greatly increased if
the WIMP mass is close to half the mass of a potential $s-$channel
resonance $R$. Naively this can allow the cross section to
(nearly) saturate the unitarity limit, if one is right on resonance.
In fact the situation is not so simple \cite{Griest:1990kh}, since the
annihilation cross section has to be thermally averaged: because WIMPs
still have sizable kinetic energy around the decoupling temperature,
this average smears out the resonance. In the MSSM the potentially
relevant resonances for heavy WIMPs are the heavy neutral Higgs
bosons; in particular, neutralino annihilation through exchange of the
CP--odd Higgs $A$ can occur from an $S-$wave initial state
\cite{Drees:1992am}. However, the neutralino coupling to Higgs bosons
is suppressed by gaugino--higgsino mixing; it will thus only be close
to full strength if the higgsino and gaugino mass parameters are {\em
  both} close to $M_A/2$.

While the new Higgs superfield $\hat S$ is a singlet under the SM
gauge group, it is charged under $U(1)'$. This forbids an $\hat S^3$
term in the superpotential. Hence the quartic scalar interaction of
this field is determined uniquely by its $U(1)'$ charge. As a result,
the mass of the physical, CP--even Higgs boson $h_3$ is {\em
  automatically} very close to that of the $Z'$ boson, in the relevant
limit $M_{Z'} \gg M_Z$. Hence for
$M_{\tilde \nu_{R,1}} \simeq M_{Z'}/2$ the annihilation cross section
of the lightest right--handed sneutrino $\tilde \nu_{R,1}$ is enhanced
by {\em two} resonances. Out of those, the exchange of $h_3$ is more
important since it can be accessed from an $S-$wave initial state.
For a complex scalar, $Z'$ exchange is accessible only from a $P-$wave
initial state, which suppresses the thermally averaged cross section.
Notice that the $h_3 \tilde \nu_{R,i} \tilde \nu^*_{R,i}$ coupling
contains terms that are proportional to the VEV of $s$, which sets the
scale of the $Z'$ mass; for $M_{\tilde \nu_{R,1}} \simeq M_{Z'}/2$
this dimensionful coupling therefore does not lead to a suppression of
the cross section. Finally, the couplings of $h_3$ to the doublet
Higgs bosons can be tuned by varying a trilinear soft breaking
term. This gives another handle to maximize the thermally averaged
$\tilde \nu_{R,1}$ annihilation cross section in the resonance region.

\section{Minimizing the Relic Abundance of the 
Right-Handed Sneutrino}   \label{section3}
\indent

As described in the Introduction, we want to find the upper bound on
the mass of the lightest RH sneutrino $\tilde \nu_{R,1}$ from the
requirement that it makes a good thermal WIMP in standard cosmology.
As well known \cite{kt}, under the stated assumptions the WIMP relic
density is essentially inversely proportional to the thermal average
of its annihilation cross section into lighter particles; these can be
SM particles or Higgs bosons of the extended sector. The upper bound
on $M_{\tilde \nu_{R,1}}$ will therefore be saturated for combinations
of parameters that maximize the thermally averaged
$\tilde \nu_{R,1} \tilde \nu_{R,1}^*$ annihilation cross section. 

All relevant couplings of the RH sneutrinos are proportional to
the $U(1)'$ gauge coupling $g'$. In particular, two RH sneutrinos can
annihilate into two neutrinos through exchange of a $U(1)'$ gaugino.
This, and similar reactions where one or both particles in the initial
and final state are replaced by antiparticles, are typical electroweak
$2 \rightarrow 2$ reactions without enhancement factors. They will
therefore not allow RH sneutrino masses in the multi--TeV range.

In contrast, $\tilde \nu_{R,1} \tilde \nu_{R,1}^*$ annihilation
through $Z'$ and scalar $h_3$ exchange can be resonantly enhanced if
$M_{\tilde \nu_{R,1}} \simeq M_{Z'}/2$; recall that
$M_{h_3} \simeq M_{Z'}$ is automatic in our set--up, if $h_3$ is
mostly an SM singlet, as we assume. Note that the $Z'$ exchange can
only contribute if the sneutrinos are in a $P-$wave. This suppresses
the thermal average of the cross section by a factor $\geq 7$. For
comparable couplings, $h_3$ exchange, which is depicted in
Fig.~\ref{Fig2}, is therefore more important.

\begin{figure}[h!]
\centering
\includegraphics[width=0.4\textwidth]{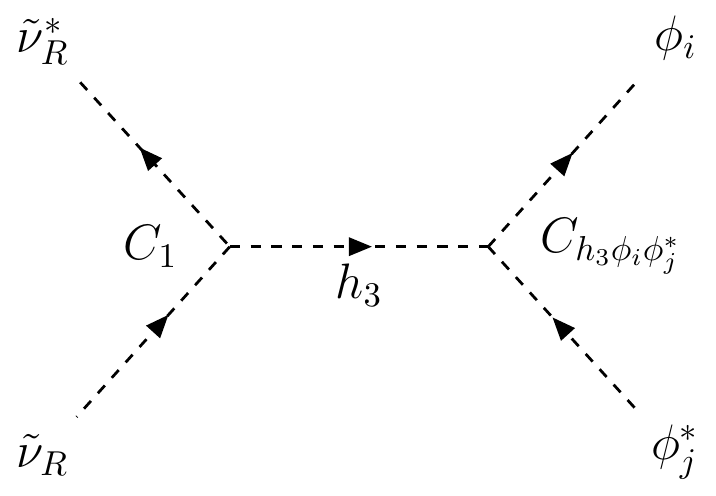}
\caption{Main annihilation process for the annihilation of RH
  sneutrinos. The final state can contain both physical Higgs
  particles and the longitudinal components of the weak $W$ and $Z$
  gauge bosons, which are equivalent to the corresponding would--be
  Goldstone modes.} \label{Fig2}
\end{figure}

In the $h_3$ resonance region the annihilation cross section scales like
\begin{equation} \label{eq 3.1}
\sigma_{\rm ann} \propto \frac {(Q'_{N^C})^2} {M^2_{\tilde{\nu}_{R,1}}}\, .  
\end{equation}
Since the $h_3 \tilde \nu_{R,1} \tilde \nu_{R,1}^*$ coupling, denoted
by $C_1$ in Fig.~\ref{Fig2}, originates from the $U(1)'$ $D-$term, it
is proportional to the product of $S$ and $N^C$ charges:
\begin{equation}  \label{eq 3.3}
\left| C_1 \right| \simeq  g'^2 \left| Q'_{N^C} Q'_S v_s \right| \,.
\end{equation}
These charges are determined uniquely once the angle $\theta_{E_6}$
has been fixed. The denominator in eq.(\ref{eq 3.1}) results from
dimensional arguments, using the fact that there is essentially only
one relevant mass scale once the resonance condition has been
imposed.\footnote{The couplings $C_1$ and $C_{h_3 \phi_i \phi_j^*}$ in
  Fig.~\ref{Fig2} carry dimension of mass. They are dominated by the
  VEV $v_s$, which is proportional to $M_{Z'} \simeq M_{h_3}$, and
  hence to $M_{\tilde \nu_{R,1}}$ if the resonance condition is
  satisfied.}

Note that near the resonance the annihilation cross section is
effectively only ${\cal O}(\alpha')$, not ${\cal O}(\alpha'^2)$, where
$\alpha' = g'^2/(4\pi)$. The $\tilde \nu_{R,1} \tilde \nu_{R,1}^*$
annihilation cross section is then larger than typical
co--annihilation cross sections, if the latter are not resonantly
enhanced. Even co--annihilation with a superparticle that can also
annihilate resonantly (e.g., a higgsino--like neutralino) will not
increase the effective annihilation cross section, but will increase
the effective number of degrees of freedom per dark matter particle
$g_\chi$. As a result, we find that co--annihilation {\em reduces} the
upper bound on $M_{\tilde \nu_{R,1}}$. For example, if all three RH
sneutrinos have the same mass, the upper bound on this mass decreases
by a factor of $\sqrt{3}$, since the annihilation cross section has to
be increased by a factor of $3$ in order to compensate the increase of
$g_\chi$. We therefore require that the lightest neutralino is at
least $20\%$ heavier than $\tilde \nu_{R,1}$.

As noted earlier, the initial--state coupling $C_1$ in Fig.~\ref{Fig2}
is essentially fixed by $\theta_{E_6}$. The upper bound on
$M_{\tilde \nu_{R,1}}$ for given $\theta_{E_6}$ can therefore be found
by optimizing the final state couplings. We find that the relic
density is minimized if the effective final state coupling $C_2$,
defined more precisely below, is of the same order as $C_1$. This can
be understood as follows. For much larger values of $C_2$ the width of
$h_3$ increases, which reduces the cross section. On the other hand,
since the peak of the thermally averaged cross section is reached for
$M_{\tilde \nu_{R,1}}$ slightly below $M_{h_3}/2$
\cite{Griest:1990kh},
$h_3 \rightarrow \tilde \nu_{R,1} \tilde \nu_{R,1}^*$ decays are
allowed, and dominate the total $h_3$ width if $C_2 \ll C_1$; in this
case increasing $C_2$ will clearly increase the cross section,
i.e. reduce the relic density.

The only sizable couplings of the singlet--like Higgs state $h_3$ to
particles with even $R-$parity (i.e., to particles possibly lighter
than the LSP $\tilde \nu_{R,1}$) are to members of the Higgs doublets.
$h_3$ couples to $H_u$ and $H_d$ through the $U(1)'$ $D-$term, with
contributions $\propto g'^2 Q'_S Q'_{H_u, H_d} v_s$; through $F-$terms
associated to the coupling $\lambda$, with contributions
$\propto \lambda^2 v_s$; and through a trilinear soft breaking term,
with contributions $\propto T_\lambda$. In the decoupling limit
$M_A^2 \gg M_Z^2$ the relevant couplings are given by:
\begin{eqnarray}
C_{h_3 H^+ H^-} \simeq C_{h_3 h_2 h_2 } \simeq C_{h_3 A A} \simeq &-& i \Big[ 
g'^2 \left( \cos^2\beta Q'_{H_u} + \sin^2\beta Q'_{H_d} \right) Q'_Sv_s   
\nonumber \\
&+& v_s \lambda^2 + \frac{\sin(2\beta)} {\sqrt{2}} T_\lambda \Big]\,;
 \label{eq 3.4}    \\
C_{h_3 G^+ G^-} \simeq C_{h_3 h_1 h_1} \simeq C_{h_3 G^0 G^0 } \simeq &-&i  \Big[ 
g'^2 \left( \sin^2\beta Q'_{H_u} + \cos^2\beta Q'_{H_d} \right) Q'_S v_s  
\nonumber \\
& + & v_s \lambda^2 - \frac{\sin(2\beta)}{\sqrt{2}} T_\lambda \Big]\,;
 \label{eq 3.5} \\
C_{h_3 H^+ G^-} \simeq C_{h_3 h_2 h_1} \simeq C_{h_3 A G^0} \simeq &-&i \Big[ 
g'^2 \frac{ \sin(2\beta)} {2} \left( Q'_{H_u} - Q'_{H_d} \right) Q'_S v_s 
\nonumber \\
& - & \frac{\cos(2\beta)}{\sqrt{2}} T_\lambda \Big]\,. \label{eq 3.6}
\end{eqnarray}
Since $M_{h_3} \gg v$, at scale $M_{h_3}$ $SU(2)_{L}$ is effectively
unbroken. The couplings of $h_3$ to two members of the heavy doublet
containing the physical states $H^\pm, h_2$ and $A$ therefore are
all the same, see eq.(\ref{eq 3.4}), as are the couplings to the light
doublet containing $h_1$ and the would--be Goldstone modes $G^0$ and
$G^\pm$, see eq.(\ref{eq 3.5}); finally, eq.(\ref{eq 3.6}) describes
the common coupling to one member of the heavy doublet and one member
of the light doublet. Of course, the would--be Goldstone modes are not
physical particles; however, again since $M_{h_3} \gg v$ the
production of physical longitudinal gauge bosons can to very good
approximation be described as production of the corresponding
Goldstone states. This is the celebrated equivalence theorem
\cite{Cornwall:1974km}.\footnote{Due to the effective restoration
of $SU(2)_{L}$ at scale $M_{h_3}$ the total decay width of $h_3$, which
determines the total annihilation cross section via $h_3$ exchange,
can still be computed from eqs.(\ref{eq 3.4}) to (\ref{eq 3.6}) even
if the decoupling limit is not reached; the dependence on the
mixing between the CP--even states drops out after summing over
all final states.}

We find numerically that the $\tilde \nu_{R,1}$ relic density is minimized
when $h_3$ decays into two members of the heavy Higgs doublet are
allowed. From eqs.(\ref{eq 2.17}) and (\ref{eq 2.175}) we see that
this requires
\begin{equation}  \label{eq 3.2}
\frac{ \sqrt{2}  T_\lambda v_s } {\sin{2\beta}} <  
\frac{1}{4} g'^2 \left( Q'_S \right)^2 v_s^2  \quad\quad\quad \Rightarrow 
\quad\quad\quad T_\lambda < \frac {g'^2 \left( Q'_S \right)^2 
\sin{2\beta}} {4 \sqrt{2}} v_s\,.
\end{equation}
This implies that the singlet--like state is indeed the heaviest physical
Higgs boson. 

The contribution of the RH sneutrino annihilation channels that appear in Fig.~\ref{Fig2} to obtain acceptable relic densities is also affected by the decay width of the singlet Higgs. In order to take this into account, we define an effective coupling squared which is the sum over all annihilation channels of the product of $\vert C_{h_{3}\phi_{i}\phi_{j}} \vert^{2}$ and the kinematic square-root factor that appears in the decay width $\Gamma_{h_{3}\phi_{i}\phi_{j}}$. It is given by

\begin{eqnarray}
C_{2}^{2} &= & \left(\vert C_{h_{3}H^{+}H^{-}} \vert^{2} \sqrt{1-\frac{4 M_{H^{+}}^{2}}{M_{h_{3}}^{2}}}\right)+\frac{1}{2} \left(\vert C_{h_{3}h_{2}h_{2}} \vert^{2} \sqrt{1-\frac{4 M_{h_{2}}^{2}}{M_{h_{3}}^{2}}}\right)+ \frac{1}{2} \left(\vert C_{h_{3}A A} \vert^{2} \sqrt{1-\frac{4 M_{A}^{2}}{M_{h_{3}}^{2}}}\right) \nonumber \\
& + & \left(\vert C_{h_{3}G^{+}G^{-}} \vert^{2} \sqrt{1-\frac{4 M_{W^{+}}^{2}}{M_{h_{3}}^{2}}}\right)+ \frac{1}{2}\left(\vert C_{h_{3}h_{1}h_{1}} \vert^{2} \sqrt{1-\frac{4 M_{h_{1}}^{2}}{M_{h_{3}}^{2}}}\right)+ \frac{1}{2} \left(\vert C_{h_{3}G^{0}G^{0}} \vert^{2} \sqrt{1-\frac{4 M_{Z}^{2}}{M_{h_{3}}^{2}}}\right)  \nonumber \\
& + & 2\left(\vert C_{h_{3}H^{+}G^{-}} \vert^{2} \sqrt{1-\frac{(M_{H^{+}}+M_{W^{+}})^{2}}{M_{h_{3}}^{2}}}\sqrt{1-\frac{(M_{H^{+}}-M_{W^{+}})^{2}}{M_{h_{3}}^{2}}}\right)  \nonumber \\
& + & \left(\vert C_{h_{3}h_{2}h_{1}} \vert^{2} \sqrt{1-\frac{(M_{h_{2}}+M_{h_{1}})^{2}}{M_{h_{3}}^{2}}}\sqrt{1-\frac{(M_{h_{2}}-M_{h_{1}})^{2}}{M_{h_{3}}^{2}}}\right) \nonumber \\
& + & \left(\vert C_{h_{3}A G^{0}} \vert^{2} \sqrt{1-\frac{(M_{A}+M_{Z})^{2}}{M_{h_{3}}^{2}}}\sqrt{1-\frac{(M_{A}-M_{Z})^{2}}{M_{h_{3}}^{2}}}\right).     \nonumber   
\end{eqnarray}
Since we are working in the s-decoupling limit, we have $M_{A}\approx M_{H^{+}}\approx M_{h_{2}}$ and we can neglect the masses of $W^{+}, Z^{0}$ and $h_{1}$. We can now define a simplified effective final--state coupling $C_2$ for the diagram
shown in Fig.~\ref{Fig2}:
\begin{equation} \label{eq 3.8}
C_2 = \sqrt{ 2 \left| C_{h_3 h_2 h_2} \right|^2 \sqrt{ 1 - 
\frac{4 M_{h_2}^2} {M_{h_3}^2} } + 2 \left| C_{h_3 h_1 h_1} \right|^2
+ 4 \left| C_{h_3 h_2 h_1} \right|^2 \left( 1 - \frac{ M_{h_2}^2} {M_{h_3}^2} 
\right)} \,.
\end{equation}
Here we have included the kinematic factors into the effective
coupling, using the same mass $M_{h_2}$ for all members of the heavy
Higgs doublet and ignoring $M_{h_1}, M_W$ and $M_Z$, which are much
smaller than $M_{h_3}$. The numerical coefficients originate from
summing over final states: $H^+ H^-, \, A A$ and $h_2 h_2$ for the
first term, where the last two final states get a factor $1/2$ for
identical final state particles; $G^+ G^-, \, G^0 G^0$ and $h_1 h_1$
for the second term, again with factor $1/2$ in front of the second
and third contribution; and $G^+ H^-,\, G^- H^+,\, G^0 A$ and
$h_1 h_2$ for the third term.

Since the contribution from $h_3$ exchange is accessible from an
$S-$wave initial state, it peaks for DM mass very close to $M_{h_3}/2$
where one needs quite small velocity to get exactly to the pole
$s=M_{h_3}^{2}$; at such a small velocity, the $Z'$ exchange
contribution, which can only be accessed from a $P-$wave initial
state, is quite suppressed. As a consequence, near the peak of the
thermally averaged total cross section the $h_3$ exchange processes
always contributes more than $90\%$ to the total, whereas the $Z'$
exchange contribution shrinks as we approach the peak. The latter
reaches its maximum at a larger difference between $M_{Z'}$ and
$2M_{\tilde \nu_{R,1}}$, but its contribution exceeds $10\%$ of the
total only if $2M_{\tilde \nu_{R,1}}$ is at least $3\%$ below $M_{Z'}$,
or else above the resonance. Note also that the annihilation into
pairs of SM fermions via $Z'$ exchange is completely determined by
$\theta_{E_6}$. In principle we could contemplate annihilation into
exotic fermions, members of $\mathbf{27}$ of $E_6$ that are
required for anomaly cancellation, as noted in Sec.~\ref{section5.2}. However, the contribution from the SM fermions already sums to an effective final
state coupling which is considerably larger than the initial state
coupling; this helps to explain why the $Z'$ contribution is always
subdominant. Adding additional final states therefore reduces the $Z'$
exchange contribution to the $\tilde \nu_{R,1}$ annihilation cross
section even further. This justifies our assumption that the exotic
fermions are too heavy to affect the calculation of the
$\tilde \nu_{R,1}$ relic density.

Finally, all other processes of the model contribute at most $1\%$ to
the thermally averaged total cross section in the resonance region.
This shows that the parameters that describe the rest of the spectrum
are irrelevant to our calculation, as long as $\tilde \nu_{R,1}$ is
the LSP and sufficiently separated in mass from the other
superparticles to avoid co--annihilation. These parameters were
therefore kept fixed in the numerical results presented below.


\section{Numerical Results}
\indent

We are now ready to present numerical results. We will first describe
our procedure. Then we discuss two choices for $\theta_{E_6}$, i.e.
for the $U(1)'$ charges, before generalizing to the entire range of
possible values of this mixing angle.

\subsection{Procedure}
\label{subsection4.1}
\indent

We have used the {\tt Mathematica} package {\tt SARAH} \cite{Staub:2008uz,
  Staub:2013tta, Staub:2015kfa} to generate routines for the precise
numerical calculation of the spectrum with {\tt SPheno}
\cite{Porod:2003um, Porod:2011nf}. This code calculates by default the
pole masses of all supersymmetric particles and their corresponding
mixing matrices at the full one--loop level in the
$\overline{{\rm DR}}$ scheme. {\tt SPheno} also includes in its
calculation all important two--loop corrections to the masses of
neutral Higgs bosons \cite{Goodsell:2014bna, Goodsell:2015ira,
  Braathen:2017izn}.  The dark matter relic density and the dark
matter nucleon scattering cross section relevant for direct detection
experiments are computed with {\tt MicrOMEGAs-4.2.5}
\cite{Belanger:2014vza}. The mass spectrum generated by {\tt SPheno}
is passed to {\tt MicrOMEGAs-4.2.5} through the SLHA+ functionality
\cite{Belanger:2010st} of {\tt CalcHep} \cite{Pukhov:2004ca,
  Belyaev:2012qa}. The numerical scans were performed by combining the
different codes using the {\tt Mathematica} tool {\tt SSP}
\cite{Staub:2011dp} for which {\tt SARAH} already writes an input
template.

{\tt SARAH} can generate two different types of templates that can be
used as input files for {\tt SPheno}. One is the high scale input,
where the gauge couplings and the soft SUSY breaking parameters are
unified at a certain GUT scale and their renormalization group (RG)
evolution between the electroweak, SUSY breaking and GUT scale is
included. The other one is the low scale input where the gauge
couplings, VEVs, superpotential and soft SUSY breaking parameters of
the model are all free input parameters that are given at a specific
renormalization scale near the sparticle masses, in which case no RG
running to the GUT scale is needed. In this template the SM gauge
couplings are given at the electroweak scale and evolve to the SUSY
scale through their RGEs. The dark matter phenomenology of a model in
the WIMP context is usually well studied at low energies; moreover,
acceptable low energy phenomenology for both the $U(1)_{\psi}$ and the
$U(1)_{\eta}$ model in the limit where the singlet Higgs decouples
works much better with nonuniversal boundary conditions
\cite{Langacker:1998tc}. Finally, a bound that is valid for general
low--scale values of the relevant parameters will also hold (but can
perhaps not be saturated) in constrained scenarios.

In our work we therefore define the relevant free parameters of the
UMSSM directly at the SUSY mass scale, which is defined as the
geometric mean of the two stop masses. We created new model files for
different versions of the UMSSM to be used in {\tt SARAH} and {\tt
  SPheno} where all the $U(1)'$ charges are written in terms of the
$U(1)$ mixing angle $\theta_{E_6}$ using eq.(\ref{eq 1.1}).

Our goal is to find the upper bound on the mass of the lightest RH
sneutrino, and therefore on $M_{Z'} \simeq M_{h_3}$. We argued in
Sec.~\ref{section3} that co--annihilation would weaken the bound. We
therefore have to make sure that all other superparticles are
sufficiently heavy so that they do not play a role in the calculation
in the relic density. The precise values of their masses are then
irrelevant to us. We therefore fix the soft mass parameters of the
gauginos and sfermions to certain values well above
$M_{\tilde \nu_{R,1}}$; recall from eq.(\ref{neut1}) that this implies
an {\em upper} bound on the mass $M_4$ of the $U(1)'$ gaugino. As
noted in Sec.~\ref{section5.2} we set ${\bf Y_\nu} = 0$, since the small
values of the neutrino masses force them to be negligible for the
calculation of the relic density. We also set most of the scalar
trilinear couplings to zero, except the top trilinear coupling $T_t$
which we use together with $\tan\beta$ and $M_3$ to keep the SM Higgs
mass in the range $125 \pm 3$ GeV, where the uncertainty is dominated
by the theory error \cite{Carena:2013ytb}. Since we are interested in
superparticle masses in excess of $10$ TeV, the correct value of
$M_{h_1}$ can be obtained with a relatively small value of
$\tan\beta$, which we also fix.

As already noted in the previous Section, all relevant interactions
of $\tilde \nu_{R,1}$ scale (either linearly or quadratically) with 
the $U(1)'$ gauge coupling $g'$. Since our set--up is inspired by
gauge unification, we set this coupling equal to the $U(1)_Y$ coupling
in GUT normalization, i.e.
\begin{equation} \label{gprime}
g' = \sqrt{ \frac{5}{3}} g_1\,.
\end{equation}
Note also that the charges in Table~1 are normalized such that
$\sum \left( Q'_\psi \right)^2 = \sum \left( Q'_\chi \right)^2 =
\frac{3}{5} \sum Y^2$, where the sum runs over a complete 
$\mathbf{27}-$dimensional representation of $E_6$ \cite{London:1986dk}.
We will later comment on how the upper bound on $M_{\tilde \nu_{R,1}}$
changes when $g'$ is varied.

Recalling that we work in a basis where the matrix
${\mathbf m}^2_{\tilde N^C}$ is diagonal, with $m^2_{\tilde N^C,11}$
being its smallest element, the remaining relevant free parameters are
thus:
\begin{equation} \label{eq 4.1}
m^2_{\tilde{N}^C, 11}, \quad v_s, \quad  \lambda, \quad T_\lambda \quad \text{and} 
\quad  \theta_{E_6}\,.
\end{equation}
All these parameters are related to the extended sector that the UMSSM
has in addition to the MSSM. Since the mixing angle $\theta_{E_6}$
defines the $U(1)'$ gauge group, we want to determine the upper bound
on the mass of the lightest RH sneutrino as a function of
$\theta_{E_6}$. We will see below that this will also allow to derive
the absolute upper bound, valid for all versions of the UMSSM.

From the discussion of the previous Section we know that the first two
of the parameters listed in (\ref{eq 4.1}) are strongly correlated by
the requirement that $M_{\tilde \nu_{R,1}}$ is close to
$M_{Z'}/2$. More precisely, the minimal relic density is found if the
RH sneutrino mass is very roughly one $h_3$ decay width below the
nominal pole position, the exact distance depending on the couplings
$C_1$ and $C_2$; this shift from the pole position is due to the
finite kinetic energy of the sneutrinos at temperatures around the
decoupling temperature \cite{Griest:1990kh}.

The parameters $\lambda$ and $T_\lambda$ have to satisfy some bounds. First,
requiring the mass of the $SU(2)_{L}$ higgsinos to be at least $20\%$ larger
than $M_{Z'}/2$ leads to the lower bound
\begin{equation} \label{eq 4.2}
\lambda > 0.85  g' |Q'_S| \,,
\end{equation}
where we have used eqs.(\ref{eq 1.4}) and (\ref{eq 2.10}). Moreover,
$T_\lambda$ has to satisfy the upper bound (\ref{eq 3.2}), so that pairs of
the heavy $SU(2)_{L}$ doublet Higgs bosons can be produced in $\tilde \nu_{R,1}$
annihilation with $M_{\tilde \nu_{R,1}} \simeq M_{Z'}/2$. Having fixed $\tan\beta$ and $T_\lambda$, the effective final state
coupling $C_2$ defined in eq.(\ref{eq 3.8}) depends only on $\lambda$,
which is constrained by eq.(\ref{eq 4.2}); fortunately this still leaves
us enough freedom to vary $C_2$ over a sufficient range.

The bound on the lightest RH sneutrino mass for a given value of
$\theta_{E_6}$ can then be obtained as follows. We start by choosing
some value of $M_{h_3} \simeq M_{Z'}$ in the tens of TeV range. Note
that this fixes the coupling $C_1$, since we have already fixed $g'$
and $\theta_{E_6}$ and hence the charge $Q'_{N^C}$. We then minimize
the relic density for that value of $M_{h_3}$ by varying the
soft--breaking contribution to the sneutrino mass and $\lambda$; as
noted in Sec.~3, the minimum is reached when the physical RH sneutrino
mass is just slightly below $M_{Z'}/2$, and $C_2$ is close to the initial state
coupling $C_1$ of eq.(\ref{eq 3.3}). If the resulting relic density
$(\Omega h^2)_1$ is very close to the measured value of
eq.(\ref{relden}), we have found the upper bound on $M_{Z'}$ and hence
on $M_{\tilde \nu_{R,1}}$. Otherwise, we change the value of
$M_{h_3}$ by the factor $\sqrt{0.12 / (\Omega h^2)_1}$, and repeat the
procedure. Since the minimal relic density to good approximation
scales like $M_{h_3}^2$, see eq.(\ref{eq 3.1}), this algorithm
converges rather quickly.

\subsection{The $U(1)_{\psi}$ Model}  
\label{subsection4.2}

We illustrate our procedure first for $U(1)' = U(1)_\psi$, where the
$U(1)'$ charge of the RH sneutrinos is relatively small (in fact, the
same as for all SM (s)fermions). We choose the SUSY breaking scale to
be 18 TeV and we fix $\tan{\beta}=1.0$, $M_3=18$ TeV, and
${\bf m^2_{\tilde Q}} = {\bf m^2_{\tilde U^C}} = {\bf m^2_{\tilde
    D^C}} = 2 \times 10^8$ GeV$^2\cdot {\bf 1}$,
${\bf m^2_{\tilde L}}={\bf m^2_{\tilde E^C}} = 2.25 \times 10^{8}$
GeV$^2 \cdot {\bf 1}$,
$\left(m^2_{\tilde N^C}\right)_{22} = 2.2 \times 10^8$ GeV$^2$,
$\left(m^2_{\tilde N^C}\right)_{33} = 2.3 \times 10^{8}$ GeV$^2$. To
keep $M_{h_{1}}$ close to 125 GeV, the top trilinear coupling took
values in the following range ${T_{u,33}}= [-55, -33]$ TeV; recall
that the physical squared sfermion masses also receive $D-$term
contributions, which amount to $M^2_{Z'}/8$ in this model.

In this model the two Higgs doublets have the same $U(1)'$ charge, and
the product $Q'_{H_u} Q'_S$ is negative. As a result, the $\lambda^2$
and the $g'^2$ terms in the diagonal couplings given in eqs.(\ref{eq
  3.4}) and (\ref{eq 3.5}) tend to cancel, while the contribution
$\propto g'^2$ to the off--diagonal couplings given in eq.(\ref{eq
  3.6}) vanishes.  The contributions involving these off--diagonal
couplings are therefore subdominant. The largest contribution usually
comes from final states involving two heavy $SU(2)_{L}$ doublet Higgs
bosons, but the contributions from two light states (including the
longitudinal modes of the gauge bosons) are not much
smaller. Moreover, due to this cancellation we need relatively large
values of $\lambda$; the numerical results shown below have been
obtained by varying it in the range from $0.32$ to $0.46$.

Figure \ref{Fig3}a depicts the relic abundance of the RH sneutrino as
a function of $M_{\tilde{\nu}_{R,1}}$ for different values of the mass
of the singlet Higgs boson. All the curves show a pronounced minimum
when $M_{\tilde{\nu}_{R,1}}$ is very close to but below
$M_{h_{3}}/2$. The blue and the green curves are for $v_s = 59$ TeV
and thus have the same coupling $C_1$ and (approximately) the same
mass of the singlet Higgs, but the blue curve has a smaller value of
$C_2$. This reduces the width of $h_3$ as well as the annihilation
cross section away from the resonance, and therefore leads to a
narrower minimum.

\begin{figure}[h!]
\begin{subfigure}[b]{0.5\linewidth}
\centering\includegraphics[width=8.2cm, height=6cm]{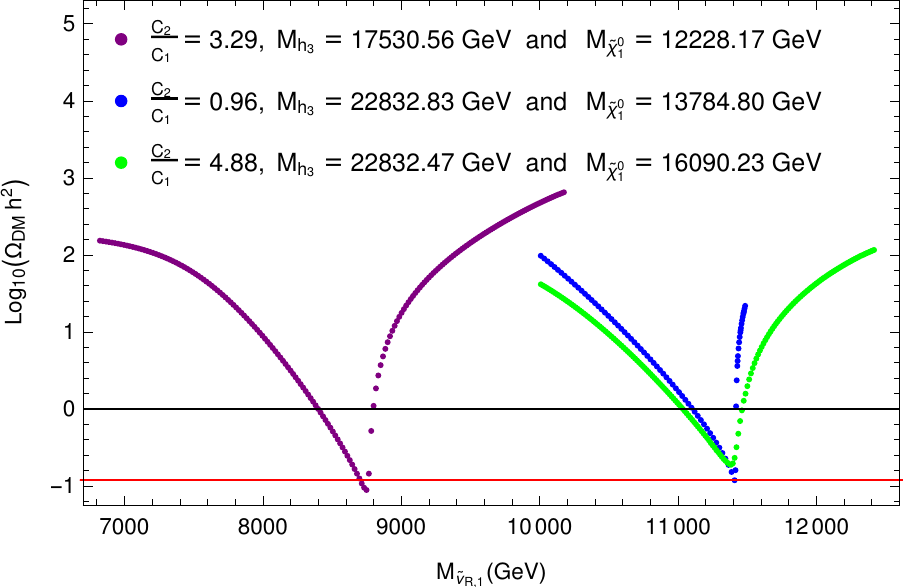}
\caption{}
\end{subfigure}\hfill
\begin{subfigure}[b]{0.58\linewidth}
\centering\includegraphics[width=7.0cm, height=6cm]{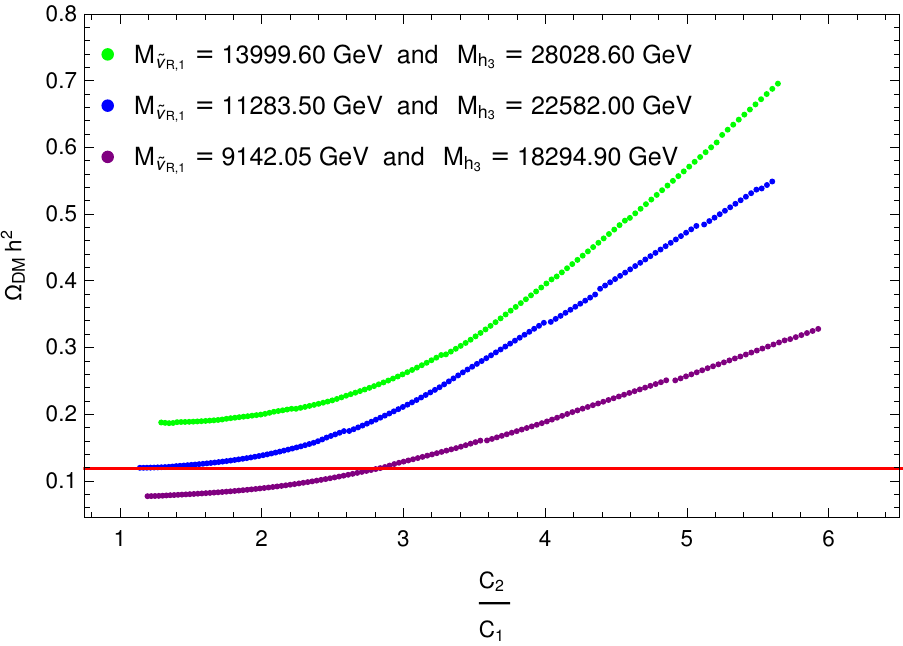}
\caption{}
\end{subfigure}   
\caption{Relic density as a function of
  $M_{\tilde{\nu}_{R,1}}$ (left) and its dependence on the ratio of
  couplings $\frac{C_{2}}{C_{1}}$ (right) for different singlet Higgs
  masses. The red lines correspond to the limits on the dark matter
  abundance obtained by the Planck Collaboration,
  $\Omega_{\text{DM}} h^2 = 0.1188\pm 0.0010$.
\label{Fig3}}
\end{figure}

In figure \ref{Fig3}b we show the dependence of the relic density on
the ratio of couplings $C_2 / C_1$ for fixed mediator masses close to
the resonance. This confirms our expectations from the previous
Section: if $C_2$ is significantly larger than $C_1$, the relic
density increases with $C_2$ because the increase of the mediator
decay width over--compensates the increased coupling strength in the
total annihilation cross section. If $C_{2}\ll C_{1}$ the width of the
mediator is dominated by mediator decays into
$\tilde \nu_{R,1} \tilde \nu^*_{R,1}$; hence increasing $C_2$ reduces
the relic density because it increases the normalization of the
annihilation cross section. Note that the relic density curve is
fairly flat over some range of $C_2 / C_1$. Moreover, the optimal
choice of $C_2/C_1$ also depends somewhat on how far
$M_{\tilde \nu_{R,1}}$ is below $M_{h_3}/2$. Altogether, for given
$M_{h_3}$ there is an extended $1-$dimensional domain in the
$(M_{\tilde \nu_{R,1}}, C_2/C_1)$ plane over which the relic density is
quite close to its absolute minimum. This simplifies our task of
minimization. Note also that we calculate the annihilation cross
section only at tree--level; a change of the predicted relic density
that is smaller than a couple of percent is therefore not really
physically significant.

The parameters of the blue curve in Fig.~\ref{Fig3}b in fact are
very close to those that maximize $M_{\tilde \nu_{R,1}}$ within the
$U(1)_\psi$ model, under the assumption that $\tilde \nu_{R,1}$ was in
thermal equilibrium in standard cosmology.
$M^{\rm max}_{\tilde{\nu}_{R,1}} \simeq 11.5$ TeV corresponds to an
upper bound on $M_{h_3}$ and $M_{Z'}$ of about $23.0$ TeV. This is
clearly beyond the reach of the LHC, and might even stretch the
capabilities of proposed $100$ TeV $pp$ colliders.

Recall that all left--handed SM (anti)fermions have the same
$U(1)_\psi$ charge. As a result, in the absence of $Z - Z'$ mixing the
$Z' f \bar f$ couplings are purely axial vector couplings, for all SM
fermions $f$. $Z'$ exchange can therefore only contribute to
spin--dependent WIMP--nucleon scattering in this model. Since our WIMP
candidate doesn't have any spin, $Z'$ exchange does not contribute at
all. Once $Z-Z'$ mixing is included, $Z$ exchange contributes a term
of order
$M_{\tilde \nu_{R,1}} M_N \sin \alpha_{ZZ'} / M_Z^2 \propto M_{\tilde
  \nu_{R,1}} M_N / M_{Z'}^2$
to the matrix element for $\tilde \nu_{R,1} N$ scattering, while the
mixing--induced $Z'$ exchange contribution is suppressed by another
factor $M_Z^2 / M_{Z'}^2$; here $M_N$ is the mass of the
nucleon. There is also a small contribution from the light SM--like
Higgs boson $h_1$, which is very roughly of order $M_N^2 /
M_{h_1}^2$. As a result the scattering cross section on nucleons is very small,
below $10^{-13}$ pb for the scenario that maximizes
$M_{\tilde \nu_{R,1}}$. For the given large WIMP mass, this is not
only several orders of magnitude below the current bound, but also
well below the background from coherent neutrino scattering
(``neutrino floor'').

\subsection{The $U(1)_\eta$ Model} 
\label{subsection4.3}

We now consider a value of $\theta_{E_6}$ with a larger $U(1)'$ charge
of the right--handed neutrino superfields. This increases the coupling
$C_1$ for given $M_{Z'}$, and thus the $\tilde \nu_{R,1}$ annihilation
cross section for given masses, which in turn will lead to a weaker
upper limit on $M_{\tilde \nu_{R,1}}$ from the requirement that the
$\tilde \nu_{R,1}$ relic density not be too large.

In our analysis we therefore choose the SUSY breaking scale to be 50
TeV and we fix $\tan\beta = 2.2$, and
${\bf m^2_{\tilde Q}} = 1.28 \times 10^9$
GeV$^{2} \cdot{\bf 1}, {\bf m^2_{\tilde U^C}} = 1.45 \times 10^9$
GeV$^{2} \cdot{\bf 1}$, ${\bf m^2_{\tilde D^C}} = 3.0 \times 10^9$
GeV$^{2} \cdot {\bf 1}$, ${\bf m^2_{\tilde L}} = 3.0 \times 10^9$
GeV$^{2} \cdot{\bf 1}$, ${\bf m^2_{\tilde E^C}} = 1.28 \times 10^9$
GeV$^{2} \cdot {\bf 1}$,
$\left(m^2_{\tilde N^C}\right)_{22} = -4.0 \times 10^8$ GeV$^2$,
$\left(m^2_{\tilde N^C}\right)_{33} = -3.9 \times 10^8$ GeV$^2$. To
keep $M_{h_{1}}$ close to 125 GeV, the top trilinear coupling took
values in the range ${T_{u,33}}= [-130, -114]$ TeV.  In this case the
$U(1)'$ $D-$term contributions are positive for
$\tilde Q, \, \tilde u^C$ and $\tilde \nu_R$, but are negative for
$\tilde L$ and $\tilde d^C$.

\begin{figure}[h!]
\begin{subfigure}[b]{0.5\linewidth}
\centering\includegraphics[width=7.8cm, height=6cm]{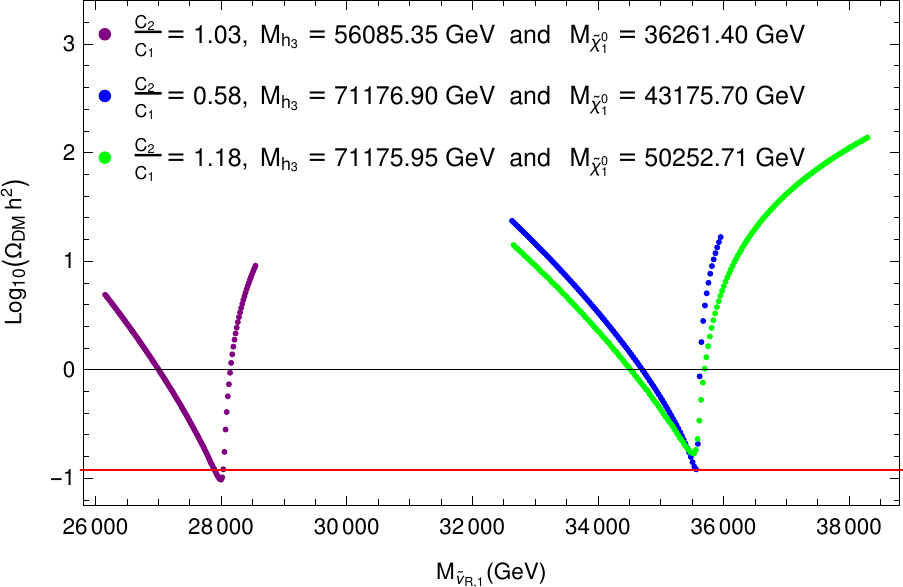}
\caption{}
\end{subfigure}\hfill
\begin{subfigure}[b]{0.55\linewidth}
\centering\includegraphics[width=7.0cm, height=6cm]{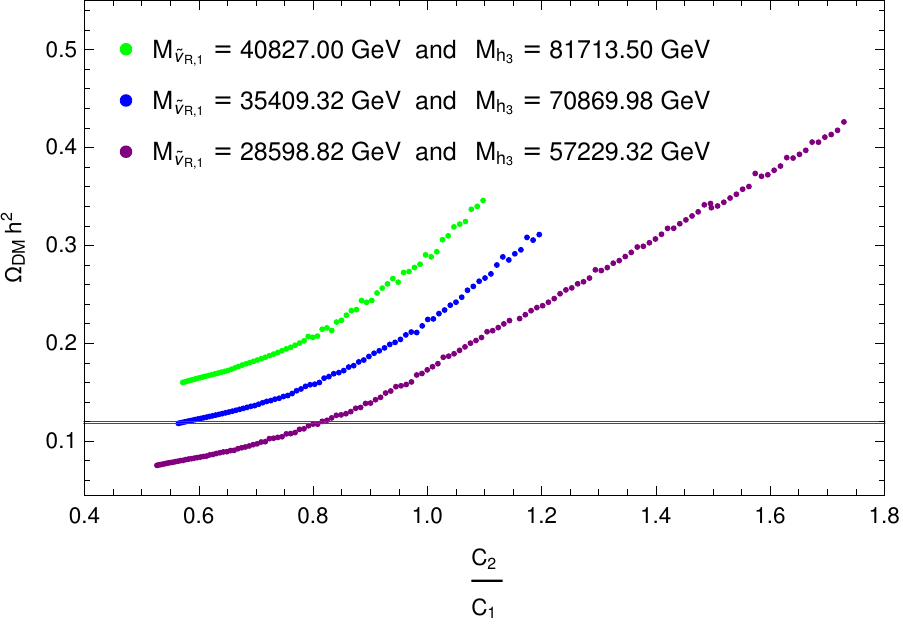}
\caption{}
\end{subfigure}   
\caption{As in Fig.~\ref{Fig3}, but for the $U(1)_\eta$ model. \label{Fig4}}
\end{figure}

In this model the two Higgs doublets have different $U(1)'$ charges;
hence there is a sizable gauge contribution to the off--diagonal
couplings of eq.(\ref{eq 3.6}). The Higgs doublet charges again have
the opposite sign as the charge of $S$, leading to cancellations
between the $\lambda^2$ and $g'^2$ terms in the diagonal couplings
(\ref{eq 3.4}) and (\ref{eq 3.5}). This cancellation is particularly
strong for the coupling to two light states, so that for the
interesting range of $\lambda$ the most important final states involve
two heavy $SU(2)_{L}$ doublets, although final states with one light and
one heavy boson are also significant. Partly because of this, and partly
because the coefficients of the $g'^2$ terms are smaller than in the
$U(1)_\psi$ model, smaller values of the coupling $\lambda$ are
required; the numerical results below have been obtained with
$\lambda \in [0.260, 0.352]$.
  
In Fig.~\ref{Fig4} we again show the dependence of the relic density on
the mass of the lightest RH sneutrino (left) and on the ratio of couplings
$C_2/C_1$ (right). The qualitative behavior is similar to that in
the $U(1)_\psi$ model depicted in Fig.~\ref{Fig3}, but clearly much larger
values of $M_{\tilde \nu_{R,1}}$ are now possible, the absolute upper bound
being near $35$ TeV (see the blue curves). The corresponding $Z'$ mass
of about $70$ TeV is definitely beyond the reach of a $pp$ collider 
operating at $\sqrt{s} = 100$ TeV

Since $Q'_Q = Q'_{U^C} \neq Q'_{D^C}$ in this model, there is no
vector coupling of the $Z'$ to up quarks, but such a coupling does
exist for down quarks. Hence now the $Z'$ exchange contribution to the
matrix element for elastic scattering of $\tilde \nu_{R,1}$ on
nucleons is comparable to that of $Z$ exchange once $Z - Z'$ mixing
has been included, and the $h_1$ exchange contribution has roughly the
same size as in the $U(1)_\psi$ model. The total $\tilde \nu_{R,1} N$
scattering cross sections are again below $10^{-13}$ pb, for
parameters near the upper bound on $M_{\tilde \nu_{R,1}}$. Since our
WIMP candidate is now even heavier than in the $U(1)_\psi$ model, this
is even more below the current constraints as well as below the
neutrino floor.

\subsection{The General UMSSM}  
\label{subsection4.4}

In this subsection we investigate in more detail how the upper bound
on $M_{\tilde \nu_{R,1}}$ depends on $\theta_{E_6}$. To this extent we
have applied the procedure outlined in subsec.~\ref{subsection4.1},
and applied to two specific $U(1)'$ models in subsecs.~\ref{subsection4.2}
and \ref{subsection4.3}, to several additional $U(1)'$ models, each
with a different value of $\theta_{E_6}$.

\begin{figure}[h!]
\centering
\includegraphics[width=0.8\textwidth]{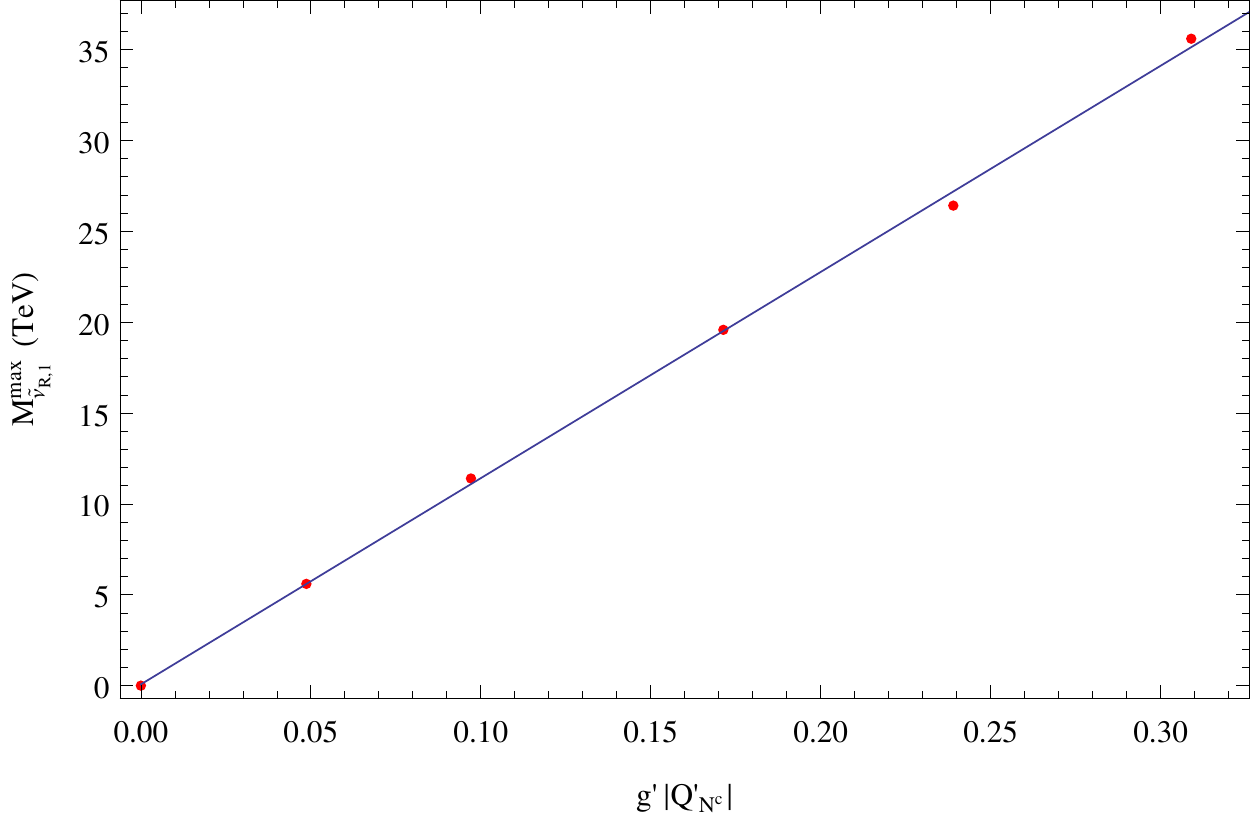}
\caption{The upper limit on $M_{\tilde \nu_{R,1}}$ derived from the
relic density as a function of $\vert Q'_{N^C} \vert$. The straight line shows a linear fit to the six numerical results.
\label{Fig5} }
\end{figure}

The results are shown in Fig.~\ref{Fig5}, where we plot the upper bound
on the mass of the lightest RH sneutrino as a function of the absolute value
of the product $g' Q'_{N^C}$. In order of increasing $\vert Q'_{N^C} \vert$, the 
six red points correspond to the following choices of $\theta_{E_{6}}$:
\[
\\ \biggl\{ \arctan\sqrt{15}, \frac{ \left( \arctan\sqrt{15} + \frac{\pi}{2}
\right) } {2}, \frac{\pi}{2}, \frac{ \left( \arctan\sqrt{\frac{3}{5}} + 
\arctan \left[ \frac{7} {\sqrt{15}} \right] \right) } {2} ,
\arctan\sqrt{\frac{3}{5}}, -\arctan\sqrt{\frac{5}{3}} \biggr\}\,.
\]
Note that the first point has a vanishing $U(1)'$ charge for the $N^C$
superfields, i.e. the resonance enhancement of the annihilation cross
section does not work in this case.  We checked that the cross section
for elastic $\tilde \nu_{R,1} N$ scattering is well below the
experimental bound for all other points.

Evidently the upper bound on $M_{\tilde \nu_{R,1}}$ scales essentially
linearly with $Q'_{N^C}$; recall that $g'$ has been fixed to
$\sqrt{5/3} g_1$ here. This linear dependence can be understood as
follows. The $h_3 \tilde \nu_{R,1} \tilde \nu_{R,1}^*$ coupling can be
written as $g' Q'_{N^C} M_{Z'} \simeq 2 g' Q'_{N^C} M_{\tilde \nu_{R,1}}$.
Moreover, we saw above that the maximal sneutrino mass is allowed if
the effective final--state coupling $C_2$ is similar to $C_1$; it is
therefore also proportional to $Q'_{N^C}$. Therefore at the point
where the bound is saturated, the $h_3$ decay width scales like
$|C_1|^2 / M_{h_3} \propto g'^2 (Q'_{N^C})^2 M_{\tilde \nu_{R,1}}$,
where we have again used that near the resonance all relevant masses
are proportional to $M_{\tilde \nu_{R,1}}$. Note finally that for a
narrow resonance -- such as $h_3$, for the relevant parameter choices
-- the thermal average over the annihilation cross section scales like
$1/(M_{h_3} \Gamma_{h_3})$ \cite{Griest:1990kh}. Altogether we thus have
\begin{equation} \label{scaling}
\langle \sigma v \rangle \propto \frac{ |C_1 C_2|^2 } { M_{h_3}
\Gamma_{h_3} M^4_{\tilde \nu_{R,1}} }
\propto  \frac {g'^2 (Q'_{N^C})^2} {M^2_{\tilde \nu_{R,1}}} \,.
\end{equation}
The linear relation between the upper bound on $M_{\tilde \nu_{R,1}}$
and $Q'_{N^C}$ then follows from the fact that the thermally averaged
annihilation cross section essentially fixes the relic density.

Note that here $Q'_{N^C}$ always comes with a factor $g'$; indeed, for a
$U(1)$ gauge interaction only the product of gauge coupling and charge
is well defined. The linear dependence of the bound on $M_{\tilde \nu_{R,1}}$
on $Q'_{N^C}$ for fixed $g'$ depicted in Fig.~\ref{Fig5} can therefore also be
interpreted as linear dependence of the bound on the product $g' Q'_{N^C}$.
A fit to the points in Fig.~\ref{Fig5} gives:
\begin{equation} \label{fit}
M^{\rm max}_{\tilde \nu_{R,1}} = ( 0.071 + 113.477 g' |Q'_{N^C}| ) \
{\rm TeV}\,.
\end{equation}
This is the central result of this work.

The highest absolute value of $|Q'_{N^{c}}|$ in the UMSSM is about
$0.82$, which is saturated for
$\theta_{E_6} = -\arctan\left[\frac{1}{\sqrt{15}}\right]$. Using the
linear fit of eq.(\ref{fit}) and $g' = \sqrt{5/3} g_1 = 0.47$ leads to
an absolute upper bound on $M_{\tilde \nu_{R,1}}$ in unifiable
versions of the UMSSM of about $43.8$ TeV. This corresponds to an
absolute upper bound on the $Z'$ mass of about $87.6$ TeV.

Finally, we recall from eq.(\ref{th_range}) that for $\theta_{E_6}$
between $-\arctan\sqrt{15}$ and $0$ one needs a negative squared soft
breaking mass in order to have $M_{\tilde \nu_{R,1}} \simeq M_{Z'}/2$.
Since the $\hat N^C$ superfields appear in the superpotential (\ref{eq
  1.3}) only multiplied with the tiny couplings ${\bf Y_\nu}$, this
superpotential will not allow to generate negative squared soft
breaking masses for sneutrinos via renormalization group running
starting from positive values at some high scale. If we insist on
positive squared soft breaking mass for all $\tilde \nu_R$ fields the
upper bound on $|Q'_{N^C}|$ is reduced to $\sqrt{5/8} \simeq 0.79$, in
which case the bound on $M_{\tilde \nu_{R,1}}$ is reduced to about
$42$ TeV. We note, however, that the $\hat N^C$ superfields can have
sizable couplings to some of the exotic color triplets that reside in
the ${\mathbf 27}$-dimensional representation
\cite{Hewett:1988xc}. Recalling that at least some of these exotic
fermions are usually required for anomaly cancellation it should not
be too difficult to construct a UV complete model that allows negative
squared soft breaking terms for (some) $\tilde \nu_R$ at the SUSY mass
scale.

\section{Prospects for Detection}
\label{newsection}

Clearly spectra near the upper bound presented in the previous subsection
are not accessible to searches at the LHC, nor even to a proposed 100 TeV
$pp$ collider.

As already noted for the $U(1)_\eta$ and $U(1)_\psi$ models the
$\tilde \nu_{R,1}$ nucleon scattering cross section is very small. The
very large $Z'$ mass suppresses the $Z'$ exchange contribution; as we saw
in subsec.~\ref{subsectionGaugeBosonsU1} it also suppresses $Z-Z'$ mixing, so that the
$Z$ exchange contribution also scales like $M_{Z'}^{-2}$. The
contribution due to the exchange of the singlet--like Higgs boson
($h_3$ in our analysis) is suppressed by the very large value of
$M_{h_3}$ as well as the tiny $h_3 q \bar q$ couplings, which solely
result from mixing between singlet and doublet Higgs bosons. Finally,
the contribution from the exchange of the doublet Higgs bosons, in
particular of the 125 GeV state $h_1$, is suppressed by the small size
of the $h_1 \tilde \nu_{R,1} \tilde \nu_{R,1}^*$ coupling, which is of
order $g' v \ll M_{\tilde \nu_{R,1}}$, as well as the rather small
$h_1 q \bar q$ couplings, which are much smaller than gauge
couplings. As a result, the $\tilde \nu_{R,1}$ nucleon scattering
cross section, and hence the signal rate in direct WIMP detection
experiments, is well below the neutrino--induced background; recall
that this ``neutrino floor'' increases $\propto M_{\tilde \nu_{R,1}}$
since the WIMP flux, and hence the event rate for a given cross
section, scales $\propto 1/M_{\tilde \nu_{R,1}}$.

The best chance to test these scenarios therefore comes from indirect
detection. Naively one expects the cross section for annihilation from
an $S-$wave initial state to be essentially independent of
temperature, in which case the correct thermal relic density implies
$\langle \sigma v \rangle \simeq 2.4 \cdot 10^{-26} {\rm \text{ cm}}^3/{\rm \text{s}}
\simeq 0.8 {\rm \text{ pb}}\cdot \text{c}$ \cite{Steigman:2012nb, Drees:2015exa}.
However, as pointed out in \cite{Ibe:2008ye, Guo:2009aj} this can
change significantly in the resonance region; here the thermally
averaged annihilation cross section can be significantly higher in
today's universe than at the time of WIMP decoupling.

\begin{figure}[h!]
\centering
\includegraphics[width=0.8\textwidth]{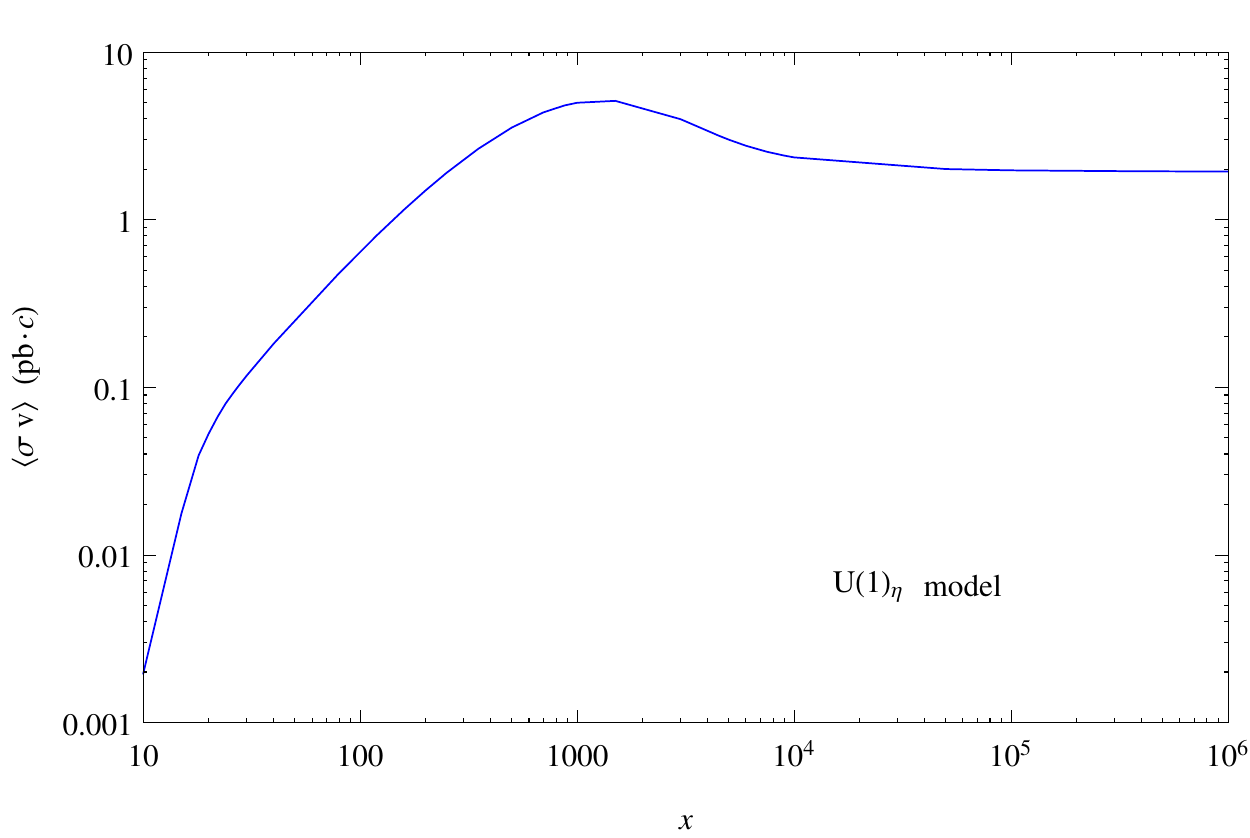}
\caption{Thermally averaged cross section as a function of the scaled
  inverse temperature $x \equiv M_{\tilde \nu_{R,1}}/T$ for the parameters
  of the $U(1)_\eta$ model that saturate the upper bound on the sneutrino
  mass. Nominal decoupling occurs at $x=x_F=27.2$, whereas in today's
  galaxies $x \sim 10^6$.
\label{Fig6} }
\end{figure}

This is illustrated in Fig.~\ref{Fig6} for the parameter choice that
saturates the upper bound on $M_{\tilde \nu_{R,1}}$ in the $U(1)_\eta$
model.  Here we show the thermally averaged
$\tilde \nu_{R,1} \tilde \nu_{R,1}^*$ annihilation cross section times
relative velocity as function of the scaled inverse temperature
$x = M_{\tilde \nu_{R,1}}/T$.\footnote{The total $\tilde \nu_{R,1}$
  annihilation rate also receives a contribution from
  $\tilde \nu_{R,1} \tilde \nu_{R,1} \rightarrow \nu \nu$ annihilation
  via neutralino exchange in the $t-$ and $u-$channels. However, since
  this contribution is not resonantly enhanced, it can safely be
  neglected.} We see that for a quite extended range of temperatures
around the decoupling temperature, $\langle \sigma v \rangle$ grows
almost linearly with $x$. This is because $M_{\tilde \nu_{R,1}}$ is
only slightly below the nominally resonant value $M_{h_3}/2$; by
reducing the temperature the fraction of the velocity distribution
that falls within approximately one $h_3$ decay width of the pole
therefore at first increases.

Today's relic density is essentially inversely proportional to the
``annihilation integral'', defined as \cite{Griest:1990kh}
\begin{equation} \label{J}
  J(x_F) = \int_{x_F}^\infty \frac { \langle \sigma v \rangle } {x^2} dx\,.
\end{equation}
An annihilation cross section that grows significantly for $x > x_F$
therefore has to be compensated by a smaller value of $\langle \sigma
v \rangle(x_F)$ in order to keep the relic density constant. As a result,
in our scenarios the annihilation cross section at decoupling is actually
significantly {\em smaller} than for typical $S-$wave annihilation.

Because for parameters that saturate the upper bound on $M_{\tilde{\nu}_{R,1}}$ the right-handed sneutrino mass is
somewhat below $M_{h_3}/2$, for very large $x$, i.e. very small
temperature, the thermally averaged annihilation cross section starts to
decrease again. However, for the parameters of Fig.~\ref{Fig6} it
asymptotes to a value that is still about three times larger than the
``canonical'' thermal WIMP annihilating from an $S-$wave initial state. As
shown in refs. \cite{Ibe:2008ye, Guo:2009aj} this enhancement factor
strongly depends on $2 M_{\tilde \nu_{R,1}} - M_{h_3}$; it can be even larger
for slightly smaller sneutrino masses that are even closer to $M_{h_3}/2$.

The WIMP annihilation rate in today's universe scales like the square
of the WIMP number density. This means that the flux of annihilation
products scales like $1/M^2_{\tilde \nu_{R,1}}$; for parameters
(nearly) saturating our upper bound on the sneutrino mass it is thus
too small to be detectable by space--based observatories like FermiLAT
\cite{Ackermann:2015zua}, simply because of their small size.  Recall
also that our sneutrinos annihilate into (longitudinal) gauge or Higgs
bosons, and thus mostly into multi--hadron final states. This leads to
a continuous photon spectrum which, for parameters near the upper
bound on the sneutrino mass, extends well into the TeV region. Photons
of this energy can be detected by Cherenkov telescopes on the ground,
via their air showers. Note also that the astrophysical cosmic ray
background drops even faster than $E^{-2}$ with increasing energy $E$
of the cosmic rays; the signal to background ratio therefore actually
improves with increasing WIMP mass. Indeed, simulations show that at
least for a favorable distribution of dark matter particles near the
center of our galaxy, the continuum photon flux of multi--TeV WIMPs
annihilating with the canonical thermal cross section should be
detectable by the Cherenkov Telescope Array \cite{Carr:2015hta}.

\section{Partial conclusions}  
\label{section5}
\indent

We found that even within minimal cosmology, and fixing the $U(1)'$ gauge
strength to be equal to that of the hypercharge interaction of the
(MS)SM (in GUT normalization), $\tilde \nu_{R,1}$ masses of tens of
TeV are possible.  For given $U(1)'$ charges the bound on
$M_{\tilde \nu_{R,1}}$ is saturated if $\tilde \nu_{R,1}$ can
annihilate resonantly through the exchange of both the new $Z'$ gauge
boson and of the new Higgs boson $h_3$ associated with the spontaneous
breaking of $U(1)'$; note that $M_{Z'} \simeq M_{h_3}$ {\em
  automatically} in this model. Scalar $h_3$ exchange is more
important since $Z'$ exchange can only occur from a $P-$wave initial
state.  The $h_3 \tilde \nu_{R,1} \tilde \nu_{R,1}^*$ coupling is
fixed by the $U(1)'$ charge $Q'_{N^C}$ of the right--handed neutrinos,
but the $h_3$ couplings to the relevant final states, involving
$SU(2)_{L}$ doublet Higgs bosons as well as longitudinal $W$ and $Z$
bosons, can be tuned independently, allowing a further maximization of
the annihilation cross section. In our analysis we used $SU(2)_{L}$ doublet Higgs
bosons as well as longitudinal $W$ and $Z$ bosons as final
states. While the light $SU(2)$ doublet Higgs states, including the
longitudinal $W$ and $Z$ modes, are always accessible, we could have
replaced the heavy Higgs doublet in the final state by some exotic
fermions which in most cases are required to cancel anomalies. The
only requirement is that the effective final state coupling of $h_3$
should be tunable to values close to its coupling to
$\tilde \nu_{R,1}$. Since the $Z'$ exchange contribution is basically
fixed by $\theta_{E_6}$, and non--resonant contributions are negligible
for $M_{\tilde\nu_{R,1}} \sim M_{Z'}/2$, most of the many free
parameters of this model, which describe the sfermion and gaugino
sectors, are essentially irrelevant to us. The only requirement is
that these superparticles are sufficiently heavy to avoid
co--annihilation, which would increase the relic density in our case.

\chapter{Conclusions}
\label{FINALCONCLUSIONS}
\indent

In this thesis we have studied many open questions related to new physics that goes beyond the SM. We covered the following topics: topologically massive mediators and spin-dependent potentials, dark matter thermal production in the early Universe and supersymmetric extensions of the SM. Our work was firstly dedicated in finding modifications of the inverse square law possibly related to the existence of a fifth fundamental force and secondly in studying theoretically well-motivated extensions of the SM that are able to solve most of its open issues where we focused on the dark matter problem. Here we give a deeper summary of the main contributions of this thesis.

\section*{Part I: Interparticle Potentials}
\indent

In the first part of this thesis we discussed how macroscopic scalar potentials can be obtained by microscopic interactions between fermionic sources and spin-1 bosons in the limit of low momentum transfer. We calculated the potentials with different classes of couplings for two cases: the well-known Proca case where the mediator obtain its mass via an explicit breaking of the gauge symmetry and the $\left\{A_{\mu}, B_{\nu \kappa} \right\}$-system where a 2-rank tensor and a 4-vector are connected via a topological coupling in such a way that can describe an on-shell massive spin-1 particle that acquires mass without breaking the $U(1)$ gauge symmetry that governs the electromagnetic interactions.

The calculation we have performed is based on the quantum field-theoretical scattering amplitude in the non-relativistic limit, and the potential obtained - which can be interpreted as an operator - is also suitable to be introduced in the Schr\"odinger equation as a time-independent perturbation to the full Hamiltonian. This is a reasonable approach if these corrections are relatively small, which is to be expected, given that the standard quantum mechanical/QED results are in good agreement with experiments.

At last, but not less interesting, one can note that it is possible to assign certain $CP$-transformation properties to the fields $A_{\mu}$ and $B_{\mu\nu}$ so that the topological mass term in eq.\ \eqref{L_B} violates $CP$. This would induce an electric dipole moment (EDM) if we couple our model to fermionic fields. Following the procedure employed by Mantry {\it et al} \cite{Mantry:2014zsa} in the context of axions, one could also use information from the EDM to find further bounds on the coupling constants and the mass of the intermediate spin-1 boson.

\section*{Part II: Supersymmetric Dark Matter}
\indent

In Part II, after reviewing the theoretical framework of SUSY and explaining how both supersymmetric and non-supersymmetric extensions of the SM can be phenomenologically explored in great details using {\tt SARAH} \cite{Staub:2008uz,Staub:2009bi,Staub:2010jh,Staub:2012pb,Staub:2013tta,Staub:2015kfa}, we studied the dark matter phenomenology of the Standard Model singlet (``right-handed'') sneutrino $\tilde \nu_R$ in a class of $U(1)'$ extensions of the Minimal Supersymmetric Standard Model (UMSSM) that originate from the breaking of the $E_6$ gauge group. 

We found that the final upper bound on $M_{\tilde \nu_{R,1}}$ is
essentially proportional to the product $g' |Q'_{N^C}|$, where $g'$ is
the $U(1)'$ gauge coupling. Within the context of theories unifiable
into $E_6$ this leads to an absolute upper bound on
$M_{\tilde \nu_{R,1}}$ of about $43.8$ TeV. This is the highest mass of a good thermal dark matter candidate in standard cosmology that has so far been found in an explicit model. Hence, our central result show that supersymmetry can provide a viable thermal dark matter candidate within standard cosmology with masses of the order of tens of TeV without needing to have co-annihilation with the NLSP. In other words, in this
fairly well motivated set--up we can find a thermal WIMP candidate
with mass less than a factor of three below the bound derived from
unitarity \cite{Griest:1989wd}. This is to be contrasted with an upper
bound on the mass of a neutralino WIMP in the MSSM of about 8 TeV for
unsuppressed co--annihilation with gluinos \cite{Ellis:2015vaa}. In a
rather more exotic model featuring a WIMP residing in the quintuplet
representation of $SU(2)$ a WIMP mass of up to $9.6$ TeV is allowed
\cite{Cirelli:2009uv}. 

Of course, this mechanism requires some amount of finetuning: the mass
of the WIMP needs to be just below half the mass of the $s-$channel
mediator. We find that typically the predicted WIMP relic density
increases by a factor of $2$ when the WIMP mass is reduced by between
$1$ and $3\%$ from its optimal value. In contrast, the recent proposal
to allow thermal WIMP masses near $100$ TeV via non--perturbative
co--annihilation requires finetuning to less than $1$ part in $10^5$
\cite{Fukuda:2018ufg}.

We also note that our very heavy WIMP candidates have very small
scattering cross sections on nuclei, at least two orders of magnitude
below the neutrino floor. This shows that both collider searches and
direct WIMP searches are still quite far away from decisively probing
this reasonably well motivated WIMP candidate. On the other hand, we
argued that indirect signals for WIMP annihilation might be detectable
by future Cherenkov telescopes. Our analysis thus motivates extending
the search for a continuous spectrum of photons from WIMP annihilation
into the multi--TeV range.

While the result (\ref{fit}) has been derived within UMSSM models that
can emerge as the low--energy limit of $E_6$ Grand Unification, it
should hold much more generally. To that end $g' |Q'_{N^C}|$ should be
replaced by $g_{\chi \chi \phi}/m_\phi$, where $\chi$ is a complex
scalar WIMP annihilating through the near resonant exchange of the
real scalar $\phi$, $g_{\chi \chi \phi}$ being the (dimensionful)
$\chi \chi^* \phi$ coupling. In order to saturate our bound the
couplings of $\phi$ to the relevant final states should be tunable
such that the effective final state coupling, which we called $C_2$ in
Sec.~\ref{section3}, should be comparable to the initial--state coupling
$g_{\chi \chi \phi}$. In this case the algorithm we used to find the
upper limit on $M_{\tilde \nu_{R,1}}$, see subsec.~\ref{subsection4.1}, can directly be
applied to finding the upper bound on $M_\chi$. We finally note that
$M_\chi$ can be increased by another factor of $\sqrt{2}$ if $\chi$
is a real scalar.

In this work we also predicted the existence of a new massive gauge boson $Z'$ whose mass is in agreement with the lower limits obtained by the ATLAS searches for $Z'$ signals based on the analysis of dielectron and dimuon final states \cite{Aaboud:2017buh}. Collider searches for heavy $Z'$ bosons have become very popular because they focus on the possible appearance of narrow dilepton resonances. Additionally, Ref.\cite{Duan:2018akc} shows that $U(1)'$ extensions of the MSSM can also provide solutions to the $R_{K}$ and $R_{K^{\ast}}$ anomalies that appear in rare decays of B meson \cite{Aaij:2014ora,Aaij:2017vbb}, which are strongly deviated from the SM predictions and cannot be explained by the MSSM.


\begin{appendices}
\chapter{Currents in the non-relativistic approximation}    \label{AppendixA}
\indent

In the following we present a brief summary of the conventions and main decompositions 
employed in the calculations carried out in the previous Sections.

\section{Basic conventions}
\indent

The basic spinors used to compose the scattering amplitude are the positive energy solutions to the Dirac equation in momentum space \cite{Ryder}, namely
\begin{equation} 
u(p) = \left( \begin{array}{c} \xi \\ \frac{\vec{\sigma} \cdot 
\vec{p}}{2m} \, \xi\end{array} \right) 
\end{equation}
where $ \xi = \left( \begin{array}{c} 1 \\ 0\end{array} \right) $ or 
$ \xi = \left(\begin{array}{c} 0 \\ 1 \end{array} \right) $ for spin-up and -down, respectively. Above we have assumed the non-relativistic limit $E + m \approx 2m$. The orthonormality relation $\xi'^\dagger _r \xi_s = \delta_{rs}$ is supposed to hold and we will usually suppress spinor indices. 

The gamma matrices are chosen as
\begin{equation} 
\gamma^0 = \left( \begin{array}{cc} 1 & 0 \\ 0 & -1 \end{array} \right) 
\, \; \, \text{and} \; 
\, \; \gamma^i = \left( \begin{array}{cc} 0 & \sigma_i \\ - \sigma_i & 0 \end{array} 
\right),
\end{equation}
and the metric and Levi-Civita symbol are defined so that $\eta^{\mu \nu} = \textrm{diag}(+,-,-,-)$ and $\epsilon^{0123} = +1$, respectively. We adopt natural units $\hbar = c = 1$ throughout.


\section{Current decompositions}

\indent

In order to calculate the spin-dependent potentials, it is useful to have the 
non-relativistic limit of the source currents, where we assume 

\begin{enumerate}
\item[{\bf 1)}] $ |\vec{p}|^{2}/m^2  \sim  \mathcal{O}\left(v^2 \right) 
\, \rightarrow \, 0$
\item[{\bf 2)}] Small momentum transfer: $ |\vec{q}|^{2}/m^2 
\, \rightarrow \, 0$
\item[{\bf 3)}] The cross product tends to zero if $|\vec{p}|/m $ and $|\vec{q}|/m$ are small. Energy-momentum conservation implies $\vec{p} \cdot \vec{q} = 0$ 
\end{enumerate}

Here, we show the results of the main 
fermionic currents. We adopt the parametrization for the first current (i.e., first vertex), 
following Fig.\ \eqref{Fig3.1}. We denote the generators of the boosts and rotations by
\begin{equation} 
\Sigma^{\mu \nu} \equiv - \frac{i}{4} \, \left[ \gamma^\mu , \gamma^\nu 
\right],
\end{equation}
and $\langle \sigma_i\rangle \,\equiv \xi'^\dagger \, \sigma_i \, \xi$. In the Dirac representation, $\gamma_5$ is given by
\begin{equation} 
\gamma_5 = \left( \begin{array}{cc}  0 & 1 \\ 1 & 0 \end{array} \right). 
\end{equation}

Making use of the Dirac spinor conjugate, $\bar{u} \equiv u^\dagger \gamma^0$, we have the following set of identities, omitting the coupling constants: 

\begin{enumerate}

\item[1)] Scalar current $(S)$:
\begin{equation} \bar{u}(p + q/2) \, u(p - q/2) \approx \delta \, . \label{scalar_c} 
\end{equation}


\item[2)] Pseudo-scalar current $(PS)$:

\begin{equation}  
\bar{u}(p + q/2) \, i \gamma_5 \, u(p - q/2) =  
- \frac{i}{2m} \, \vec{q} \cdot \langle \vec{\sigma}\rangle\,
 \label{pseudo_scalar_c}  
 \end{equation}


\item[3)] Vector current $(V)$:
\begin{equation} 
\bar{u}(p + q/2) \, \gamma^\mu \, u(p - q/2), 
\end{equation}

\begin{enumerate}


\item[{\bf 3i)}] For $\mu = 0$,

\begin{equation} 
\bar{u}(p + q/2) \, \gamma^0 \, u(p - q/2) \approx  \delta 
\label{vector_c_1} 
\end{equation}
\item[{\bf 3ii)}] For $\mu = i$,
\begin{equation} 
\bar{u}(p + q/2) \, \gamma^i \, u(p - q/2) = 
 \frac{\vec{p}_i}{m} \, \delta - \frac{i}{2m} \, \epsilon_{ijk} \, \vec{q}_j \, 
 \langle \sigma_k \rangle  \label{vector_c_2} 
 \end{equation}
 
\end{enumerate}


\item[4)] Pseudo-vector current $(PV)$:
 \begin{equation} 
 \bar{u}(p + q/2) \gamma^\mu \gamma_5  u(p - q/2)  
 \end{equation}
 
\begin{enumerate}

\item[{\bf 4i)}] For $\mu = 0$,

\begin{equation} 
\bar{u}(p + q/2) \, \gamma^0 \, \gamma_5 \, u(p - q/2) =
  \frac{1}{m} \, \langle \vec{\sigma}\rangle \cdot \vec{p}   
 \label{pseudo_vector_c_1} 
 \end{equation}
 
\item[{\bf 4ii)}] For $\mu = i$,
\begin{equation} 
\bar{u}(p + q/2) \, \gamma^i \, \gamma_5 \, u(p - q/2) \, \approx \,
  \langle \sigma_i\rangle   \label{pseudo_vector_c_2} 
\end{equation}

\end{enumerate}


\item[5)] Tensor current $(T)$:
\begin{equation} 
\bar{u}(p + q/2) \, \Sigma^{\mu \nu} \, u(p - q/2)  
\label{tensor_c_0} 
\end{equation}

\begin{enumerate}

\item[{\bf 5i)}] For $\mu = 0 \, $ and $\, \nu = i \,$, 

\begin{equation} 
\bar{u}(p + q/2) \, \Sigma^{0 i} \, u(p - q/2) = 
  \frac{1}{2m} \, \epsilon_{ijk} \, \vec{p}_j \, \langle \sigma_k\rangle  + 
\frac{i}{4m} \, \delta \, \vec{q}_i  \label{tensor_c_1}   
\end{equation}

\item[{\bf 5ii)}] For $\mu = i \,$ and $ \, \nu = j \,$,
\begin{equation} 
\bar{u}(p + q/2) \, \Sigma^{ij} \, u(p - q/2) \approx  
   - \frac{1}{2} \, \epsilon_{ijk} \langle \sigma_k\rangle    \label{tensor_c_2} \end{equation}
   
\end{enumerate}


\item[6)] Pseudo-tensor current $(PT)$:

\begin{equation} 
\bar{u}(p + q/2) \, i \, \Sigma^{\mu \nu} \, \gamma_5 \, u(p - q/2) 
\end{equation}

\begin{enumerate}

\item[{\bf 6i)}] For $\mu = 0 \, $ and $ \, \nu = i$,

\begin{equation} 
\bar{u}(p + q/2) \, i \, \Sigma^{0i} \, \gamma_5 \, u(p - q/2) \approx  
 \frac{1}{2} \langle \sigma_i\rangle  \label{pseudo_tensor_c_1}  
 \end{equation}
 
\item[{\bf 6ii)}] For $\mu = i \, $ and $ \, \nu = j $

\begin{eqnarray}
\bar{u}(p + q/2) \, i \, \Sigma^{ij} \, \gamma_5 \, u(p - q/2) & = &  \frac{1}{2m} \, \left( \vec{p}_i \langle \sigma_j\rangle - \vec{p}_j \langle \sigma_i\rangle \right)  + \\ \nonumber
& + & \frac{i}{4m} \, \delta \, \epsilon_{ijk}  \, \vec{q}_k  \label{pseudo_tensor_c_2} 
\end{eqnarray}  

\end{enumerate}
\end{enumerate}

In the manipulations above, we have kept the $rs$ indices implicit in the $\delta_{rs}$, as adopted in the main text, pointing out only the particle label. Due to momentum conservation and our choice of reference frame (CM), the second current (or second vertex) can be obtained by performing the changes $q \rightarrow - q$ and $p \rightarrow - p$ in the first one.


\chapter{Spin operators}    \label{AppendixB}
\indent

The spin operators satisfy the following algebra:

\begin{equation} 
\left( P^1_b + P^1_e \right)_{\mu \nu  , \, \rho \sigma } = \frac{1}{2} 
\left(\eta_{\mu \rho} \eta_{\nu \sigma} - \eta_{\mu \sigma} \eta_{\nu \rho} 
\right) \equiv 1_{\mu \nu  , \, \rho \sigma }^{a.s.} \end{equation}

\begin{equation} 
\left( P^1_b \right)_{\mu \nu  , \, \alpha \beta} 
\left( P^1_b \right)^{\alpha \beta}_{  \, \; \; \, , \, \rho \sigma} = 
\left( P^1_b \right)_{\mu \nu  , \, \rho \sigma} \end{equation} 

\begin{equation} 
\left( P^1_e \right)_{\mu \nu  , \, \alpha \beta} 
\left( P^1_e \right)^{\alpha \beta}_{  \, \; \; \, , \, \rho \sigma} = 
\left( P^1_e \right)_{\mu \nu  , \, \rho \sigma} \end{equation} 

\begin{equation} 
\left( P^1_b \right)_{\mu \nu  , \, \alpha \beta} 
\left( P^1_e \right)^{\alpha \beta}_{  \, \; \; \, , \, \rho \sigma} = 0 
\end{equation}

\begin{equation} 
\left( P^1_e \right)_{\mu \nu  , \, \alpha \beta} 
\left( P^1_b \right)^{\alpha \beta}_{  \, \; \; \, , \, \rho \sigma} = 0.
\end{equation}

We notice that the mixing term between $A_\mu$ and 
$B_{\mu \nu}$ introduces a new operator, $S_{\mu \nu \kappa} \equiv \epsilon_{\mu \nu \kappa \lambda} 
\, \partial^\lambda$, which is {\it not} a projector, since
\begin{equation} 
\epsilon^{\mu \nu \alpha \beta} \, A_\mu \partial_\nu B_{\alpha 
\beta} = \frac{1}{2} \left[A^\mu \, S_{\mu \kappa \lambda } B^{\kappa \lambda } 
-  B^{\kappa \lambda } \, S_{\kappa \lambda \mu } \, A^\mu \right], 
\end{equation}
so that we need to study the algebra of $S_{\mu \nu \kappa}$ with the projectors 
$(\ref{proj_B_1})$ and $(\ref{proj_B_2})$, giving us

\begin{equation} S_{\mu \nu \alpha} S^{\alpha \kappa \lambda} = - 2\Box 
\left( P^1_b \right)^{ \, \; \, \; \, \; \kappa \lambda}_{ \mu \nu ,} \end{equation}

\begin{equation} \left( P^1_b \right)_{\mu \nu , \, \alpha \beta } S^{\alpha \beta 
\kappa} = S_{\mu \nu}^{\; \, \; \, \; \kappa} \end{equation}

\begin{equation} S^{\kappa \alpha \beta} 
\left( P^1_b \right)_{\alpha \beta ,}^{  \, \; \; \, \; \, \; \mu \nu} = S^{\kappa \mu 
\nu} \end{equation}

\begin{equation} \left( P^1_e \right)_{\mu \nu , \, \alpha \beta } S^{\alpha \beta 
\kappa} = 0 \end{equation}

\begin{equation} 
S^{\kappa}_{\; \, \; \, \alpha \beta} 
\left( P^1_e \right)^{ \alpha \beta , \, \mu \nu } = 0 \end{equation}

\begin{equation} 
S_{\mu \alpha \beta} S^{\alpha \beta}_{\, \; \, \; \nu} = - 2 \Box 
\theta_{\mu \nu} \label{S_S_theta}.
\end{equation}

The possibility to obtain a closed algebra is not only desirable, but very important, in order to complete the inversion of the matrix in eq.\ \eqref{matrix_op}.

\end{appendices}

\endgroup

\end{document}